\pdfoutput=1
\documentclass[12pt]{article}
\usepackage{graphicx}
\usepackage{amssymb}
\usepackage{amsmath}
\usepackage{amsfonts}

\usepackage{xcolor}
\usepackage{url}
\usepackage{float}
\usepackage{float}

\usepackage{pifont}

\usepackage{lineno}

\usepackage{soul}
\usepackage[english]{babel}
\usepackage[T1]{fontenc}
\usepackage{lmodern}
 \usepackage[utf8]{inputenc}  
          \usepackage{bbm}

\usepackage{hyperref}
\hypersetup{
  colorlinks   = true, 
  urlcolor     = blue, 
  linkcolor    = black, 
  citecolor   = black 
}

\usepackage[makeroom]{cancel}

\usepackage{enumitem}

\usepackage[numbers,sort&compress]{natbib}

\let\OLDthebibliography\thebibliography
\renewcommand\thebibliography[1]{
  \OLDthebibliography{#1}
  \setlength{\parskip}{4pt}
  \setlength{\itemsep}{0pt plus 0.3ex}
}

\usepackage{caption}
\usepackage[showframe=false,hmargin=0.8in,vmargin=1.0in,headheight=0.0in]{geometry}

\usepackage[tight]{units}

\graphicspath{{plots/}}

\newcommand{\gsim}
{\;\raisebox{-.3em}{$\stackrel{\displaystyle >}{\sim}$}\;}

\newcommand\tb{\tan\beta}

\newcommand\LP{\left(}
\newcommand\RP{\right)}

\newcommand\ReDiag{\mathop{
  \raise .5pt\hbox{[}
  \widetilde{\mathrm{Re}}
  \raise .5pt\hbox{]}}}
\newcommand\ReOffDiag{\mathop{
  \raise .5pt\hbox{$\llbracket$}
  \widetilde{\mathrm{Re}}
  \raise .5pt\hbox{$\rrbracket$}}}

\newcommand\MA{M_A}

\newcommand{\mphi}{m_\phi}

\newcommand\ino[1]{\widetilde\chi_{#1}}

\newcommand\neu[1]{\ino{#1}^0}

\newcommand\refeq[1]{Eq.~(\ref{#1})}

\newcommand\refta[1]{Tab.~\ref{#1}}
\newcommand\refse[1]{Sect.~\ref{#1}}

\newcommand\citere[1]{Ref.~\cite{#1}}
\newcommand\citeres[1]{Refs.~\cite{#1}}
\newcommand\refap[1]{App.~\ref{#1}}

\newcommand{\tev}{\,\, \mathrm{TeV}}
\newcommand{\gev}{\,\, \mathrm{GeV}}

\newcommand\fb{\ensuremath{\mbox{fb}}}
\newcommand\ab{\ensuremath{\mbox{ab}}}

\newcommand\ifb{\ensuremath{\,\fb^{-1}}}
\newcommand\iab{\ensuremath{\,\ab^{-1}}}

\newcommand{\br}{\text{BR}}

\newcommand{\sig}{\sigma}

\def\reffi#1{\mbox{Fig.~\ref{#1}}}

\def\Ga{\Gamma}
\def\ga{\gamma}
\def\De{\Delta}

\def\la{\lambda}

\def\gmin2{\ensuremath{(g-2)_\mu}}

\definecolor{Orange}{named}{orange}
\definecolor{Purple}{named}{purple}
\definecolor{Lightblue}{cmyk}{0.9,0.1,0.1,0.3}
\definecolor{dgelborange}{cmyk}{0.,0.3,0.5, 0.}
\definecolor{Lila}{rgb}{0.5,0.,1}
\definecolor{Darkgreen}{rgb}{0.,.7,0.2}

\newcommand{\tautau}{\ensuremath{\tau^+ \tau^-}}

\newcommand{\cattbar}{\ensuremath{c_{\HHOdd t\bar{t}}}}
\newcommand{\chttbar}{\ensuremath{c_{\HHEven t\bar{t}}}}
\newcommand{\ttbar}{\ensuremath{t\bar{t}}}

\newcommand{\HHEven}{\ensuremath{H}}
\newcommand{\HHOdd}{\ensuremath{A}}
\newcommand{\re}{{\mathrm e}}
\newcommand{\chisqttbar}{\ensuremath{\chi^2_{\ttbar}}}
\newcommand{\chisq}{\ensuremath{\chi^2}}

\allowdisplaybreaks
\begin{document}
\thispagestyle{empty}

\def\thefootnote{\fnsymbol{footnote}}

\begin{flushright}
\mbox{}
IFT--UAM/CSIC--21-041\\
DESY 21-132
\end{flushright}

\vspace{0.5cm}

\begin{center}

{\large\sc
{\bf 
Possible indications for new Higgs bosons in the reach of\\[.3em]
the LHC: N2HDM and NMSSM interpretations}}

\vspace{1cm}

{\sc
T.~Biek\"otter$^{1}$\footnote{email: thomas.biekoetter@desy.de},
A.~Grohsjean$^1$\footnote{email: alexander.grohsjean@desy.de},
S.~Heinemeyer$^{2,3}$\footnote{email: Sven.Heinemeyer@cern.ch},\\[.5em]
C.~Schwanenberger$^{1,4}$\footnote{email: christian.schwanenberger@desy.de}~and
G.~Weiglein$^{1,4}$\footnote{email: georg.weiglein@desy.de}
}

\vspace*{.7cm}

{\sl
$^1$Deutsches Elektronen-Synchrotron DESY,
Notkestra{\ss}e 85, 22607 Hamburg, Germany

\vspace*{0.1cm}

$^2$IFT (UAM/CSIC), Universidad Aut\'onoma de Madrid,
Cantoblanco, 28048, Spain

\vspace{0.1cm}

$^3$Campus of International Excellence UAM+CSIC,
Cantoblanco, 28049, Madrid, Spain

\vspace*{0.1cm}

$^4$Universit\"at  Hamburg, Luruper Chaussee 149, D-22761 Hamburg, Germany
}

\end{center}

\vspace*{0.1cm}

\begin{abstract}
\noindent
In several searches for additional Higgs bosons at the LHC,
in particular in a CMS search exploring decays to pairs of top quarks,
$\ttbar$, and in an ATLAS search studying tau leptons, $\tau^+\tau^-$,
local excesses of about
$3\,\sig$ standard deviations or above
have been observed at a mass scale of approximately $ 400 \gev$. 
We investigate to what extent a possible signal in these
channels could be accommodated in the 
Next-to-Two-Higgs-Doublet Model (N2HDM) or the Next-to Minimal
Supersymmetric Standard Model (NMSSM).
In a second step we analyze whether such a model could be compatible
with both a signal at around $400\gev$ and $96\gev$, where the latter possibility is
motivated by observed excesses in searches for the $b \bar b$ final state at 
LEP and the di-photon final state at CMS. 
The analysis for the N2HDM reveals that the observed excesses at $400 \gev$ in the
$\ttbar$ and $\tau^+\tau^-$ channels point towards different regions of
the parameter space, while one such excess and an additional Higgs
boson at around $96\gev$ could simultaneously be accommodated. 
In the context of the NMSSM an experimental confirmation of a signal in the 
$\ttbar$ final state would favour the alignment-without-decoupling limit of the model,
where the Higgs boson at $125\gev$ could be essentially
indistinguishable from the Higgs boson of the
standard model.
In contrast, a signal in the $\tau^+\tau^-$ channel
can only be accommodated outside of this limit,
and parts of the investigated parameter space
could be probed with Higgs signal-rate
measurements at the (HL-)LHC.
\end{abstract}

\def\thefootnote{\arabic{footnote}}
\setcounter{page}{0}
\setcounter{footnote}{0}

\newpage

\section{Introduction}
\label{sec:intro}

In the year 2012 the ATLAS and CMS Collaborations have discovered a new
particle with a mass of about 
$125 \gev$~\cite{Aad:2012tfa,Chatrchyan:2012ufa,Khachatryan:2016vau}.
Within the current experimental and theoretical uncertainties 
the properties of this particle, which in the
following is denoted as $h_{125}$, agree with the predictions for the 
Higgs boson of the Standard Model~(SM) of particle physics, 
but they are also
compatible with the interpretation as a Higgs boson in 
a variety of SM extensions corresponding to different underlying
physics.
Valid parameter regions in scenarios of physics beyond the SM (BSM) featuring 
extended Higgs-boson sectors are established taking into account 
the measurements of Higgs-boson couplings, which are known
experimentally to a precision between $10\%$ and
$30\%$~\cite{CMS:2020xwi,ATLAS:2019nkf}, the existing
limits from searches for additional Higgs bosons, as well as other constraints
(see the discussion below).
Consequently, the question whether the observed scalar boson forms 
part of an extended Higgs sector is one of the science drivers for
the upcoming run of the LHC beginning in 2022
(Run~3), and beyond.

The most obvious possibility for realisations of 
extended Higgs-boson sectors is that all 
additional Higgs bosons have masses that are larger than $125 \gev$. 
But also cases where at least one of the additional Higgs bosons is lighter
than the one at $125 \gev$ are phenomenologically viable.
Thus, searches for BSM Higgs bosons ranging from very low to very
high mass scales are crucial in this context.
Among the models with extended Higgs-boson sectors the most studied ones
are models with two Higgs doublets (2HDM), possibly with an additional
(real or complex) Higgs singlet (N2HDM or 2HDMS), as well as their
supersymmetric (SUSY) counter parts, the MSSM and the NMSSM.

The searches at the LHC for BSM Higgs bosons with masses above $125 \gev$ have
shown several excesses in the data recorded between 2015 and 2018
(Run~2), which is a finding that by itself is not 
unexpected in view of the large number of searches that have been conducted.
However, it is remarkable that several searches show an excess of events 
above the background expectation
around the same mass scale of a hypothetical new Higgs boson $\phi$ of
$\mphi \approx 400 \gev$.
The various excesses each reach
the level of 
about $3\,\sig$ local significance or above,
but the global significance of any excess individually, taking
into account  the ``look elsewhere effect'', stays below $3\,\sig$. The
current experimental 
situation regarding the mentioned excesses at $\mphi \approx 400 \gev$ 
can be briefly summarized as follows
(we use here a notation in terms of the CP-even and -odd
eigenstates, $H$ and $A$, respectively, but
interpretations in terms of CP-mixed states would also be possible).

\begin{itemize}[noitemsep,topsep=0pt]
  \item[-] $A \to t\bar t$: CMS reported a local
    excess of $3.5\,\sig$ in
    their first year Run~2 data~\cite{Sirunyan:2019wph}. There is no
    corresponding Run~2 ATLAS analysis available.
  \item[-] $\phi \to \tau^+\tau^-$: ATLAS reported a local excess of
    $2.7\,\sig$ in their full Run~2 data~\cite{Aad:2020zxo}.
    The corresponding CMS analysis, using only first year
    Run~II data, does not show an excess, but also has
    substantially weaker expected sensitivities.
 \item[-] $A \to Z h_{125}$: ATLAS reported a local
    excess of $3.6\,\sig$ in their first year Run~2
    data at a mass of around $440 \gev$ in the b-quark 
    associated production channel~\cite{Aaboud:2017cxo}.
    The corresponding CMS analysis is
    consistent with the SM background
    expectation~\cite{Sirunyan:2019xls}. An updated full Run~2
    ATLAS analysis 
        for the gluon fusion (ggF) production channel also agrees with
    the SM expectation~\cite{ATLAS:2020pgp}.
\end{itemize}

\noindent
This experimental situation
(see also the discussion in \citere{Richard:2020cav})
triggers the question whether all or some of these
excesses could be fitted simultaneously.
In a recent publication~\cite{Arganda:2021yms},
in which this question was addressed
in a model-independent approach by assuming
a CP-odd state at $400\gev$
with independently adjustable
couplings to the SM particles
as the origin of both excesses,
it was found that the preferred values
of the couplings
would correspond to rather contrived scenarios
that do not lend themselves to 
an immediate interpretation within 
UV-complete models.
Within the present paper we take a different viewpoint and analyse the observed 
excesses in the context of two popular models, namely the 
N2HDM and the NMSSM. The symmetry properties of these models give rise to
correlations between the couplings of the CP-odd Higgs boson to different
SM particles, in contrast to the freely adjustable couplings that were
considered in \citere{Arganda:2021yms}.

In our investigation in this paper we will
perform a $\chi^2$ analysis taking into account the
observed excesses in the 
$pp \to A \to t \bar t$ channel at CMS and the
$pp \to \phi \to \tau^+\tau^-$
channel at ATLAS, since in both cases the observed excesses over the
background expectation
are not in direct tension with corresponding results
(in terms of search channel and integrated luminosity)
from the other collaboration. The situation is
more ambiguous for the excess in the 
$A \to Z h_{125}$ search reported by ATLAS 
(as we will discuss in
more detail in \refse{sec:excesses}). 
In view of this situation we will focus on the
parameter space that is preferred as a result of the
$\chi^2$ analysis where the $A \to Z h_{125}$ search is not included.
However,
we investigate 
whether these parameter regions would also
be compatible with a possible signal in the 
$b \bar b \to A \to Z h_{125}$ channel. 

As a second step of our analysis we take into account the possibility that a
Higgs boson of an extended Higgs sector could also be lighter than the
observed state at $125\gev$.
Searches for Higgs bosons below $125 \gev$ have been performed at
LEP~\cite{Abbiendi:2002qp,Barate:2003sz,Schael:2006cr},
the Tevatron~\cite{Group:2012zca} and the
LHC~\cite{Sirunyan:2018aui,Sirunyan:2018zut,ATLAS:2018xad}. 
It is intriguing that also 
two of those searches show a $2-3\,\sig$ local excess around the same
mass of about $96 \gev$, that is not in tension with other search limits:

\begin{itemize}[noitemsep,topsep=0pt]
\item[-] $pp \to \phi \to \ga\ga$: CMS reported a local excess of
  about $3\,\sig$ in their first year Run~2 data~\cite{Sirunyan:2018aui}, with a
    similar upward deviation of~$2\,\sigma$ local
    in the Run\,1 data at a comparable
    mass~\cite{CMS:2015ocq}. The ATLAS results based on the data of the
    first two years of Run~2~\cite{ATLAS:2018xad} are not sensitive to
    the excess.
  \item[-] $e^+e^- \to Z\,\phi \to Z\,b\bar b$: LEP reported a local excess of
    around $2\,\sig$~\cite{Biekotter:2019kde}.
\end{itemize}

\noindent
In previous analyses focussing just on the excesses at about $96\gev$
it was shown that type~II and type~IV of the N2HDM can fit both
excesses simultaneously~\cite{Biekotter:2019kde}.
This was also investigated in combination with a viable dark-matter
candidate for the case where a complex singlet instead
of a real singlet is considered~\cite{Biekotter:2021ovi}.
It was furthermore
demonstrated that SUSY models like
the NMSSM~\cite{Cao:2016uwt,Domingo:2018uim,Choi:2019yrv}
or the
$\mu\nu$SSM~\cite{Biekotter:2017xmf,Biekotter:2019gtq}
can account for the excesses at a level
of roughly $1\sigma$.
On the other hand, in the MSSM neither a $400 \gev$ Higgs boson
can be accomodated in view of the existing constraints (see the
discussion in \refse{sec:nmssm}), nor the CMS excess at around 
$96 \gev$ can be realized~\cite{Bechtle:2016kui}. Consequently,
in the present paper we focus on the N2HDM and the NMSSM. 
We address the question, after separately analyzing
possible
interpretations of the observed excesses at about $400\gev$ as described
above (for a discussion of the $t \bar t$ excess
in a general 2HDM, see \citere{Hou:2019gpn}),
whether the observed patterns at $400 \gev$ and $96
\gev$ can be described simultaneously within the considered models.

Our paper is organized as follows.
In \refse{sec:excesses} we summarize the experimental results in the various
search channels and define the various $\chi^2$ contributions
employed in our analysis. \refse{sec:n2hdm} is devoted to possible
interpretations of 
the observed excesses within the N2HDM.
We first investigate
whether the model can fit the
observed excesses at $400 \gev$ in the
$pp \to A \to t \bar t$ and $pp \to \phi \to \tau^+\tau^-$ channels 
while being in agreement with
the relevant theoretical and experimental constraints.
We find that 
the parameter regions that would be preferred by possible signals in the
two channels do not overlap with each other. On the other hand, each of the
two excesses individually can be
described very well by the
N2HDM with Yukawa structure of type~II. 
In a second step we then demonstrate that this model
can simultaneously also describe both excesses at $96 \gev$.
In \refse{sec:nmssm} we extend our analysis to the case of the NMSSM, where the
structure of the Higgs sector is more rigid than in the
N2HDM. Nevertheless, we demonstrate that also the NMSSM can fit each of
the excesses at $400 \gev$ individually, while complying with the
BSM Higgs-boson searches and the signal-rate measurements
of the Higgs boson at $125\gev$.
In this analysis, we show that a particularly
well motivated parameter region to accommodate the
$t \bar t$ excess is given by the alignment-without-decoupling 
limit of the NMSSM~\cite{Carena:2015moc}.
In this limit a light singlet-like scalar is naturally
present in the Higgs spectrum. Consequently, as a next step we
include the $96 \gev$ excesses also into the NMSSM analysis and show
that only the CMS excess can be accommodated alongside with the
$400 \gev$ excesses. In \refse{sec:prosp}
these analyses are supplemented with a
discussion on the prospects for
future experimental tests of the parameter space regions of the
N2HDM and the NMSSM that are favored by the experimental excesses.
We conclude in \refse{sec:conclusion}.
In the Appendix we provide supplementary material on different Yukawa types of the
N2HDM and a discussion of the constraints from the signal at $125\gev$ in the
alignment-without-decoupling limit of the NMSSM.

\section{Experimental situation}
\label{sec:excesses}

Current LHC Run 2 data of proton-proton collisions at a
center-of-mass energy of $13\,\unit{TeV}$ show some tantalizing local excesses 
in various channels with local significances around the
$3\,\sigma$ confidence level (C.L.) which are summarized in the following.

\subsection{Search for \texorpdfstring{\boldmath{$H,A
\to t\bar{t}$}}{HA2tt}}
A search for additional scalar or pseudoscalar
Higgs bosons decaying to a top quark pair was performed by the CMS
experiment~\cite{Sirunyan:2019wph}. The data set analyzed
corresponds to an integrated luminosity of
$35.9\,\unit{fb}^{-1}$. Final states with one or two charged leptons
were considered. The invariant mass of the 
reconstructed top quark pair system and angular variables
sensitive to the spin of the particles decaying into the top quark
pair were used to search for signatures of the \HHEven\ or \HHOdd\ bosons,
taking the interference with the SM \ttbar\ background into account. 

\begin{figure}
\centering
\includegraphics[width=0.6\textwidth]{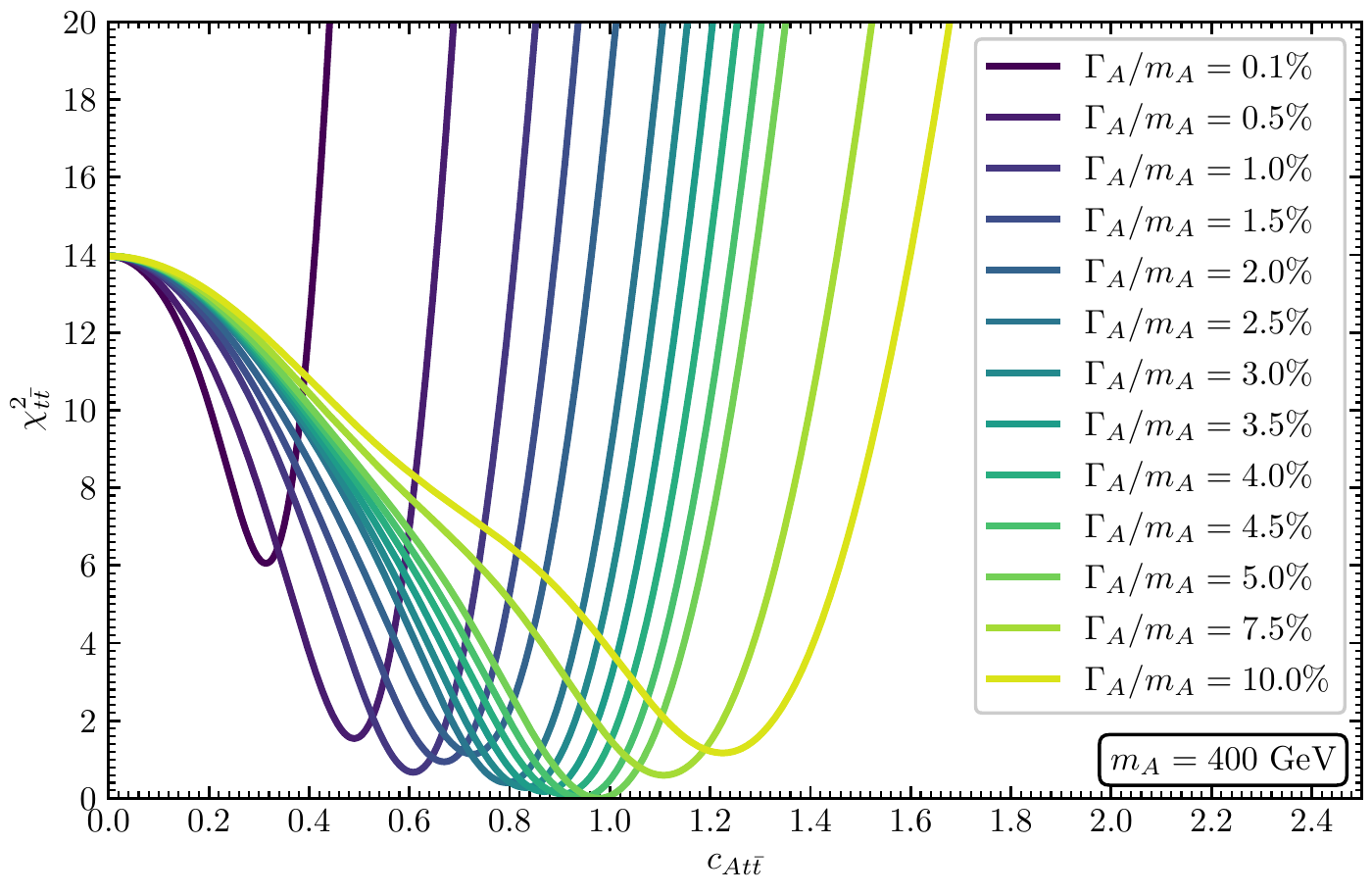}
\caption{\small $\chi^2_{t \bar t}$ as a function
of $c_{A t \bar t}$ assuming $m_A = 400\gev$ for the
different width hypotheses used in the experimental
analysis~\cite{Sirunyan:2019wph,url400cms}.
$\chi^2_{t \bar t}$ is defined relative
to the best fit point.
}
\label{Chisqtt}
\end{figure}

Constraints on the strength modifiers of the \HHEven\ttbar
(\HHOdd\ttbar) couplings \chttbar (\cattbar) were derived
as a function of the mass and width of the heavy Higgs boson \HHEven
(\HHOdd), with the mass of the bosons ranging from 400 to
$750\gev$. Here, the modifiers are defined as the coupling strengths
relative to the ones of a SM Higgs boson. 
In this study it is assumed that there is no CP violation in the Higgs
sector. As a consequence the \HHEven\ and \HHOdd\ Higgs bosons are
CP eigenstates and do not mix with each other, which means that there
are no inteference effects between \HHEven\ and \HHOdd.
The model-independent search was performed for
each CP state separately, setting the coupling modifier for the
other respective CP state to zero.
As a result, a moderate signal-like
deviation was observed for the hypothesis of a pseudoscalar Higgs 
boson with the mass ${m_\HHOdd \approx 400\gev}$ and a total relative
width of $\Gamma_\HHOdd / m_\HHOdd \approx 4.5\%$, with a local significance of $3.5 \pm
0.3$ standard deviations. This translates to a global significance of
$1.9 \, \sigma$. The excess is also consistent with slightly different masses
and widths of a possible signal, however, with less
statistical significance.
The deviation is driven by the CP-odd Higgs boson, while the impact from the
CP-even Higgs boson is small, because at a mass of ${\approx 400\gev}$
the resonant production cross section of a CP-odd Higgs boson is much
larger than that of a CP-even Higgs boson.
This can also be inferred, for instance, from Fig.~3
of~\citere{Bernreuther:1997gs}.

There is no corresponding search at
$13\,\unit{TeV}$ in this channel yet from the ATLAS Collaboration, but
there is a search for heavy pseudoscalar and scalar Higgs bosons
decaying into a \ttbar\ pair performed with an integrated luminosity
of $20.3\,\unit{fb}^{-1}$ at a center-of-mass energy of
$8\,\unit{TeV}$~\cite{Aaboud:2017hnm}. No significant deviation from
the SM prediction was observed in the \ttbar\ invariant mass spectrum
in final states with an electron or muon, large missing transverse
momentum, and at least four jets. However, dileptonic final states and angular
variables were not utilized for this analysis, and
cross section limits were set only for masses ranging
from $500\gev$ to $750\gev$. Consequently, a combined analysis of
the ATLAS and CMS results is not possible.

The \chisq-distribution associated to the CMS
analysis~\cite{url400cms,Sirunyan:2019wph}, \chisqttbar\, is shown in
\reffi{Chisqtt} as a function of the coupling strength \cattbar\ for
the case of a pseudoscalar Higgs boson with an assumed mass of
$m_{\HHOdd}=400 \gev$ and different values of the width,
$\Gamma_{\HHOdd}$, ranging from 0.1\% to 25\% of its mass.\footnote{It 
should be noted that in Fig.~7 of \citere{Sirunyan:2019wph} there is a
typo 
in the label of the vertical axis, which should read 
$-\ln\left(L(g_{At \bar t}) / L_{\mathrm{SM}}\right)$.}
Width values between $1\%$ and $10\%$ have a \chisq\ value close
or below 1. The \chisq\ is given by twice the negative logarithm of
the likelihood function 
\begin{linenomath}
  \begin{equation}
    \label{Eq:likelihood}
    \begin{gathered}
\chi^2_{\ttbar}
  = - 2 \cdot \ln\LP  L(\cattbar; m_{\HHOdd} , \Gamma_{\HHOdd},
  \vec{\nu}) \over L_{\rm max} \RP\,, \\[.3em]
      L(\cattbar; m_{\HHOdd} , \Gamma_{\HHOdd}, \vec{\nu}) = \left(\prod_i
      \frac{\lambda_i^{n_i}(\cattbar; m_{\HHOdd},
        \Gamma_{\HHOdd}, \vec{\nu})}{n_i!}\,
      \re^{-\lambda_i(\cattbar; m_{\HHOdd} ,
        \Gamma_{\HHOdd}, \vec{\nu})}\right) \, G(\vec{\nu}) ,\\[.3em]
      \lambda_i(\cattbar; m_{\HHOdd}, \Gamma_{\HHOdd}, \vec{\nu}) = 
      \cattbar^4\,  s_{R,i}^\HHOdd(m_\HHOdd,
            \Gamma_\HHOdd, \vec{\nu}) + \cattbar^2\,
            s_{I,i}^\HHOdd(m_\HHOdd, \Gamma_\HHOdd, \vec{\nu}) +
            b_i(\vec{\nu})
      , 
    \end{gathered}
  \end{equation}
\end{linenomath}

\noindent
with $b_i$ denoting the combined background yield in a given bin $i$
of the invariant \ttbar\ mass and spin analyzing distribution,
$s_{R,i}^\HHOdd$ and $s_{I,i}^\HHOdd$ the signal yields in a given bin for
the resonant and interference part, respectively, $\vec{\nu}$
the vector of nuisance parameters on which the signal and background
yields generally depend, and $n_i$ the observed yield in data.
The external constraints on the nuisance parameters
are taken into account in the likelihood via a product of
corresponding probability density functions, $G(\vec{\nu})$.   
The \chisqttbar\ distribution is normalized by
$L_{\rm max} = L(\cattbar = 0.94; m_{\HHOdd} = 400 \gev , \Gamma_{\HHOdd}
= 4.5\% \ m_A ,
  \vec{\nu}_{\rm max}) $ such that it vanishes for the most likely choice of
  mass, width, coupling strength and nuisance parameters,
  $\vec{\nu}_{\rm max}$, providing the best description of the data.

\subsection{Search for \texorpdfstring{\boldmath{$\phi
\to \tautau$}}{phi2ll}}
A search for heavy neutral Higgs bosons decaying into two tau
leptons was performed by the ATLAS Collaboration utilizing data corresponding
to an integrated luminosity of
$139\,\unit{fb}^{-1}$~\cite{Aad:2020zxo}. The search covered the mass
range $0.2$--$2.5\,\unit{TeV}$ of the heavy 
resonance,  decaying into $\tautau$ with at least one $\tau$ lepton decaying into final states with
hadrons. 
The natural width of the scalar boson is assumed to be negligible
compared to the experimental resolution.
Upper limits on the production cross section times branching fraction
for a Higgs boson $\phi$ produced
via gluon-gluon fusion (ggF) and
$b$-quark associated production ($b \bar b\phi$) were derived.
The limits were calculated from a statistical
combination of three different final states, involving one $\tau$ lepton
decaying into a neutrino and hadrons and one $\tau$ lepton decaying into
neutrinos and an electron ($e\tau_h$) or into neutrinos and a muon
($\mu\tau_h$), and involving two $\tau$ leptons decaying into a
neutrino and hadrons each ($\tau_h\tau_h$). 
For ggF a local excess was observed in the
data at the $2.2 \,\sigma$ C.L.\
at $m_{\phi} = 400\,\unit{GeV}$,
while for $b \bar b\phi$ production a local
excess at the $2.7 \,\sigma$ C.L.\ at $m_{\phi} = 400\,\unit{GeV}$ was found.
We have performed a
combined statistical analysis of the excesses in both
production modes assuming $m_{\phi} = 400\gev$
in terms of a
two-dimensional likelihood function as provided by
ATLAS~\cite{Aad:2020zxo,Aad:2020zxoLink}.
The point with the highest excess is located
at $\sigma(gg \to \phi) \times \br(\phi \to \tautau) = 20.19 \,\fb$ and
$\sigma(b\bar b \to \phi) \times \br(\phi \to \tautau) = 38.37 \,\fb$.

In our N2HDM and NMSSM analyses
all points receive a contribution $\chi^2_{\tautau}$
relative to the best-fit point
according to the two-dimensional likelihood function.
In both models it is in principle possible
that more than one Higgs boson can
contribute to the observed excess.
Thus, we define the signal as the incoherent
sum of all Higgs bosons with
masses in the range $(400\pm 40)\gev$,
and use the $\chi^2_{\tautau}$ values of
the sum according to the experimentally
measured likelihood function for $m_\phi = 400\gev$.
It should be noted, however, that
in the N2HDM only the CP-odd Higgs boson with a mass of
$m_A \approx 400\gev$ can give rise to a sizable
contribution to the excess. This is due to the fact that for the case where 
a second (CP-even)
Higgs boson $h_2$ has a similar mass (i.e.\
$360\gev \leq m_{h_2} \leq 440\gev$)
the applied constraints require that $h_2$ has to be an almost pure 
gauge singlet in the parameter space region that is relevant for
the $\tautau$ excess, and its contributions to the signal cross
section are therefore suppressed
compared to the ones of the CP-odd state.

The corresponding search of the CMS Collaboration for additional neutral
Higgs bosons in the 
$\tautau$ final state was done on a dataset corresponding to an
integrated luminosity of
$35.9\,\unit{fb}^{-1}$~\cite{Sirunyan:2018zut}. Model-independent limits were
set on the product of the production cross section times 
the branching fraction for the decay into $\tau$ leptons for both the 
ggF and the $b \bar b \phi$ production modes.
The search for heavy
resonances was performed over the mass range
$0.09$--$3.2\,\unit{TeV}$, assuming a narrow
width. Final states involving one $\tau$ lepton decaying into
neutrinos and an electron and one $\tau$ lepton decaying into neutrinos
and a muon ($e\mu$) were considered as well as $e\tau_h$, $\mu\tau_h$ and
$\tau_h\tau_h$ final states. No excess over
the SM background was found in this analysis.
However, due to the lower amount of data included,
the expected and observed cross section
limits of the CMS analysis are weaker
than the ones from ATLAS utilizing the full Run~2 dataset.
Thus, the signal intepretation of the ATLAS excess at around $400\gev$
at the best-fit point is compatible with the expected (observed)
upper limits from the CMS analysis at the level of $1.1 (1.9) \sigma$.

\subsection{Search for \texorpdfstring{\boldmath{$A
\to Zh_{125}$}}{A2Zh}}
\label{secAZh}
The ATLAS Collaboration performed a search for new resonances decaying
into a $Z h_{125}$ 
in $\nu\bar{\nu}b\bar{b}$ and $\ell^+\ell^-b\bar{b}$ final states, where
$\ell^\pm = e^\pm$ or $\mu^\pm$~\cite{Aaboud:2017cxo}. The data used 
corresponds to a total integrated luminosity of
$36.1\,\unit{fb}^{-1}$. The search was conducted by examining the
reconstructed invariant mass distribution of $Zh$ candidates for evidence
of a localised excess in the mass range of $220\,\unit{GeV}$ up to
$5\,\unit{TeV}$. The results of the search for the CP-odd
scalar boson $A$ in two-Higgs-doublet models
were interpreted in terms of constraints
on the ggF and $b$-quark associated ($b\bar{b}A$) production
cross-sections times the branching fraction of $A \to Zh_{125}$
and the branching fraction of $h_{125} \to b\bar{b}$.

In the search for $b\bar{b}A$ production, an excess of events was
observed for a resonance mass around $440\,\unit{GeV}$, mainly driven
by dimuon final states with three or more $b$-tags required. The local
significance of this excess with respect to the background-only
hypothesis was estimated to be $3.6\,\sigma$, and the global
significance, accounting for the look-elsewhere effect, was estimated to
be $2.4\,\sigma$. In the search for ggF production, agreement between the
data and the background-only hypothesis was found in a preliminary
update analysing the full Run 2 data set corresponding
to an integrated luminosity of $139\,\unit{fb}^{-1}$~\cite{ATLAS:2020pgp}.

The respective analysis performed by the CMS Collaboration for a heavy
pseudoscalar boson $A$ decaying to $Z h_{125}$ similarly
considered final states where the 
Higgs boson decays to a bottom quark and antiquark, and the $Z$ boson
decays either into a pair of electrons, muons, or
neutrinos~\cite{Sirunyan:2019xls}. The 
analysis was performed using a data sample corresponding to an
integrated luminosity of $35.9\,\unit{fb}^{-1}$. Exclusion limits were
set in the context of 2-Higgs-doublet models in the $A$ boson mass
($m_A$) range between $225$ and $1000\,\unit{GeV}$ on the ggF and
$b\bar{b}A$ production
cross-sections times the branching fraction of $A \to Zh_{125}$
and the branching
fraction of $h_{125} \to b\bar{b}$. In this search, however, the data was
found to be consistent with the background expectations over the whole
range of the reconstructed resonance mass $m_A$. 

In view of this somewhat 
ambiguous experimental situation we do not
incorporate the excess in the $A \to Zh_{125}$ channel into our
$\chi^2$ analysis. We rather
investigate whether the parameter regions in the N2HDM and the NMSSM
that are favoured as a result of the other observed excesses discussed in this
section could give rise to a possible signal also in the 
$A \to Zh_{125}$ channel.
For a discussion of the ATLAS excess in the $A \to Zh_{125}$
channel in the context of the 2HDM 
see \citere{Ferreira:2017bnx}, and in the context of
a general NMSSM see \citere{Coyle:2018ydo}.

\subsection{Searches for Higgs bosons in the low mass region}
\label{sec:excesses96}

As discussed above, in the context of searches for additional Higgs
bosons it is important to take into account also the mass region below the
observed state at $125\gev$.
Higgs bosons with relatively low masses can be searched
for at the LHC via diphoton resonant searches.
CMS found a deviation between the expected and
observed exclusion limits at a mass of about $96\gev$ in
the Run~1 data set, with a local significance of about
$2\,\sigma$~\cite{CMS:2015ocq}. Remarkably, the updated analysis including
the first Run~2 data found an excess of about $3\,\sigma$ at
a comparable mass~\cite{Sirunyan:2018aui}. Taking into account the 
accumulated data at 7, 8 and $13\tev$ center-of-mass
energy, CMS reported a signal interpretation of
the excesses, corresponding to a signal strength of
\begin{equation}
\mu_{\rm CMS}^{\rm exp} = 0.6 \pm 0.2 \ .
\end{equation}
Here the signal strength $\mu_{\rm CMS}$ is
defined, as usual, as the resonant signal cross
section assuming a scalar particle around $96\gev$,
normalized to the expected signal cross section
assuming a SM Higgs boson at the same mass.

Similar searches in the diphoton final state
have also been performed by
ATLAS using $80\ifb$ of the $13\tev$
dataset~\cite{ATLAS:2018xad}. Here, only a very small
excess of events over the SM background has
been found around $96\gev$.
However, overall the resulting upper limits
on the cross sections found by ATLAS
are substantially above
the corresponding CMS exclusion limits.
Consequently, the ATLAS result cannot exclude the signal
interpretation of the CMS excess.

The CMS excesses have gained considerable interest
in the literature due to the fact that it
is consistent with another excess found
already at the Large Electron Positron collider
(LEP). At LEP, Higgs bosons in this mass range could be searched
for also in hadronic final states due to the
much lower backgrounds. Here an excess was reported
in the search $e^+ e^- \rightarrow Z H \rightarrow
Z b \bar b$ at a mass of about $ 98\gev$,
with a local significance of
$2.3\,\sigma$~\cite{Barate:2003sz}.
The signal strength for the excess was found to
be~\cite{Cao:2016uwt}
\begin{equation}
\mu_{\rm LEP}^{\rm exp} = 0.117 \pm 0.057 \ .
\end{equation}
The mass resolution of the LEP excess is rather
limited due to the hadronic final state, such that
both the LEP and the CMS excesses could have
a common origin. Various different scenarios to
accommodate both excesses 
at about $96\gev$ have already been discussed in the literature.
\footnote{A recent overview about the various BSM models
that can explain both the CMS and the LEP excesses
can be found in \citere{Biekotter:2020cjs}.}
In particular, possible 
interpretations comprise 
type~II and type~IV of the N2HDM~\cite{Biekotter:2019kde} and
SUSY models like
the NMSSM~\cite{Cao:2016uwt,Domingo:2018uim,Choi:2019yrv} and the
$\mu\nu$SSM~\cite{Biekotter:2017xmf,Biekotter:2019gtq}, where the SUSY models
can account for the excesses at a level
of roughly $1\sigma$.
In the present paper we will first focus on the 
observed excesses at
$\approx 400\gev$ 
and then investigate whether a simultaneous 
realization of also the excesses at $96\gev$
is possible.

In order to quantify
how well the excesses are fitted
for each parameter
point in our numerical analysis, we define
the $\chi^2$ regarding the excesses at
$96\gev$ via
\begin{equation}
\chi^2_{96} =
\frac{(\mu_{\rm CMS} - 0.6)^2}{0.2^2} +
\frac{(\mu_{\rm LEP} - 0.117)^2}{0.057^2} \, .
\end{equation}
Here $\mu_{\rm CMS}$ and $\mu_{\rm LEP}$
are the theoretical predictions. For the SM, for
which we have set ${\mu_{\rm CMS}=\mu_{\rm LEP}=0}$,
we find a penalty of $\chi^2_{{\rm SM},96} = 13.2$
from these contributions.

\section{N2HDM interpretation}
\label{sec:n2hdm}
In this section we will discuss possible
realizations of the described
excesses in the Higgs searches within the
context of the
N2HDM. We start by introducing the model
and its parameters, followed
by a discussion of the relevant theoretical
and experimental constraints.
The numerical analysis will be divided
into two parts. First, we discuss how the
excesses at $400\gev$ can be accommodated,
and whether a simultaeneous realization of
both the $t \bar t$ and the $\tautau$
excesses is possible.
In a second step, we additionally take into
account the excesses at $96\gev$ in order
to address the question whether these can be
realized in the parameter region in which
one of the excesses at $400\gev$ is
accommodated.

\subsection{Model definitions}
\label{sec:defn2hdm}
The N2HDM contains two doublet scalar fields
$\Phi_{1,2}$ and a real scalar
singlet field $\Phi_S$~\cite{Chen:2013jvg}.
In the physical basis,
the Higgs particle sector consists of a
total of three CP-even Higgs bosons $h_i$, a
CP-odd state $A$ and two charged
Higgs bosons $H^\pm$.
In order to avoid tree-level flavor-changing
neutral currents (FCNC) a $Z_2$ symmetry is assumed
as in the usual 2HDM, which is only softly broken by
the terms $
- m_{12}^2 \left(
\Phi^\dagger_1 \Phi_2 + \Phi^\dagger_2 \Phi_1
\right)$
appearing in the Lagrangian.
In addition, the scalar potential respects
a second $Z_2$ symmetry, under which only
the singlet field $\Phi_S$ is charged.
This symmetry is however broken spontaneously
when the singlet field has a non zero vacuum expectation value (vev),
${\langle \Phi_S \rangle = v_S}$, which we
will assume throughout the analysis.
The breaking of the electroweak gauge symmetry
originates from the vevs of the doublet
fields $\langle \Phi_{1,2} \rangle =
v_{1,2} / \sqrt{2}$, where
$\sqrt{v_1^2 + v_2^2} = v \approx 246\gev$.
A crucial parameter
for the analysis of the experimental excesses is
the ratio of the doublet vevs given by
\begin{equation}
\tan\beta = \frac{v_2}{v_1} \ .
\end{equation}
The three real neutral components
of $\Phi_{1,2,S}$ mix to form the mass
eigenstates $h_{1,2,3}$ with masses
$m_{h_1} < m_{h_2} < m_{h_3}$.
The mixing in the CP-even scalar sector
can be described by the three mixing angles
$\alpha_{1,2,3}$, defining the 
matrix
\label{eqmmatrix}
\begin{equation}
R = \begin{pmatrix}
c_{\alpha_{1}}c_{\alpha_{2}} & s_{\alpha_{1}}c_{\alpha_{2}} & s_{\alpha_{2}}\\
-\left(c_{\alpha_{1}}s_{\alpha_{2}}s_{\alpha_{3}}+s_{\alpha_{1}}c_{\alpha_{3}}\right)
&
c_{\alpha_{1}}c_{\alpha_{3}}-s_{\alpha_{1}}s_{\alpha_{2}}s_{\text{\ensuremath{\alpha_{3}}}}
&
c_{\alpha_{2}}s_{\alpha_{3}}\\
-c_{\alpha_{1}}s_{\alpha_{2}}c_{\alpha_{3}}+s_{\alpha_{1}}s_{\alpha_{3}} &
-\left(c_{\alpha_{1}}s_{\alpha_{3}}+s_{\alpha_{1}}s_{\alpha_{2}}c_{\alpha_{3}}\right)
& c_{\alpha_{2}}c_{\alpha_{3}}
\end{pmatrix} \ .
\end{equation}
At least in principle, the properties of
each of the Higgs bosons $h_i$ 
could be such that it can be identified with the SM-like
Higgs boson at $125\gev$. The masses of the
CP-odd Higgs boson
and the charged Higgs bosons
are denoted by $m_A$ and $m_{H^\pm}$,
respectively.

Assigning consistent charges under the $Z_2$ symmetry to the 
fermions yields the typical four types of Yukawa structures that
are familiar from the 2HDM. Depending on the type, the couplings of
the scalar particles to the 
fermions will be different. The relevant
parameters defining the strength of the
couplings normalized to the one of a
hypothetical SM Higgs boson of the same mass
can be given in terms of $\tan\beta$
and the mixing angles $\alpha_i$
of the CP-even sector.

As already mentioned above, in a first step we will analyze whether
the N2HDM can realize the excesses at $400\gev$.
We will go a step further in the subsequent analysis and investigate
whether and how the excesses at $400\gev$ and
at $96\gev$ can be realized simultaneously.
Our approach for covering
the relevant
parameter space will be discussed in detail
in \refap{n2hdmstrategy}.
Before, we briefly summarize the relevant
theoretical and experimental constraints
that we take into account in our analysis.

\subsection{Theoretical and experimental constraints}
\label{n2hdmconstraints}
Numerous theoretical constraints on the N2HDM
parameter space have to be taken into account
in order
to exclude unphysical parameter configurations.
To assure the presence of a viable electroweak minimum,
described by the vevs $v_1$, $v_2$ and $v_S$
as explained in \refse{sec:defn2hdm},
we demand that the scalar potential is bounded
from below.
The necessary conditions are given
in terms of the quartic scalar couplings
which define the behavior of the potential
for large field values~\cite{Klimenko:1984qx}.
Furthermore, we verified that the electroweak minimum
is either the global minimum of the scalar
potential, or, in case other deeper minima
exist, that the lifetime of the electroweak minimum
is large compared to the age of the observable
universe. Furthermore, in order to verify that the perturbative
treatment of the model is adequate, we checked
that the perturbative unitarity constraints
are fulfilled. These are also given in terms
of the quartic scalar couplings and set upper
limits on their absolute values~\cite{Muhlleitner:2016mzt}.
All the theoretical constraints mentioned
here have been applied using the public code
\texttt{ScannerS}~\cite{Coimbra:2013qq,
Ferreira:2014dya,Costa:2015llh,
Muhlleitner:2016mzt,Muhlleitner:2020wwk}
in combination with the public code
\texttt{EVADE}~\cite{Hollik:2018wrr,Ferreira:2019iqb}
for the calculation of the
lifetime of metastable electroweak minima.

Besides the mentioned theoretical constraints,
we have also included the most relevant experimental constraints on
the N2HDM parameter space
into our analysis:\\
\textit{(i)} We perform a $\chi^2$ test regarding
the measurements of the signal rates of 
$h_{125}$ using the public code
\texttt{HiggsSignals v.2.6.0}~\cite{Bechtle:2013xfa,
Stal:2013hwa,Bechtle:2014ewa,Bechtle:2020uwn}.
In the following, we will denote the result
of the \texttt{HiggsSignals} test
by $\chi^2_{125}$, which is constructed taking
into account $n_{\rm obs} = 107$ observables.
The value of $\chi^2_{125}$ for each N2HDM
parameter point will be compared with the SM
prediction $\chi^2_{\mathrm{SM},125} = 84.42$.\\
\textit{(ii)} We verify for each parameter point
that none of the N2HDM scalars is excluded
by searches for additional Higgs bosons at
the LHC, the Tevatron and at LEP by making use
of the public code
\texttt{HiggsBounds v.5.9.0}~\cite{Bechtle:2008jh,
Bechtle:2011sb,Bechtle:2013gu,Bechtle:2013wla,
Bechtle:2015pma,Bechtle:2020pkv}.
For the considered parameter point
this code selects the most sensitive search
channel for each Higgs boson based on the
expected experimental sensitivity. It then
compares the model predictions for the
signal rates $\mu_{\rm theo.}$
with the observed upper limits
at the $95\%$ C.L., $\mu_{\rm excl.}$,
and rejects a point
if for one of the Higgs bosons
the ratio $r \equiv \mu_{\rm theo.} /
\mu_{\rm excl.}$ is larger than one.\\
\textit{(iii)} We include constraints from
electroweak precision observables (EWPO) in terms
of the so-called oblique parameters
$S$, $T$ and $U $~\cite{Peskin:1990zt,
Peskin:1991sw}.
Using the implementation of the code
\texttt{ScannerS} we exclude points
for which the $\chi^2$ value obtained
by comparing to the fit result of
\citere{Haller:2018nnx} deviates by more than
$2\,\sigma$. The N2HDM model predictions
for $S$, $T$ and $U$ are calculated
at the one-loop level
following \citeres{Grimus:2007if,
Grimus:2008nb}.\\
\textit{(iv)} Mainly due to the presence
of the charged Higgs boson in the N2HDM,
constraints from flavor physics are
important in some regions of parameter
space.
Since the extra field of the N2HDM
compared to the 2HDM is a gauge singlet,
it is sufficient to take over 
the bounds for most of the flavor
physcis observables from the
2HDM~\cite{Biekotter:2019kde,Muhlleitner:2020wwk}.
We apply the flavor constraints from
\citere{Haller:2018nnx}
as implemented in \texttt{ScannerS}.
They yield a lower limit $m_{H^\pm} \gtrsim 550\gev$
in both type~II and type~IV.
In type~I and type~III no such lower
limit on $m_{H^\pm}$ can be established.
In all types the flavor constraints
give rise to a lower limit on $\tb$,
where the specific value is different
in each type, and the strongest constraint is obtained
in the type~IV (N)2HDM.\footnote{Besides the
constraints from processes involving the charged Higgs
boson, also the constraints from ${B_{d,s} \rightarrow \mu^+ \mu^-}$,
involving the neutral scalar sector, can be relevant. However,
since the gauge singlet field does not
couple to the SM fermions directly,
only subleading corrections are expected
to be present in the N2HDM compared to
the limits from the 2HDM. 
We therefore apply
the 2HDM constraints also for this
observable.}

Except the $\chi^2$ result of
\texttt{HiggsSignals} regarding the
signal rates of $h_{125}$, the
experimental constraints described in
this section have been taken into account
in terms of a hard cut, i.e., 
a parameter point is regarded to be excluded if the considered
experimental constraints are not fulfilled for this point.
As will
be described in \refse{numlowII},
we combine $\chi^2_{125}$ with the
$\chi^2$ values for the different
observed excesses. In this way we quantify how well the considered models 
describe both the experimental results for $h_{125}$ 
and the potential signals
at $400\gev$, as well as in a second step 
also the potential signals
at $96\gev$.

\subsection{A Higgs boson at \texorpdfstring{\boldmath{$400\gev$}}{400gev}
in type~II}
\label{fullII}
In order to accommodate the $t \bar t$ excess
one needs values of $|c_{A t \bar t}| \gtrsim 0.5$
(see \reffi{Chisqtt}). Since in all four Yukawa types of the
N2HDM one finds $|c_{A t \bar t}| = 1 / \tan\beta$,
it is easy to understand that beyond the fact that
different constraints apply in each type, all
four Yukawa types
are very similar regarding the $t \bar t$ excess.
The situation is more complicated for the
$\tautau$ excess, where it is required that
two conditions are fulfilled: Firstly, in order
to accommodate values for the $b \bar b$ associated
production cross section of $A$ of the same size
or larger than the gluon-fusion production cross section,
the condition $|c_{A b \bar b}| \gg |c_{A t \bar t}|$
has to be fulfilled. Secondly, in order to obtain
sufficiently large values for $\br(A \to \tautau)$,
the condition $|c_{A \tautau}| \gg |c_{A t \bar t}|$
has to be fulfilled. The coupling coefficients
$c_{A b \bar b}$ and $c_{A \tautau}$ are different
in each type, either proportional to $\tan\beta$
or its inverse. One finds that only
the type~II Yukawa structure is able to satisfy
both conditions (see \refap{n2hdmstrategy} for a detailed
explanation), such that it is the only type
that can potentially accommodate the $\tautau$ excess.
In the following, we will focus our
numerical discussion on the N2HDM type~II, while 
in the appendix we will also provide a discussion 
of the N2HDM type~IV.

\begin{table}
\centering
\footnotesize
\def\arraystretch{1.5}
\begin{tabular}{cccccc}
$m_{h_a}$ & $m_{h_b}$ & $m_{h_c}$ &
    $m_A$ & $m_{H^\pm}$ & $\tan\beta$ \\
\hline
$[20,1000]$ & $125.09$ & $[20, 1000]$ &
    $400$ & $[550, 1000]$ & $[0.5, 12.5]$ \\
\hline
\hline
$c_{h_b VV}^2$ & $c_{h_b t \bar t}^2$ &
    $\mathrm{sign}(R_{b3})$ & $R_{a3}$ &
        $m_{12}$ & $v_S$ \\
\hline
$[0.6, 1.0]$ & $[0.6, 1.2]$ & $-1,1$ &
    $[-1, 1]$ & $[0, 1000]$ & $[10, 1500]$
\end{tabular}
\caption{\small Values of input parameters for the
scan in type~II for the investigation of the excesses
at~$400\gev$. The states $h_{a,b,c}$ are
automatically arranged according to the mass-ordered
notation $h_{1,2,3}$ for each point
by \texttt{ScannerS}.}
\label{n2hdmfulltbparas}
\end{table}

In this section we
start by investigating
in which parameter
regions of the type~II N2HDM the excesses at $400\gev$
can be accommodated.\footnote{In contrast to
\citere{Arganda:2021yms}, in which
this question was also addressed,
we take into account
the possibility of explaining the
local excess in the $b \bar b \to A \to Z h_{125}$
search. 
\citere{Arganda:2021yms} only considered
the cross section limits of this search
as upper limits.
Here it is
important to note that
we used as a constraint
for the ggF production mode
$gg \to A \to Zh_{125}$
the most recent cross
section limits reported by
ATLAS using the full Run~2
data set~\cite{ATLAS:2020pgp}, while
the older result~\cite{Aaboud:2017cxo}
used in \citere{Arganda:2021yms}
(based only on
$36\ifb$ of data)
gives a substantially
weaker limit.
In addition, our analysis includes the
cross section limits from the process
$pp \to t \bar t \, A \to t \bar t \, t \bar t$
as published by CMS using the full Run~2
data set~\cite{Sirunyan:2019wxt}. 
Instead, in \citere{Arganda:2021yms} a reinterpreation
of the ATLAS measurement of the SM cross section
for four-top production was
used~\cite{ATLAS:2020hrf}
by simply adding the contribution of a CP-odd Higgs boson
at $400\gev$ to the SM
cross section for four-top production.
Consequently, effects on signal acceptances etc.\ 
through the exchange of the CP-odd Higgs boson are not accounted for.}
We give in \refta{n2hdmfulltbparas} the ranges
of the free parameters used here.
The choice of the
input parameters is the default one of the
public code \texttt{ScannerS}~\cite{Coimbra:2013qq,
Muhlleitner:2020wwk} that was used
to generate the parameter points
respecting the various theoretical and
experimental constraints discussed
in \refse{n2hdmconstraints}.
Each set of
input parameters corresponds to a unique benchmark
point under the assumption
$c_{h_b t \bar t} \cdot c_{h_b V V} > 0$.
This condition arises from the constraints on the
signal rates of
the state $h_{125}$
(here $h_{125} = h_b$,
as defined in \refta{n2hdmfulltbparas}).
In the following, the CP even scalars will
be denoted as $h_{a,b,c}$ when referring to
the input parameters as shown in \refta{n2hdmfulltbparas},
or alternatively as $h_{1,2,3}$, where the
mass ordering $m_{h_1} < m_{h_2} < m_{h_3}$
is implied.

Motivated by the pattern of the observed excesses described in
\refse{sec:excesses}, our scan in the N2HDM is performed such that the observed
excesses at $400 \gev$ are mainly associated with the CP-odd
Higgs boson $A$ 
(a CP-even Higgs boson in this mass
region could yield only a very small contribution to the signal cross
section as a consequence of the constraints and the
required large singlet component),
and we therefore fix $m_A = 400 \gev$ in the scan. The scan
range of $\tan\beta = [0.5, 12.5]$ is chosen in view of our qualitative
discussion above of the observed excesses and of the bounds from searches for
additional Higgs bosons. 
Moreover, the mass $m_{h_b}$ is fixed to $m_{h_b} = 125.09\gev$\cite{Khachatryan:2016vau}, 
and the corresponding couplings
are restricted to ranges that allow for the presence of
a Higgs boson $h_b$ that resembles the properties
of the Higgs boson that has been detected at $125\gev$.
The lower limit of $m_{H^\pm} \geq 550\gev$ is chosen in view of the
constraints from flavor physics observables
(see \refse{n2hdmconstraints}). The remaining
mass parameters were varied up to values of $1\tev$,
except for $10 \leq v_S \leq 1.5\tev$.
The mixing matrix element $R_{a3}$, see \refeq{eqmmatrix},  
is scanned over all theoretically possible values,
and both possibilities for
$\mathrm{sign}(R_{b3})$ are taken into account.
Due to the large number of free parameters,
it is not possible to cover the 
entire parameter space region defined
in \refta{n2hdmfulltbparas} sufficiently well
without applying
other theoretical prejudices
on the model parameters.
However, since the
properties of the pseudoscalar $A$ are to a large
extent determined by the parameters $m_A$
and $\tan\beta$, robust conclusions can be drawn
with respect to the realization of the excesses
by randomly sampling the parameter space,
without further theoretical restrictions,
for instance, in terms of priors.

We quantify the results of this scan
in terms
of the total $\chi^2$
arising from the comparison of the N2HDM predictions with the observed
results for the excesses at
$400\gev$ in the $t \bar t$ and $\tautau$ channels
as well as with the signal-rate
measurements of the Higgs boson at about
$125\gev$, such that
\begin{equation}
\chi^2 = \chi^2_{125} +
\chi^2_{t \bar t} +
\chi^2_{\tautau} \ .
\label{eqchisqttno96}
\end{equation}
The evaluation of the individual $\chi^2$ contributions in
\refeq{eqchisqttno96} has been described in \refse{sec:excesses} and
\refse{n2hdmconstraints}.
In our analysis we consider
all points with $\chi^2 \leq \chi^2_{\rm SM}$
as acceptable, where $\chi^2_{\rm SM}$
is evaluated via \refeq{eqchisqttno96}
assuming no signal contributions to
the excesses.
For $\chi^2_{t \bar t}$ and
$\chi^2_{\tautau}$
we remind the reader that the values are
defined relative to the best-fit
points.
One finds $\chi^2_{\mathrm{SM}, t \bar t} = 13.92$
and $\chi^2_{{\rm SM}, \tautau} = 9.99$
assuming the SM hypothesis.
All remaining theoretical and experimental constraints
are included as hard cuts, either allowing or excluding
a parameter point (see \refse{n2hdmconstraints}).
As a consequence of the condition
$\chi^2 \leq \chi^2_{\rm SM}$, the sample of
parameter points
displayed in our plots below will also contain
points with $\chi^2_{125} \approx \chi^2_{\mathrm{SM},125}$
that potentially do not describe
either of the excesses with a confidence level below
$2 \, \sigma$. However, since our
results allow
a point-by-point
comparison\footnote{It should be noted in this context that 
the identification of parameter regions with $\chi^2$ values that are lower than
the one of the SM does not automatically indicate a global statistical 
preference for the considered model as compared to the SM. We refrain from such
an analysis, which would in particular require to account for the relevant
degrees of freedom, in the present paper where our main focus is on the
description of the
observed excesses rather than on global fits of different models.}
of the individual $\chi^2$ values,
it should be clearly visible which of the displayed
the scan points yield a good description of one or both
of the excesses, i.e.~featuring
small values of
$\chi^2_{t \bar t}$ and/or
$\chi^2_{\tautau}$, while
mantaining an acceptable $\chi^2_{125}$.

\begin{figure}
\centering
\includegraphics[width=0.33\textwidth]{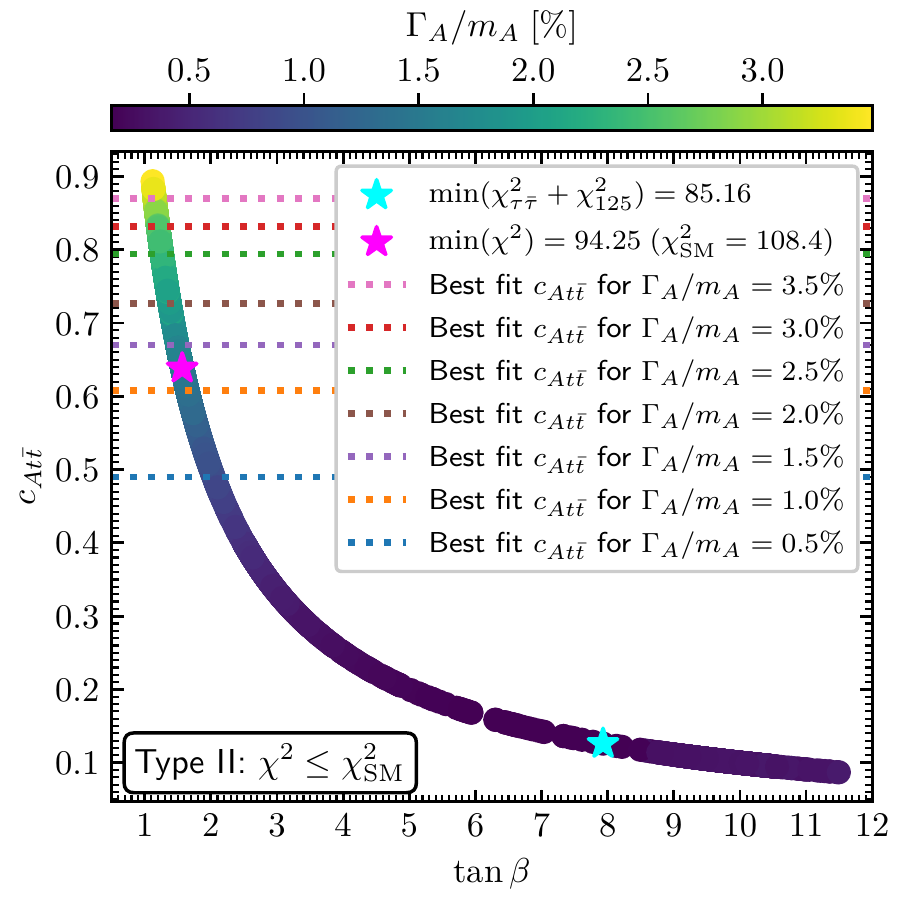}~
\hspace*{-0.6cm}~
\includegraphics[width=0.33\textwidth]{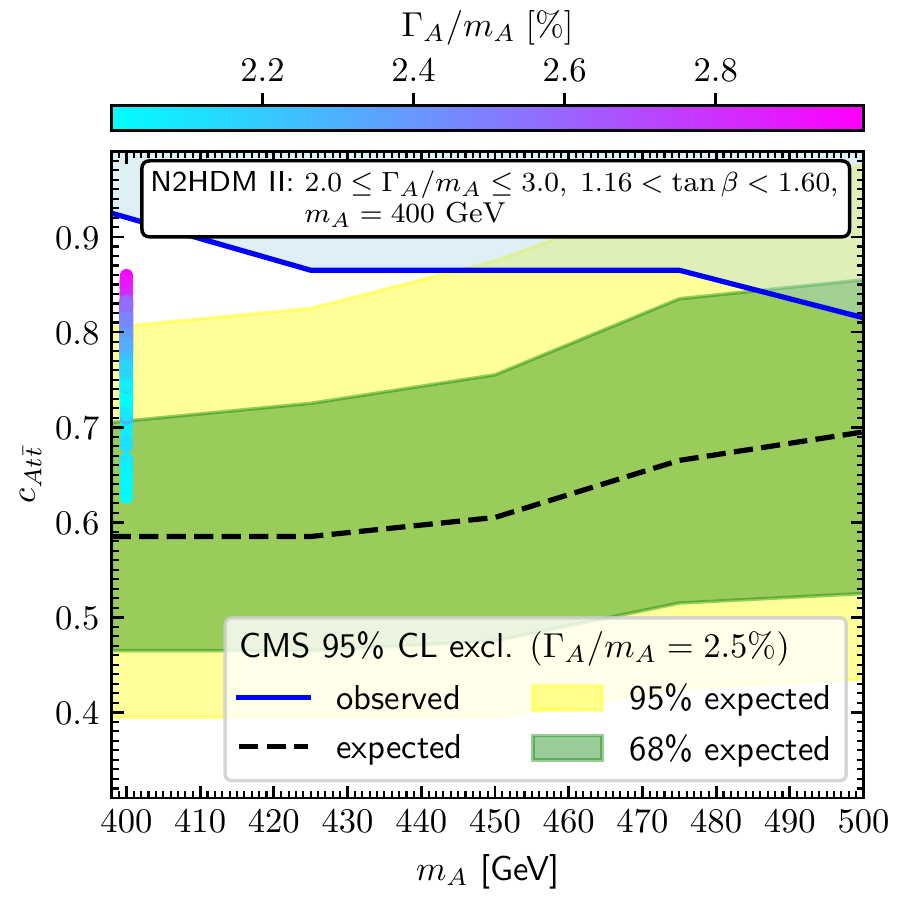}~
\hspace*{-0.6cm}~
\includegraphics[width=0.33\textwidth]{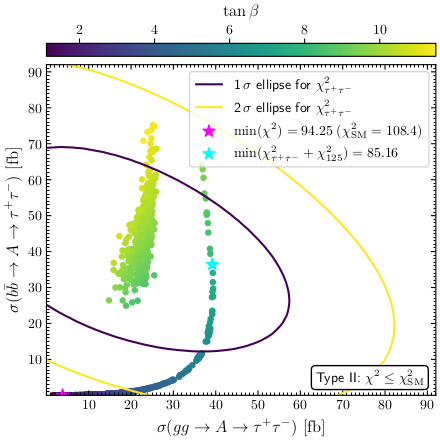}
\caption{\small Left: $c_{A t \bar t}$ in dependence of $\tan\beta$.
The colors of the points indicate the value
of $\Gamma_A / m_A$~in~\%. The dashed horizontal
lines indicate the best-fit values of $c_{A t \bar t}$
for different width hypotheses in the
experimental analysis~\cite{Sirunyan:2019wph}.
Middle:
Values of $c_{A t \bar t}$ for the parameter points
with $2.0\% \leq \Gamma_A / m_A \leq 3.0\%$ in comparison to
the observed (blue)
and expected (black dashed) upper limits
at the $95\%$ C.L.\ as well as the corresponding
$1\sigma$ (green) and $2\sigma$ (yellow)
regions around the expected limit assuming
$\Gamma_{A}/m_{A} = 2.5\%$, as
published in \citere{Sirunyan:2019wph}.
Right: Signal cross sections of $A \to \tautau$
for the $gg$ production mode on the
horizontal axis and the $b \bar b$ production mode
on the vertical axis. The colors of the points
indicate the value of $\tan\beta$. The dark blue
and the yellow lines indicate the $1\sigma$
and $2\sigma$ ellipse of
$\chi^2_{\tautau}$~\cite{Aad:2020zxo},
respectively.
In both plots the best-fit point, labelled as
$\mathrm{min}(\chi^2)$,
is indicated with a magenta star.
The point with the smallest value of
$\chi^2_{\tautau} + \chi^2_{125}$, labelled as
$\mathrm{min}( \chi^2_{\tautau} + \chi^2_{125} )$,
is indicated with a cyan star.}
\label{figGEN1}
\end{figure}

As explained in \refap{n2hdmstrategy}, the properties of
the $A$-boson
are driven by the dependence of its couplings on $\tan\beta$.
In the left plot of \reffi{figGEN1} 
$c_{A t \bar t}$ ($= 1 / \tan\beta$) is displayed as a function 
of $\tan\beta$ for different values of  $\Gamma_A/m_A$.
One can see that the best-fit values
for $c_{A t \bar t}$ showing the best agreement with the 
$t \bar t$ excess,
which are indicated for different width hypotheses by
the horizontal dashed lines,
are reached only for small values of $\tan\beta \lesssim 2$.
As discussed in \refse{sec:excesses} (see \reffi{Chisqtt}) 
the experimental excess is most pronounced for $\Gamma_A/m_A \approx 4\%$. 
This feature manifests itself in the plot by the fact that the largest
displayed values of $\Gamma_A/m_A$ correspond to the largest 
best-fit values of $c_{A t \bar t}$.
Overall, the N2HDM of type~II yields a good description of the 
$t \bar t$ excess in the region of small values of $\tan\beta$, 
$1.2 \lesssim \tan\beta \lesssim 2.5$.
The lower bound on $\tan\beta$ arises here
from the cross section limits on the process
$p p \rightarrow t \bar t (A) \rightarrow t \bar t (t \bar t)$,
as reported by CMS making use of the full
Run~2 data set~\cite{Sirunyan:2019wxt}.
For values of $\tan\beta \gsim 2.5$ we find values of
$c_{A t \bar t}$ substantially below the experimental
best-fit value even for the lowest value
of $\Gamma_A / m_A = 0.5\%$ taken into account
in the analysis. Thus, we conclude that
the $t \bar t$ excess can be well described in the
N2HDM of type~II with low $\tan\beta$, while for
$\tan\beta \gsim 2.5$ 
the compatibility with the $t \bar t$ excess is reduced.
To further illustrate the fit result
we show in the middle plot of \reffi{figGEN1}
the subset of points for which
we find $2.0\% \leq \Gamma_A / m_A \leq 3.0\%$
in the plane of $m_A$ and $c_{A t \bar t}$, where
we also indicate the expected and observed
exclusion limits from the experimental analysis
assuming a width of $\Gamma_A / m_A = 2.5\%$
as published in \citere{Sirunyan:2019wph}.
One can see that the points stretch over the
values of $c_{A t \bar t} \approx 0.8$ that provide the best
description of the $t \bar t$ excess.

In the right plot of \reffi{figGEN1} we show
the parameter points in the plane of the
signal rates regarding the $\tautau$
excess~\cite{Aad:2020zxo},
where the production cross sections were
calculated with the public code
\texttt{SusHi}~\cite{Harlander:2012pb,Harlander:2016hcx}.
Here, the colors of the points indicate
the value of $\tan\beta$.
Our scan results in two
distinct parameter regions in the displayed plane.
The origin of these regions will be discussed
below.
We also show the
$1\sigma$ and $2\sigma$ ellipses 
of
$\chi^2_{\tautau}$ with the dark blue
and the yellow lines.
One can see that, as expected following
the discussion in \refap{n2hdmstrategy}, there are only
points with larger values of $\tan\beta > 6$
inside the $1\sigma$ ellipse.
Comparing with
the region of $\tan\beta$
in which the $t \bar t$ excess can be 
well described,
we conclude that the N2HDM cannot provide a good simultaneous fit of both
excesses.

This feature can be seen more clearly
in \reffi{figGEN2}, where the
$\chi^2$ distributions of the points
regarding both excesses are displayed, together with their sum.
The values
of 
$\chi^2_{t \bar t}$ are shown in blue 
and the ones of
$\chi^2_{\tautau}$ in orange
for all
the parameter points of this scan.
In agreement with the observations described
above, low values of $\chi^2_{t \bar t} \lesssim 4$
can only be found for $\tan\beta \lesssim 2.5$.
Low values of $\chi^2_{\tautau} \lesssim 4$
are found for $\tan\beta \gtrsim 5.5$.
Consequently, the sum of both $\chi^2$
values, as indicated by the green points,
has the smallest values in both $\tan\beta$
ranges suitable for the $t \bar t$ or the
$\tautau$ excess
(but does not drop below a value of about 10),
while larger values are found
in between both ranges, at $3.5 \lesssim
\tan\beta \lesssim 5.5$.
Therefore,
it is not possible to simultaneously
accommodate both excesses in the N2HDM.

\begin{figure}
\centering
\includegraphics[width=0.68\textwidth]{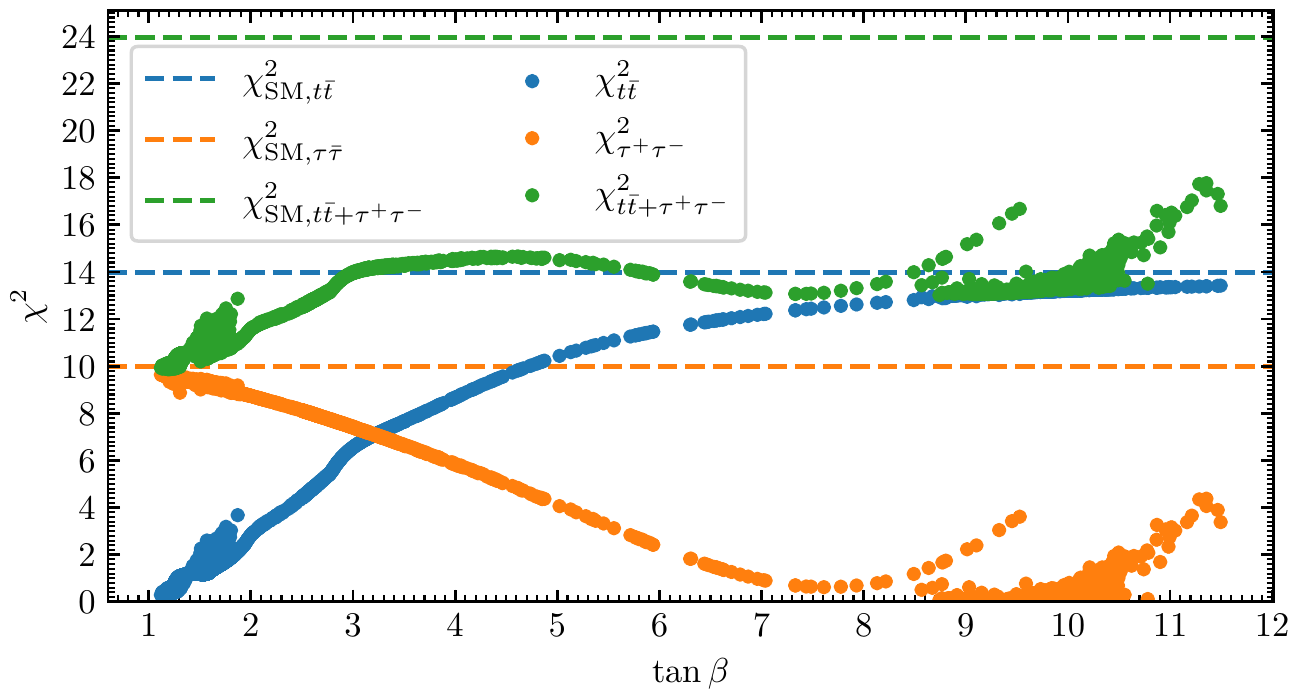}
\caption{\small
$\chi^2_{t \bar t}$ (blue) and
$\chi^2_{\tautau}$ (orange)
and $\chi^2_{t \bar t} +
\chi^2_{\tautau}$ (green)
in dependence of $\tan\beta$.
The horizontal blue, orange
and green dashed lines
indicate the value of
$\chi^2_{{\rm SM}, t \bar t}$,
$\chi^2_{{\rm SM}, \tautau}$
and 
$
\chi^2_{{\rm SM}, t \bar t} +
\chi^2_{{\rm SM}, \tautau}$, respectively.}
\label{figGEN2}
\end{figure}

The two distinct regions found in the right
plot of \reffi{figGEN1}
correspond to the two separate distributions of points that are visible
for
$\chi^2_{\tautau}$ in the region of large $\tan\beta$.
The left one corresponds to the right region of
points in \reffi{figGEN1}, which have larger
values of $\sigma(gg \rightarrow A \rightarrow \tautau)$
and do not cross the center of the ellipses.
Consequently, $\chi^2_{\tautau}$ does
not exactly
reach zero for this set of points. On the contrary, the left band
of points in \reffi{figGEN1} exactly crosses
the center of the ellipses for values
of $\tan\beta \approx 9.5$. Therefore, the 
distribution of 
points displaying $\chi^2_{\tautau}$
near this value of $\tan\beta$
has its minimum at zero.
The two different regions have their origin
in the presence (or absence) of the decay
of $A$ to a lighter Higgs boson $h_i$ and
a $Z$ boson. If such a decay is possible,
it leads to a reduction of
$\br(A\rightarrow \tautau)$.

\begin{figure}[t]
\centering
\includegraphics[width=0.48\textwidth]{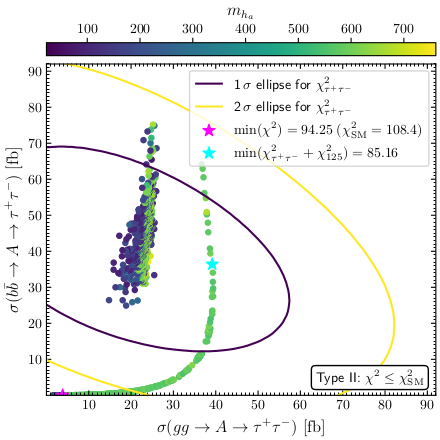}~
\includegraphics[width=0.48\textwidth]{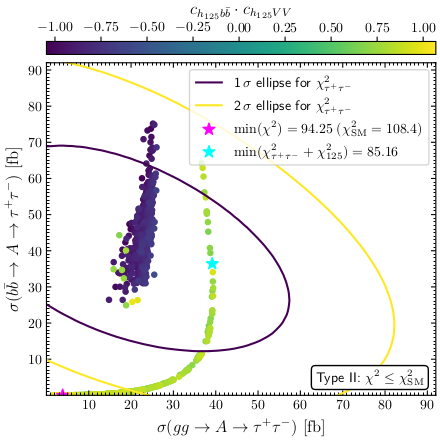}
\caption{\small
Signal cross sections of $A \to \tautau$
for the $gg$ production mode on the
horizontal axis and the $b \bar b$ production mode
on the vertical axis. The colors of the points
indicate the value of $m_{h_a}$ (left)
and $c_{h_{125} b \bar b} \cdot c_{h_{125} VV}$ (right).
The dark blue
and the yellow lines indicate the $1\sigma$
and $2\sigma$ ellipse of
$\chi^2_{\tautau}$~\cite{Aad:2020zxo},
respectively.
In both plots the best-fit point, labelled as
$\mathrm{min}(\chi^2)$,
is indicated with a magenta star.
The point with the smallest value of
$\chi^2_{\tautau} + \chi^2_{125}$, labelled as
$\mathrm{min}( \chi^2_{\tautau} + \chi^2_{125} )$,
is indicated with a cyan star.}
\label{figGEN3}
\end{figure}

The impact of such an
additional decay mode of the CP-odd Higgs
boson $A$
is illustrated in \reffi{figGEN3}, in
which we show the same as in the right
plot of \reffi{figGEN1}, but with the colors
of the points indicating the value of
$m_{h_a}$ (left) and the value of
the product $c_{h_{125} b \bar b} \cdot
c_{h_{125} VV}$ on the right.
In the left plot of \reffi{figGEN3} one
can see that all points with $m_{h_a}
< m_A - M_Z \lesssim 300\gev$,
for which the decay mode
$A \to h_a Z$ is kinematically open,
are in the left band of points.
Hence, the presence of a second light
Higgs boson $h_a$ relatively close
in mass to $125\gev$ improves the fit
to the $\tautau$ excess within the
N2HDM since it can give rise to a reduction
of $\br(A\rightarrow \tautau)$.
In addition, there are also points with
$m_{h_a} > 300\gev$ in the left band of
points. For these points the smaller
values of $\br(A\rightarrow \tautau)$
compared to the points in the right
band are achieved by sizable values of
$\br(A\rightarrow h_{125} Z)$, with $h_{125} = h_1$.
This branching ratio vanishes in the
alignment limit of the N2HDM, in which
the tree-level couplings of $h_{125}$ reduce to the SM
predictions,
and accordingly
$c_{h_{125} A Z} = 0$.
However, in the N2HDM the
so-called wrong sign Yukawa coupling
scenario can be realized. In this scenario, the absolute
values of the couplings of $h_{125}$ to fermions
are close to the SM prediction, but
$c_{h_{125} b \bar b}$ has the opposite sign
than $c_{h_{125} V V}$. This allows for the
agreement of the properties of $h_{125}$
with the measured signal rates without
vanishing couplings to the other scalars
and pseudoscalars. One can see in
the right plot of \reffi{figGEN3} that
all the points in the left band of
points that do not have $m_{h_a} < 300\gev$
are in the wrong sign Yukawa coupling regime.
For these points the presence of the
decay of $A$ into
$h_{125}$
and a $Z$ boson improves the fit to
the $\tautau$ excess.
However, such a signature 
has also been probed in direct searches.
For the $gg$ production mode (see below for a discussion of the observed
excess in the $b \bar b A$
production mode~\cite{Aaboud:2017cxo}) the limits obtained from 
the search for $gg \to A \rightarrow Z (h_{125}) \rightarrow
Z (b \bar b)$
recently reported by ATLAS~\cite{ATLAS:2020pgp} exclude
parameter points in the wrong sign Yukawa
coupling regime with values of $\tan\beta \lesssim
8.5$. This leads to the lower cut on the left
branch of points in 
\reffi{figGEN1} (right) and
\reffi{figGEN3},
and also to the 
lower bound on $\tan\beta$ in
the right branch of points
in \reffi{figGEN2}.

It should also be noted that the parameter
point with the smallest value of $\chi^2_{\tautau}
+ \chi^2_{125}$ (indicated by the cyan star) 
in \reffi{figGEN3}
lies on the right branch of points. Since this branch
is further away from the center of the ellipses,
this branch has overall slightly larger values of $\chi^2_{\tautau}$.
The fact that we nevertheless find
the cyan star in this branch indicates that
for these points 
the compatibility with the measured properties of $h_{125}$ can be better
(yielding a lower $\chi^2_{125}$)
compared to the left branch of points.
Thus, parameter points
in which the additional decay channels $A \to Z h_{1,2}$
are relevant (left branch) are associated with
mildly larger deviations of the signal rates of $h_{125}$
compared to the SM prediction. However, we checked
that the values of $\chi^2_{125}$ can also be very close
to the SM value $\chi^2_{\rm SM, 125} \approx 84$ in the left
branch.
As a consequence, the accuracy of the 
signal rate measurements of $h_{125}$ at the (HL)-LHC~\cite{Cepeda:2019klc} 
will not be sufficient to fully discriminate between the points in the two
branches.

\begin{figure}
\centering
\includegraphics[width=0.4\textwidth]{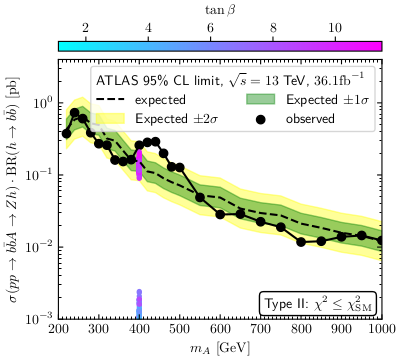}~
\includegraphics[width=0.4\textwidth]{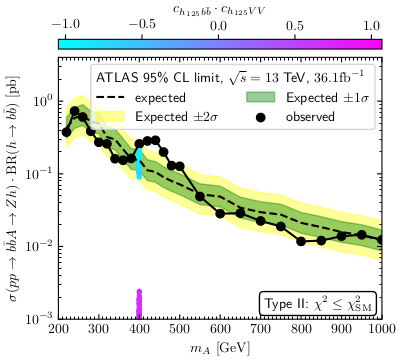}
\caption{\small 
Predicted rate for
$\sigma(b \bar b \to A \to Z  h_{125}) \times \br(h_{125} \to b \bar b)$
in comparison with the expected and observed $95\%$ confidence level
upper limits obtained 
by ATLAS~\cite{Aaboud:2017cxo}. The colors of the points indicate
the value of $\tan\beta$ (left) and
the value of the product
$c_{h_{125} b \bar b} \cdot c_{h_{125} VV}$ (right).}
\label{figGEN4}
\end{figure}

As already mentioned above,
in the N2HDM type~II
the presence of one of the decays
$A \rightarrow h_{1,2} Z$ with sizable
branching ratio 
can have the effect that 
the $\tautau$ excess is described so well that one encounters
a vanishing contribution to $\chi^2_{\tautau}$.
Regarding the points in the
wrong sign Yukawa coupling regime, it is
interesting to note that a local $3.6\,\sigma$ excess
at roughly $400\gev$ was found
by the ATLAS Collaboration
in the search $b \bar b \to
A \rightarrow h_{125} Z$~\cite{Aaboud:2017cxo}.
The corresponding search
for the $gg$ production mode has meanwhile been
updated including
the full Run~2 dataset, and no excess was found
at $m_A \approx 400\gev$.
As explained above 
(see also the discussion in \refse{secAZh})
the obtained limit 
gives rise to 
the lower bound on $\tan\beta$ for the left branch of points
in \reffi{figGEN1} (right) and
\reffi{figGEN3}.
In order to investigate whether the presence of a CP-odd Higgs boson $A$
with $m_A \approx 400\gev$ could on the one hand be
the origin of the excess
in the $b \bar b \to A \to Z h_{125}$ search
\footnote{As discussed in \citere{Ferreira:2017bnx}, the
$A \rightarrow Z h_{125}$ excess can be realized
in the wrong sign Yukawa coupling regime
of the 2HDM.}
while on the other hand
being compatible with the cross
section limits from the $g g \to A \to Z h_{125}$ search,
we show in \reffi{figGEN4}
the predicted rate for
$\sigma(b \bar b \to A \to Z  h_{125}) \times
\br(h_{125} \to b \bar b)$
for each parameter point, in combination
with the expected and observed upper limits
from the ATLAS analysis~\cite{Aaboud:2017cxo}.
The colors of the points indicate the value
of $\tan\beta$ (left) and
the value of the product
$c_{h_{125} b \bar b} \cdot c_{h_{125} VV}$ (right),
i.e., the blue points in the right plot
are in the wrong sign Yukawa coupling regime.
One can see that 
the points in the
wrong sign Yukawa coupling regime could indeed contribute to the excess
observed by ATLAS in the $b \bar b \to A$ production mode,
while 
at the same time being compatible
with the limits obtained for the $gg$ production mode. On the other
hand, for the points that are not in the wrong sign Yukawa coupling
regime in our scan we do not find a sizable contribution to the 
$\sigma(b \bar b \to A \to Z  h_{125}) \times \br(h_{125} \to b \bar b)$
rate, since the decays
$A \rightarrow h_{1,2} Z$, if kinematically open, 
are suppressed for the case where the Higgs boson in the final state is
$h_{125}$.

For the points that are not in the wrong sign Yukawa coupling regime our
analysis has revealed that the presence of a 
second Higgs boson $h_a$ with a mass below $300\gev$ can improve the
description of the $\tautau$ excess. It is tempting in this context to
entertain the possibility that this additional Higgs boson could have a mass
of about $96\gev$ and be the origin of the observed excesses in this mass
range, see the discussion in \refse{sec:excesses96}. Within the context of the
N2HDM of type~II, 
the simultaneous realization of the $\tautau$ excess and the observed
pattern around $96\gev$ leads to a very interesting scenario that is very
predictive and can serve as a 
benchmark scenario that can be probed in future experiments. We will
investigate this possibility in \refse{numhighII}.
Before turning to this discussion
we will investigate
whether the excesses at $96\gev$ can also
be accommodated in combination with
the $t \bar t$ excess. The latter, as
discussed in detail in
\refap{n2hdmstrategy}, can be realized both
in the type~II and the type~IV N2HDM.
A corresponding parameter scan
for the type~II is described in \refse{numlowII},
and for completeness the according scan in type~IV
can be found in \refap{numlowIV}.

\subsection{Higgs bosons at
\texorpdfstring{\boldmath{$96\gev$}}{96gev} and
\texorpdfstring{\boldmath{$400\gev$}}{400gev}
for low \texorpdfstring{\boldmath{$\tan\beta$}}{tb}
in type~II}
\label{numlowII}
In this section we present the analysis for the scan
in the low $\tan\beta$ regime of the type~II
N2HDM, which is dedicated
to accommodate the $t \bar t$ excess at $400\gev$
in combination with the excesses at $96\gev$.
The ranges of the free input parameters are given
in \refta{n2hdmlowtbparas},
where for the scan in the present section the $\tan\beta$ range specified
as $\tan\beta_{\rm low}$ is adopted.

\begin{table}
\centering
\def\arraystretch{1.5}
\footnotesize
\begin{tabular}{ccccccc}
$m_{h_1}$ & $m_{h_2}$ & $m_{h_3}$ &
    $m_A$ & $m_{H^\pm}$ & $\tan\beta_{\rm low}$ &
        $\tan\beta_{\rm high}$ \\
\hline
$[95,98]$ & $125.09$ & $[550, 1000]$ &
    $400$ & $[550, 1000]$ & $[0.5, 4]$ & $[6, 12]$ \\
\hline
\hline
$c_{h_2 VV, \rm low}^2$ & $c_{h_2 VV, \rm high}^2$ & $c_{h_2 t \bar t}^2$ &
    $\mathrm{sign}(R_{23})$ & $R_{13}$ &
        $m_{12}$ & $v_S$ \\
\hline
$[0.6, 0.9]$ & $[0.6, 1.0]$ & $[0.6, 1.2]$ & $-1,1$ &
    $[-1, 1]$ & $[0, 1000]$ & $[10, 1500]$
\end{tabular}
\caption{\small Values of input parameters for the
scans in the low and high $\tan\beta$ regime.}
\label{n2hdmlowtbparas}
\end{table}

We start by briefly commenting on our choice of parameter
ranges.\footnote{The requirement of a Higgs boson
at $96\gev$ in combination with the
experimental constraints fixes the mass ordering of the
CP even Higgs bosons, i.e.\
$h_{125} = h_2$. We will therefore use
the notation $h_{1,2,3}$ in the following.}
The lightest Higgs boson $h_1$ plays
the role of the candidate 
that is associated with
the excesses
at $96\gev$. Thus, the mass $m_{h_1}$
was chosen to be in this range. The SM-like
Higgs boson is $h_2$ with a mass of
$m_{h_2} = 125.09\gev$. As before,
the $t \bar t$ excess will be
described in terms of the CP-odd Higgs boson with
a mass of $m_A = 400\gev$. The scan range
of the charged Higgs-boson mass
$m_{H^\pm} = [550,1000]\gev$ is chosen due
to a lower limit of
$m_{H^\pm} \gtrsim 550\gev$ arising from the
flavor constraints. The indirect constraints
from the EWPO lead to the restriction that
the third CP-even Higgs boson should have a similar
mass $m_{h_3} \approx m_{H^\pm}$, since
all remaining Higgs bosons have fixed masses
much below the lower limit on $m_{H^\pm}$.
Thus, $m_{h_3}$ is scanned in the same mass
range as $m_{H^\pm}$.
The range of $\tan\beta =
\tan\beta_{\rm low}$ gives rise to sizable
couplings of the CP-odd Higgs boson $A$ to top quarks,
which is desired for the description of the $t \bar t$ excess
(see the discussion in \refap{n2hdmstrategy}).
The ranges for the squared couplings of the
SM-like Higgs boson $h_2$ were taken over from
\citere{Biekotter:2019kde}, where it
was shown that such values are suitable for
explaining the excesses at $96\gev$
in the N2HDM. To be precise,
these ranges allow a substantial
mixing between $h_1$ and $h_2$, such that
$h_1$ can have sizable couplings to
quarks and gauge bosons, while
the properties of $h_2$ are still
in agreement with the signal-rate measurements
of the discovered Higgs boson at around $125\gev$.
The remaining parameters
in \refta{n2hdmlowtbparas}, $R_{13}$, $m_{12}$ and $v_S$,
were scanned over
a wide range, and both possible signs of $R_{23}$ were taken into account.
We emphasize that the requirement of simultaneously explaining
the $t \bar t$ excess and the excesses at
$96\gev$ practically fixes the entire scalar
spectrum of the model. The only remaining free mass
parameter is the common mass scale of $h_3$ and
$H^\pm$. However, this scale cannot be much larger than
$m_A$ (which is fixed to $400\gev$) as a consequence of the applied
constraints (see also the discussion below).

The results of this scan will be quantified in terms
of the total $\chi^2$ given by the
contributions arising from the
fit to the excesses at $96\gev$,
the fit to the $t \bar t$ and the $\tautau$
excesses at $400\gev$, and
the signal-rates
measurements of the Higgs boson at $125\gev$, such that
\begin{equation}
\chi^2 = \chi^2_{96} + \chi^2_{125} +
\chi^2_{t \bar t} +
\chi^2_{\tautau} \ .
\label{eqchisqtt}
\end{equation}
We include here for consistency also the contribution from
the $\tautau$ excess even though no
sizable contribution to the correspoding signal
cross sections are expected in the
low $\tan\beta$ regime.
All points with $\chi^2 \leq \chi^2_{\rm SM}$
are considered to be acceptable, where $\chi^2_{\rm SM}$
is evaluated assuming no signal contributions to
the excesses. 
Their contribution to $\chi^2_{\rm SM}$ amounts to
${\chi^2_{\rm SM, 96} + \chi^2_{\rm SM, \tautau}
+ \chi^2_{\rm SM, t \bar t} = 37.2}$.
The best-fit point in our scan is determined
by the lowest $\chi^2$ value
according to \refeq{eqchisqtt}.

\begin{figure}
\centering
\includegraphics[width=0.33\textwidth]{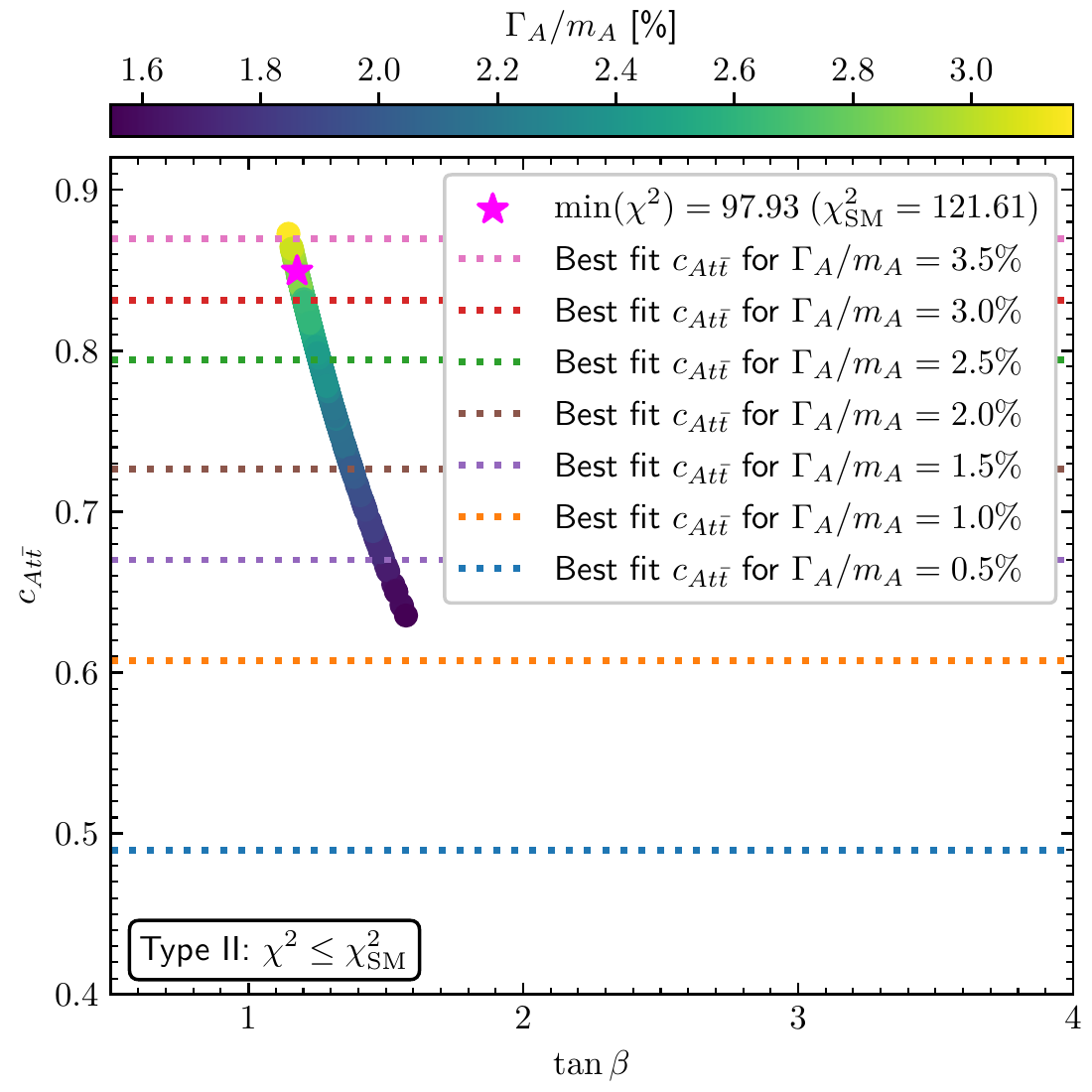}~
\hspace*{-0.6cm}~
\includegraphics[width=0.33\textwidth]{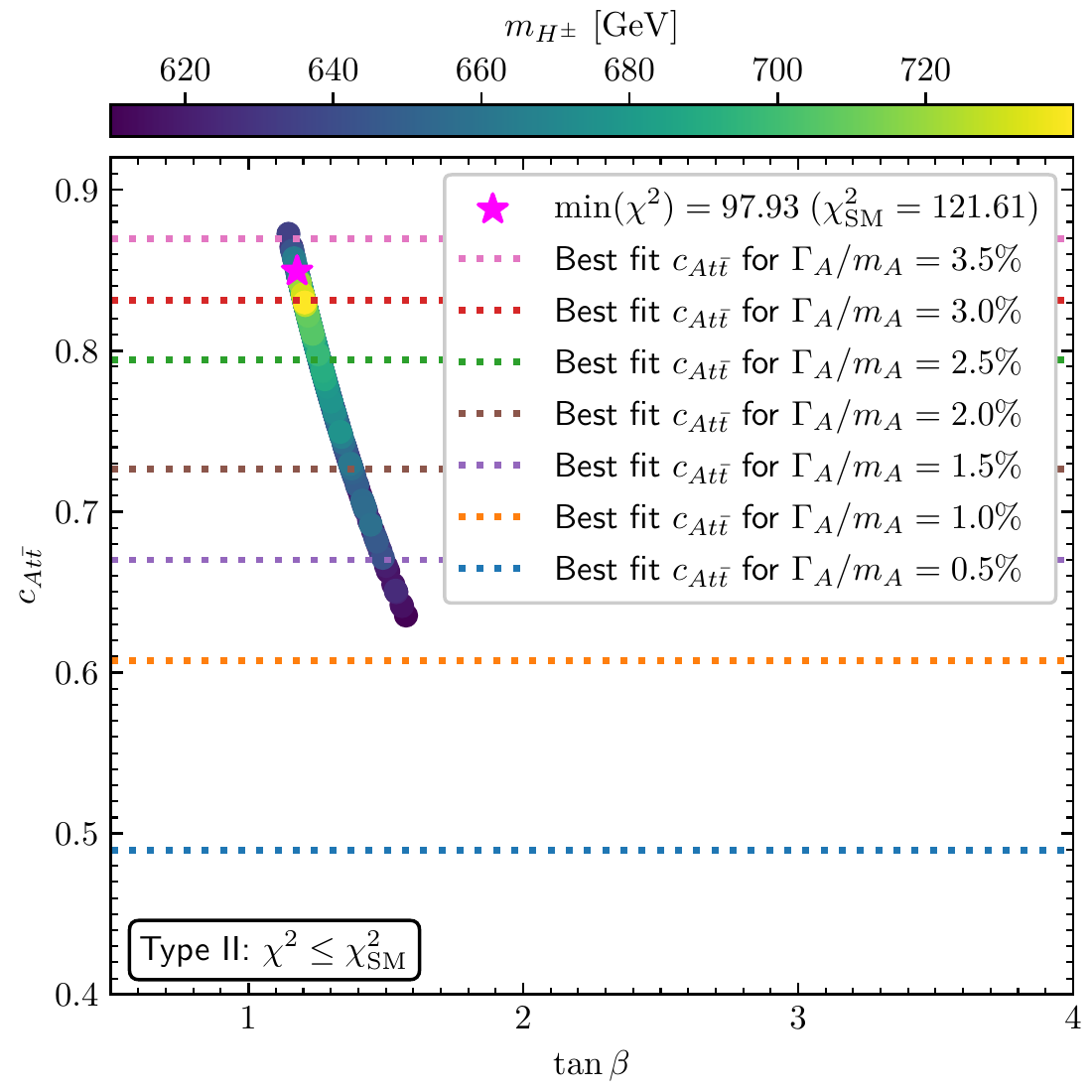}~
\hspace*{-0.6cm}~
\includegraphics[width=0.33\textwidth]{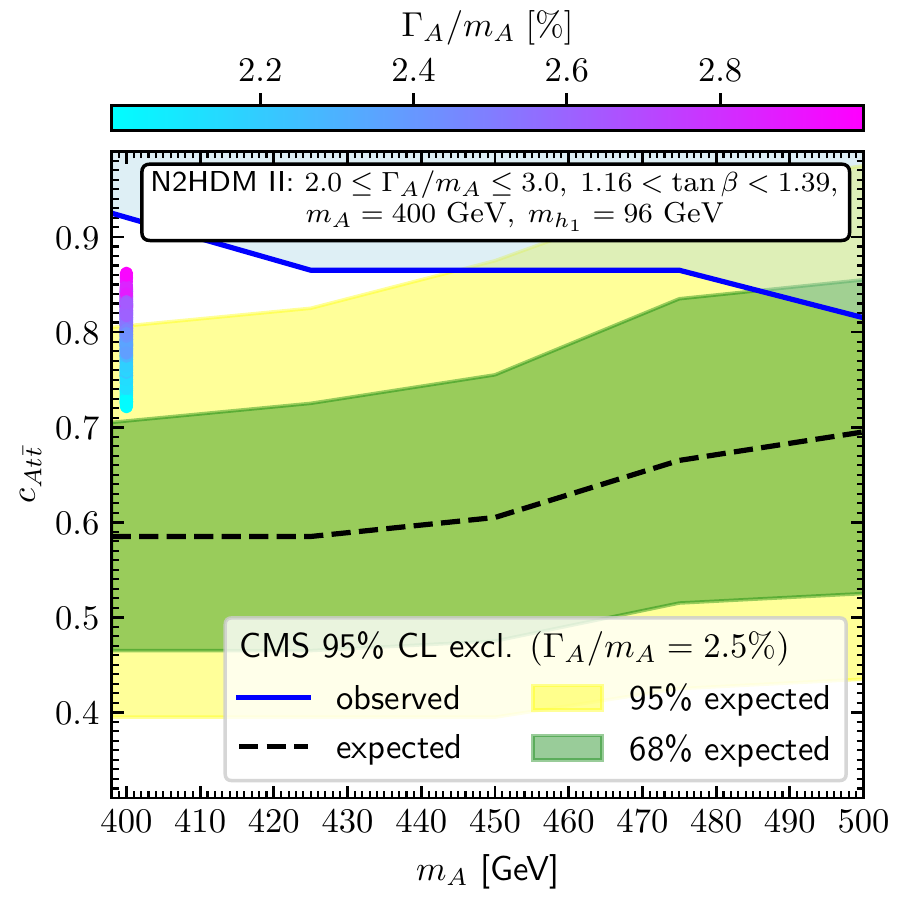}
\caption{\small
Left and center: $c_{A t \bar t}$ in dependence of $\tan\beta$.
The colors of the points indicate the values
of $\Gamma_A / m_A$ in \% (left) and
the values of $m_{H^\pm}$ (center).
The dashed horizontal
lines indicate the best-fit values of $c_{A t \bar t}$
for different width hypotheses in the
experimental analysis~\cite{Sirunyan:2019wph}.
Right:
Values of $c_{A t \bar t}$ for the parameter points
with $2.0\% \leq \Gamma_A / m_A \leq 3.0\%$ in comparison to
the observed (blue)
and expected (black dashed) upper limits
at the $95\%$ C.L.\ as well as the corresponding
$1\sigma$ (green) and $2\sigma$ (yellow)
regions around the expected limit assuming
$\Gamma_{A}/m_{A} = 2.5\%$, as
published in \citere{Sirunyan:2019wph}.}
\label{figttIItt}
\end{figure}

In \reffi{figttIItt} we show the values of the coupling
coefficient $c_{A t \bar t}$ in dependence of
$\tan\beta$ for all the parameter points with
$\chi^2 \leq \chi^2_{\rm SM}$. One can see that the
parameter points lie in a narrow range of
$1.1 \lesssim \tan\beta \lesssim 1.6$.
The preference for those low values 
of $\tan\beta$
is related to the fact that for values of
$\tan\beta \gtrsim 2$ no sizable contribution to
the $t \bar t$ excess is present. 
The feature that no points are displayed in our scan 
for higher values of $\tan\beta$ although our scan has been performed up to
$\tan\beta = 4$ is mainly due to the sampling used in our scan in combination
with the fact that many points in this region do not fulfill 
the condition $\chi^2 \leq \chi^2_{\rm SM}$.
For the allowed points, the size of the couplings
$c_{A t \bar t}$ and the
values for
$\Gamma_A / m_A$, indicated by the colors of the
points in the left plot of \reffi{figttIItt},
provide a very good fit to the $t \bar t$ excess.
In particular, for values slightly above $\tan\beta = 1$
we find $c_{A t \bar t} \approx 0.8$
and $\Gamma_A / m_A \approx 2.5\%$, 
which gives rise to values of
$\chi^2_{t \bar t} < 1$.
In the middle plot of
\reffi{figttIItt} the colors of the points indicate
the value of the charged Higgs boson mass
$m_{H^\pm}$. Here it is important to note that no
values above $m_{H^\pm} \approx 750\gev$ were found.
This upper limit arises from theoretical constraints
on the allowed values of the quartic couplings
under the condition that $m_{h_1} = 96\gev$,
$m_{h_2} = 125\gev$ and $m_A = 400\gev$.
Hence, the presence of a charged Higgs
boson, and due to the constraints on the $T$ parameter
also of the third CP-even Higgs boson $h_3$, with
masses $m_{h_3} \approx m_{H^\pm}$
below the TeV scale are firm predictions
of the scenario presented here.
In the right plot of \reffi{figttIItt}
we again show the subset of points that
feature values of $2.0\% \leq \Gamma_A / m_A \leq 3.0\%$
in comparison to the experimental expected
and observed exclusion limits~\cite{Sirunyan:2019wph}.
All of the points
lie close to the experimental best-fit value
of $c_{A t \bar t} \approx 0.8$ for a width
of $\Gamma_A / m_A = 2.5\%$.

\begin{figure}
\centering
\includegraphics[width=0.48\textwidth]{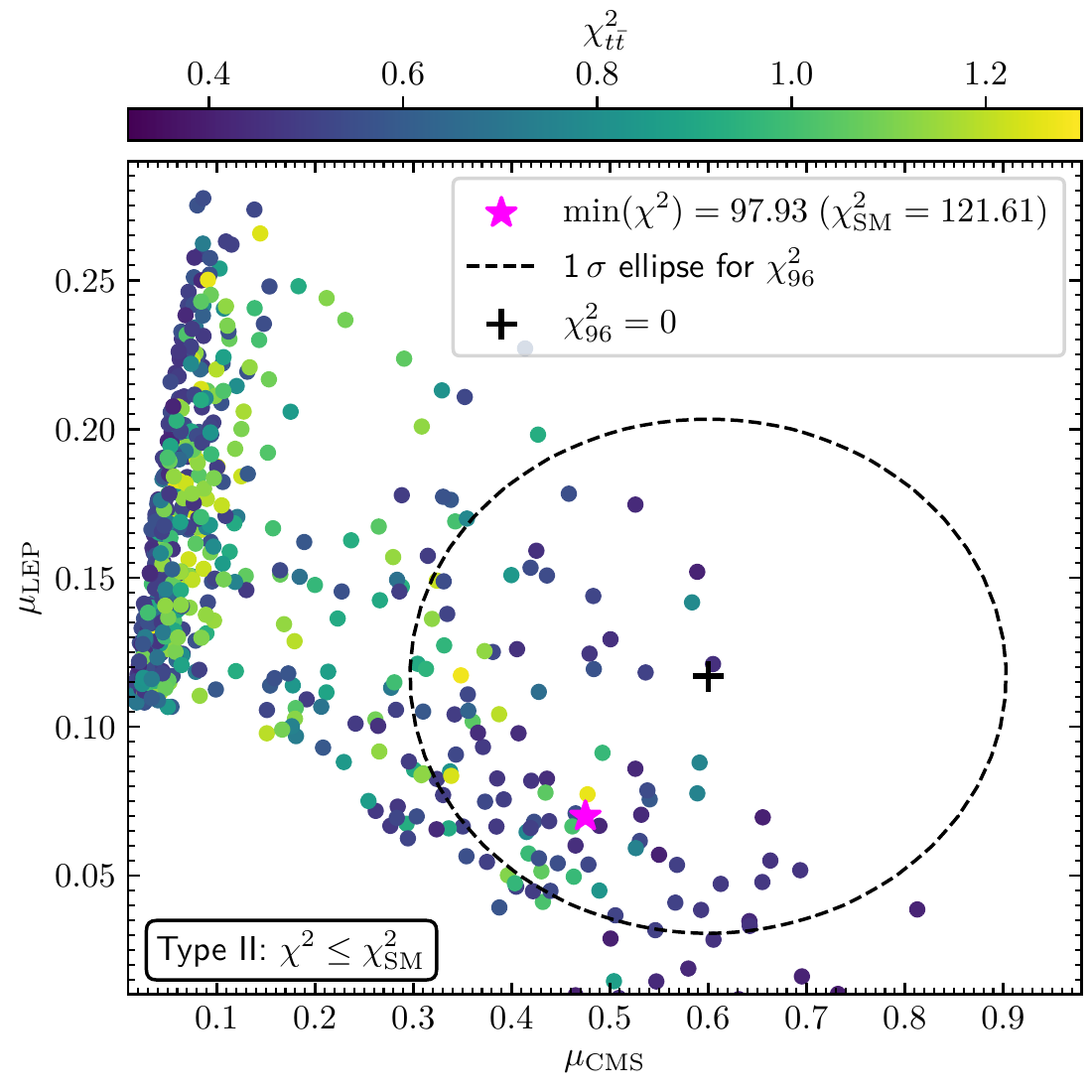}~
\includegraphics[width=0.48\textwidth]{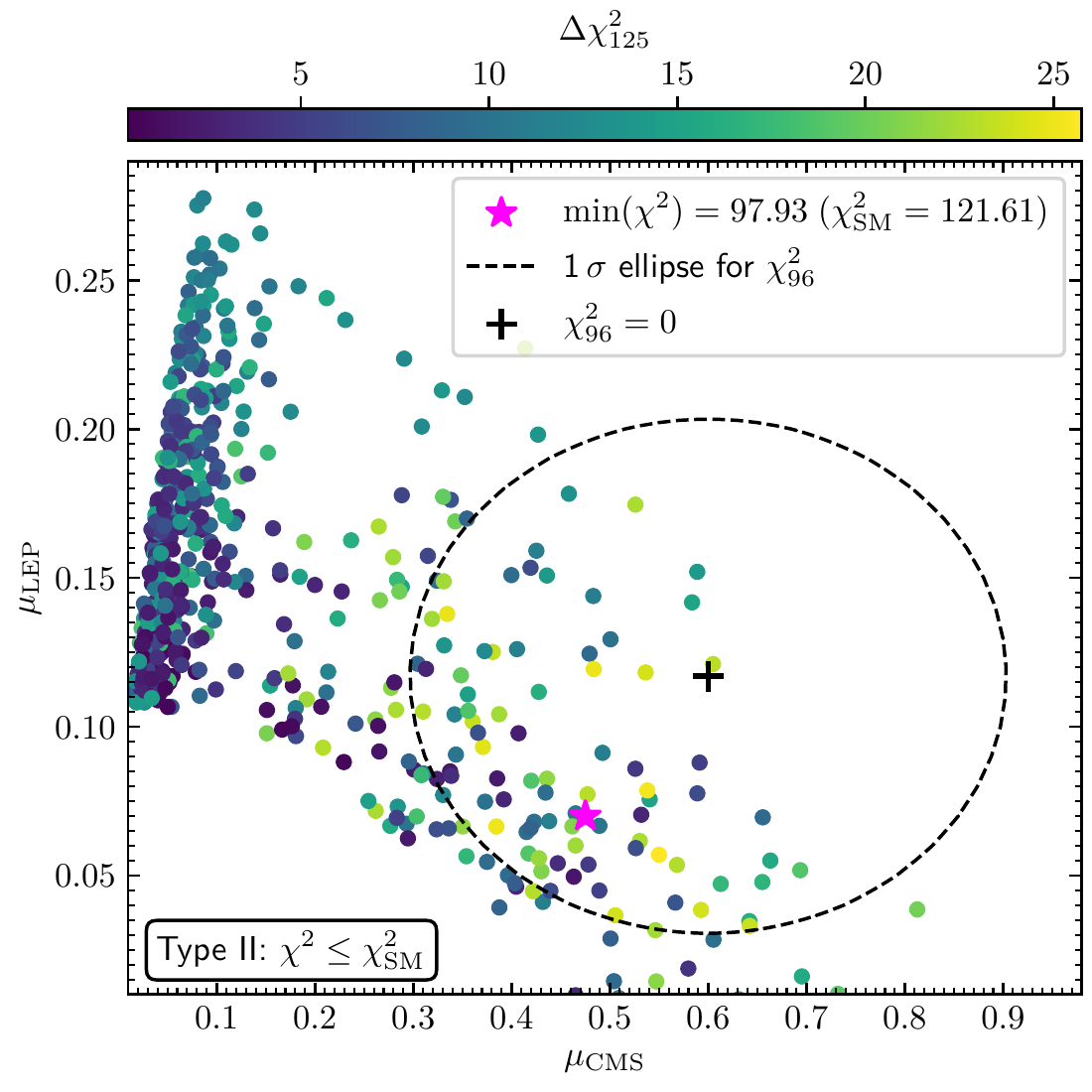}
\caption{\small 
The $\mu_{\rm CMS}$--$\mu_{\rm LEP}$ plane
for the points of the low $\tan\beta$ scan in the
type~II of the N2HDM.
The black ellipse indicates the $1\sigma$ region of
$\chi^2_{96}$ with its center marked with
a black cross.
The best-fit point is highlighted with
a magenta star.
The colors of the points indicate
$\chi^2_{t \bar t}$ in the left
plot and $\Delta \chi^2_{125}$ in the right plot.}
\label{figttII}
\end{figure}

The width of the CP-odd Higgs boson $A$ normalized to its mass ranges
between roughly $1.5\%$ and $3.5\%$, as can be seen in the left plot
of \reffi{figttIItt}. These values are substantially larger than the
ones that one would obtain within the 2HDM for a CP-odd Higgs boson
at $400\gev$.
This is due to the fact that in the considered N2HDM scenario
$A$ has additional decay channels available
into $h_{1,2}$ and a $Z$ boson.
In the experimental analysis~\cite{Sirunyan:2019wph}
it was found that
for 
values of $\Gamma_A / m_A \approx 2.5\%$
the expected
$95\%$ C.L.\ exclusion limit is at
roughly $c_{A t \bar t} \approx 0.6$,
while the observed exclusion limit varies
(depending on the precise value
of $\Gamma_A / m_A$)
between $c_{A t \bar t} \approx 0.9$ and
$c_{A t \bar t} \approx 1.1$. Thus, our scan essentially covers
the 
range of $c_{A t \bar t}$ that is associated with the observed excess,
as can be seen in \reffi{figttIItt}. This is also reflected in the
very small values of $\chi^2_{t \bar t}$, as
discussed in the following.

We show the parameter points of the
scan in the
$\mu_{\rm CMS}$--$\mu_{\rm LEP}$ plane in
\reffi{figttII}.
The color coding of the points indicates the values
of 
$\chi^2_{t \bar t}$ and
$\Delta \chi^2_{125} \equiv
\chi^2_{125} - \chi^2_{\mathrm{SM},125}$ in the
left and right plot, respectively.
One can see
that a large
fraction of points lies inside the $1\sigma$
region of $\chi^2_{96}$ while at the same time
reproducing the excess observed for $c_{A t \bar t}$,
since many points have $\chi^2_{t \bar t} \approx 0$.
Thus, we conlcude that the low $\tan\beta$
region of the type~II N2HDM is well suited for accommodating
the excesses at $96\gev$ in combination with
the $t \bar t$ excess at $400\gev$.
Moreover, we can observe
in the right plot of \reffi{figttII} that several
points have values of $\Delta \chi^2_{125}$ very close
to zero. Consequently, the fit to the
excesses can be realized without being in tension
with the signal rates measurements of the SM
Higgs boson.
The best fit point with a value
of $\chi^2 = 97.97$
is substantially below the
SM value $\chi^2_{\rm SM} = 121.61$.

We also performed
a scan in the low $\tan\beta$ region of type~IV
with the goal of explaining the $t\bar t$
excess and the excesses at $96\gev$.
We found that also this type can account
for the excesses, but with a slightly worse
fit result regarding the CMS excess and
the properties of $h_{125}$. This scan
is presented in \refap{numlowIV} for completeness
and to allow for a
comparison to the type~II results.

\subsection{Higgs bosons at
\texorpdfstring{\boldmath{$96\gev$}}{96gev} and
\texorpdfstring{\boldmath{$400\gev$}}{400gev}
for large \texorpdfstring{\boldmath{$\tan\beta$}}{tb}
in type~II}
\label{numhighII}
As was discussed
before, the type~II N2HDM
is the only candidate of the four different Yukawa
types that can potentially accommodate the
$\tautau$ excess at $400\gev$
in combination with the two excesses at
$96\gev$.
The existence of a region of
parameter space fulfilling the
relevant criteria
(see the discussion in \refap{n2hdmstrategy})
depends on the simultaneous enhancement
of $\sigma( b \bar b \rightarrow A)$ and
$\br (A \rightarrow \tautau)$, such that both the signal rates
in the $gg$ production mode and the $b \bar b$
production mode have the adequate size for
providing a good description of the data.
Here we present a scan in the
high $\tan\beta$ regime dedicated to answer the
question whether there is a $\tan\beta$ range
for which the observed excesses can be accommodated simultaneously.
The input parameters are the same as the ones in
the low $\tan\beta$ regime (as shown in \refta{n2hdmlowtbparas}), but
with considerably larger values for $\tan\beta$
in the range $6 \leq \tan\beta \leq 12$,
and the upper limit of $c_{h_2 VV} \leq 0.9$ is
removed.\footnote{For larger values of
$\tan\beta$ it is possible to achieve relatively large values
of $\mu_{\rm CMS}$ even for $c_{h_2 VV} > 0.9$.
It is then also possible to find points within the
$1\sigma$ ellipse of $\chi^2_{96}$ even though
hardly any signal is present for
the LEP excess.}

As before, we refer to the point with
the lowest $\chi^2$ value as the best-fit
point, where the same definition as
in \refeq{eqchisqtt} was used 
to obtain the total $\chi^2$ value.
Moreover, in the results of our scan we only display
points with a lower value of the overall
$\chi^2$ than the one of the SM, i.e.,
$\chi^2 \leq \chi^2_{\rm SM}$.
Note that for consistency we again include both the
contributions $\chi^2_{\tautau}$
and $\chi^2_{t \bar t}$ in the fit.
However, for the high $\tan\beta$ values
considered here no sizable contribution
to the $t \bar t$ excess is expected.
The condition $\chi^2 \leq \chi^2_{\mathrm{SM}}$
is practically unchanged
depending on whether
$\chi^2_{t \bar t}$ is taken
into account here or not, as can also
be seen in \reffi{figGEN2}, in which a difference
of $\chi^2_{\mathrm{SM},t \bar t} -
\chi^2_{t \bar t} \lesssim 2$ is visible for
$\tan\beta \geq 6$.

\begin{figure}
\centering
\includegraphics[width=0.48\textwidth]{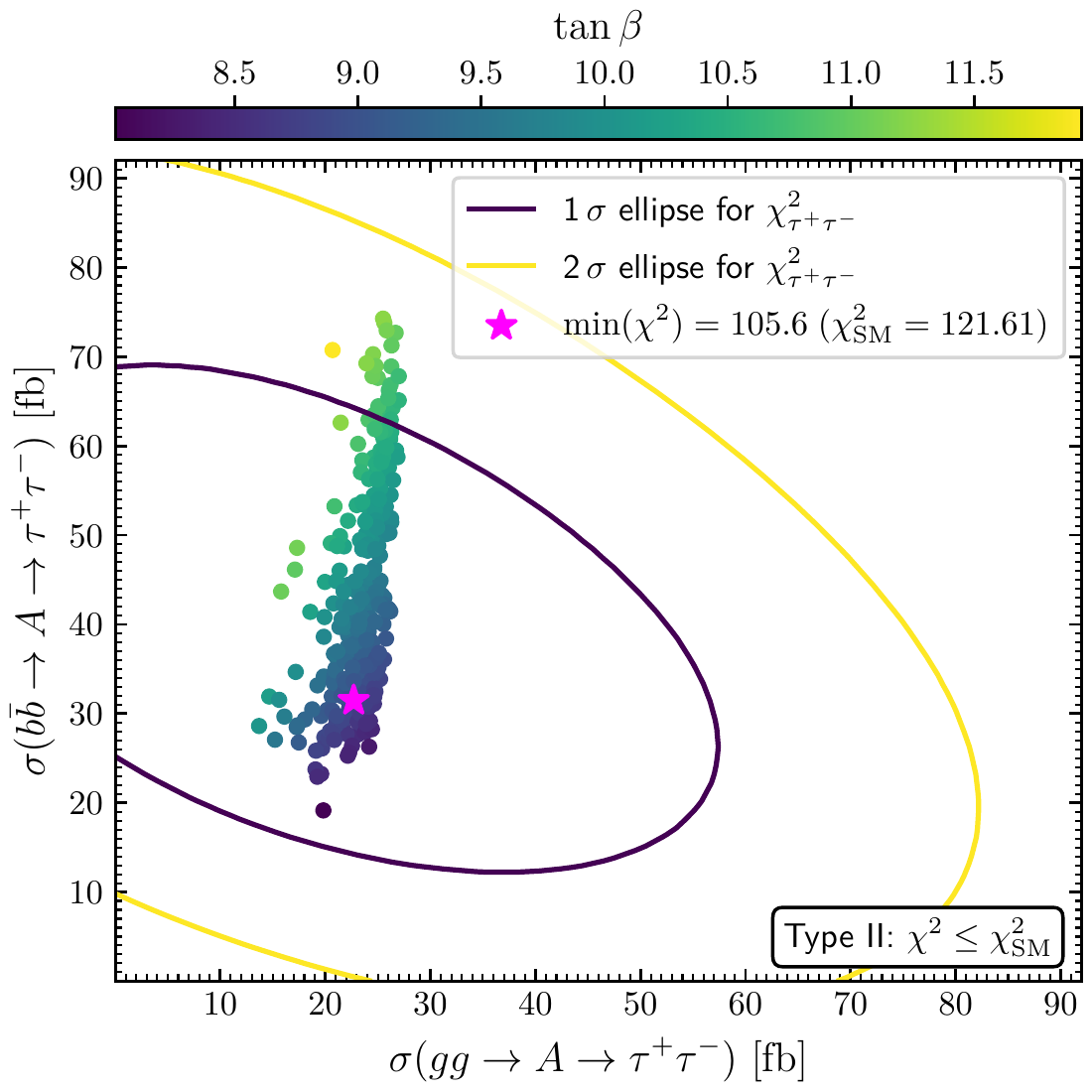}~
\includegraphics[width=0.48\textwidth]{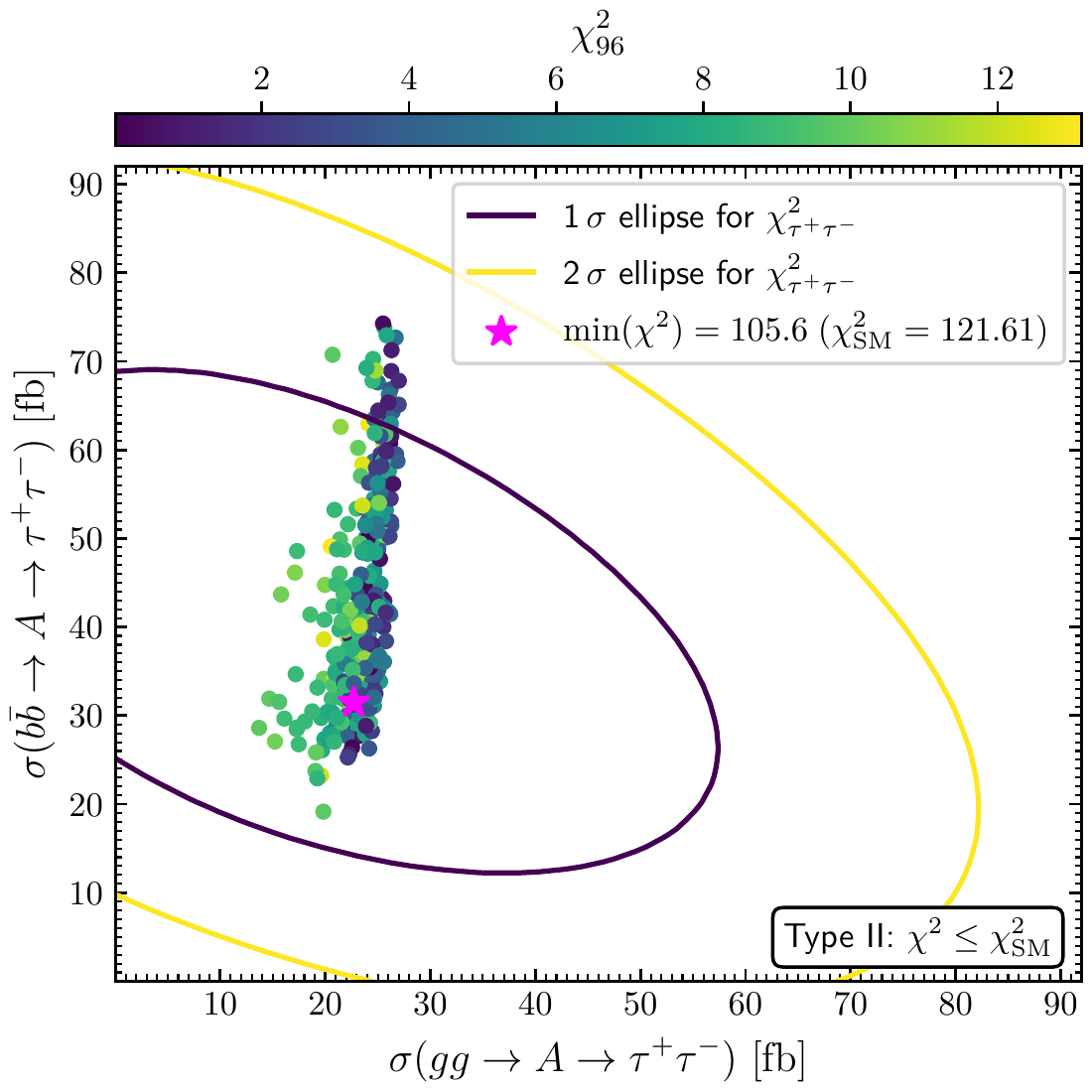}
\caption{\small
Signal cross sections of $A \to \tautau$
for the $gg$ production mode on the
horizontal axis and the $b \bar b$ production mode
on the vertical axis. The colors of the points
indicate the value of $\tan\beta$ (left)
and $\chi^2_{96}$ (right).
The dark blue
and the yellow lines indicate the $1\sigma$
and $2\sigma$ ellipse of $\chi^2_{\tautau}$, 
respectively.
The best-fit point is indicated with a magenta star.}
\label{figll1}
\end{figure}

We show the parameter points of
the type~II scan at high $\tan\beta$
in the plane of the signal cross sections regarding
the $\tautau$ excess in \reffi{figll1}
and regarding the signal rates of
the excesses at $96\gev$
in \reffi{figll2}.
In the left plot of \reffi{figll1} one can see
that, as anticipated from the couplings of the $A$ boson to fermions 
in type~II, the signal rate for the $b \bar b A$ production mode grows with
increasing values of $\tan\beta$, while
$\sigma(g g \rightarrow A \rightarrow \tautau)$ shows less sensitivity
to $\tan\beta$.
For values
of $6 \leq \tan\beta \leq 11$ the points lie
within the $1\sigma$ region regarding $\chi^2_{\tautau}$,
which is indicated by the dark blue ellipse.
Moreover, for $\tan\beta \approx 10$ the points
approximately lie at the
center of the ellipses, indicating that
the observed excesses are well described.
We emphasize that this is a non-trivial
result. Given the fixed mass spectrum
in this scenario,
and the fact that
the presence of $h_{96}$
gives rise to the additional decay channel
$A \to Z h_{96}$, the location of the band on which
the points are found in \reffi{figll1} is fixed approximately
(the band corresponds to the left branch of points
in \reffi{figGEN3}).
Hence, a type~II
N2HDM CP-odd Higgs boson $A$ with a mass of
$m_A = 400\gev$, in combination with
a CP-even Higgs boson $h_1$ at around $96\gev$,
yields a predicted pattern that precisely matches the one of the 
excesses observed
by ATLAS in the Higgs searches in the $\tautau$ final state.

\begin{figure}
\centering
\includegraphics[width=0.48\textwidth]{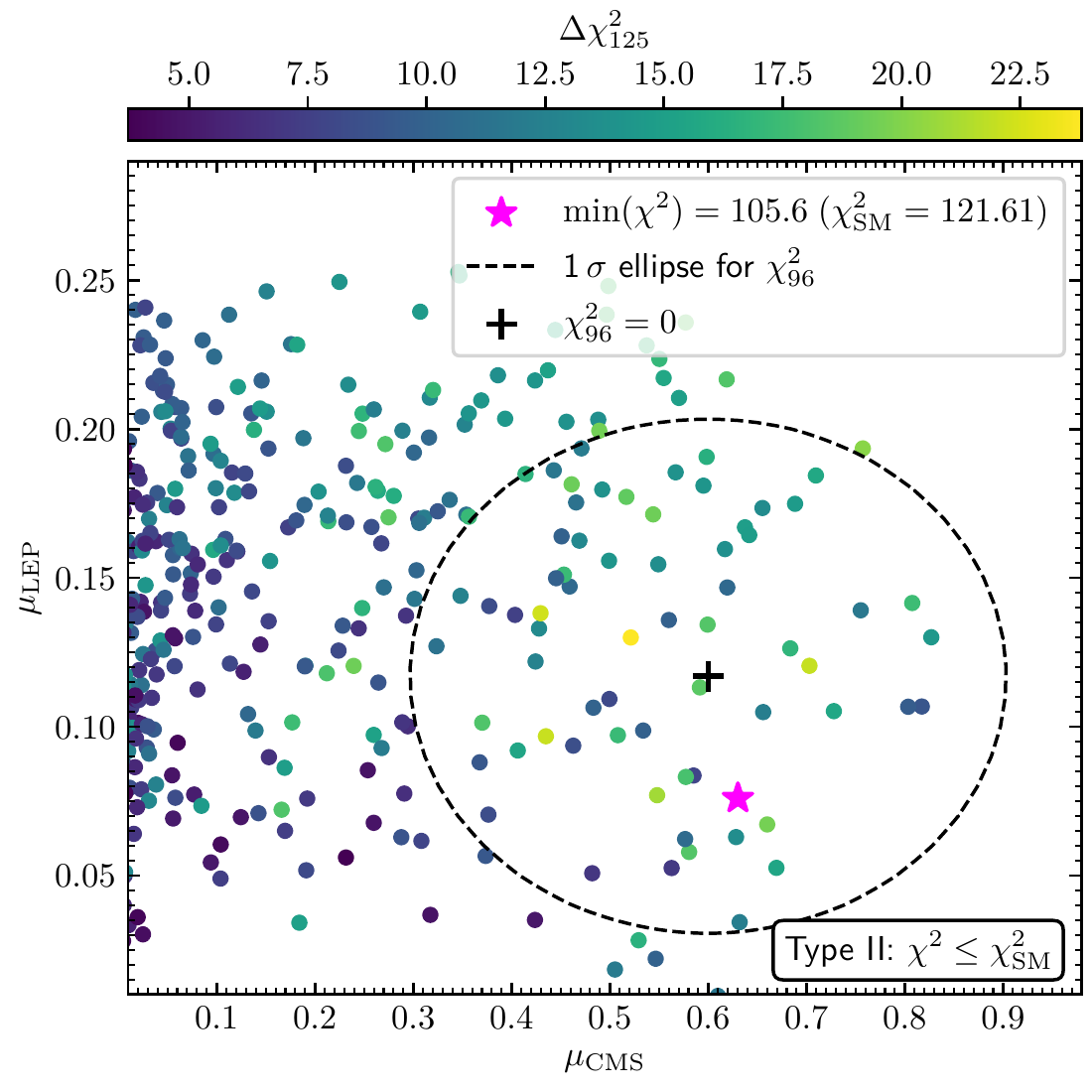}~
\includegraphics[width=0.48\textwidth]{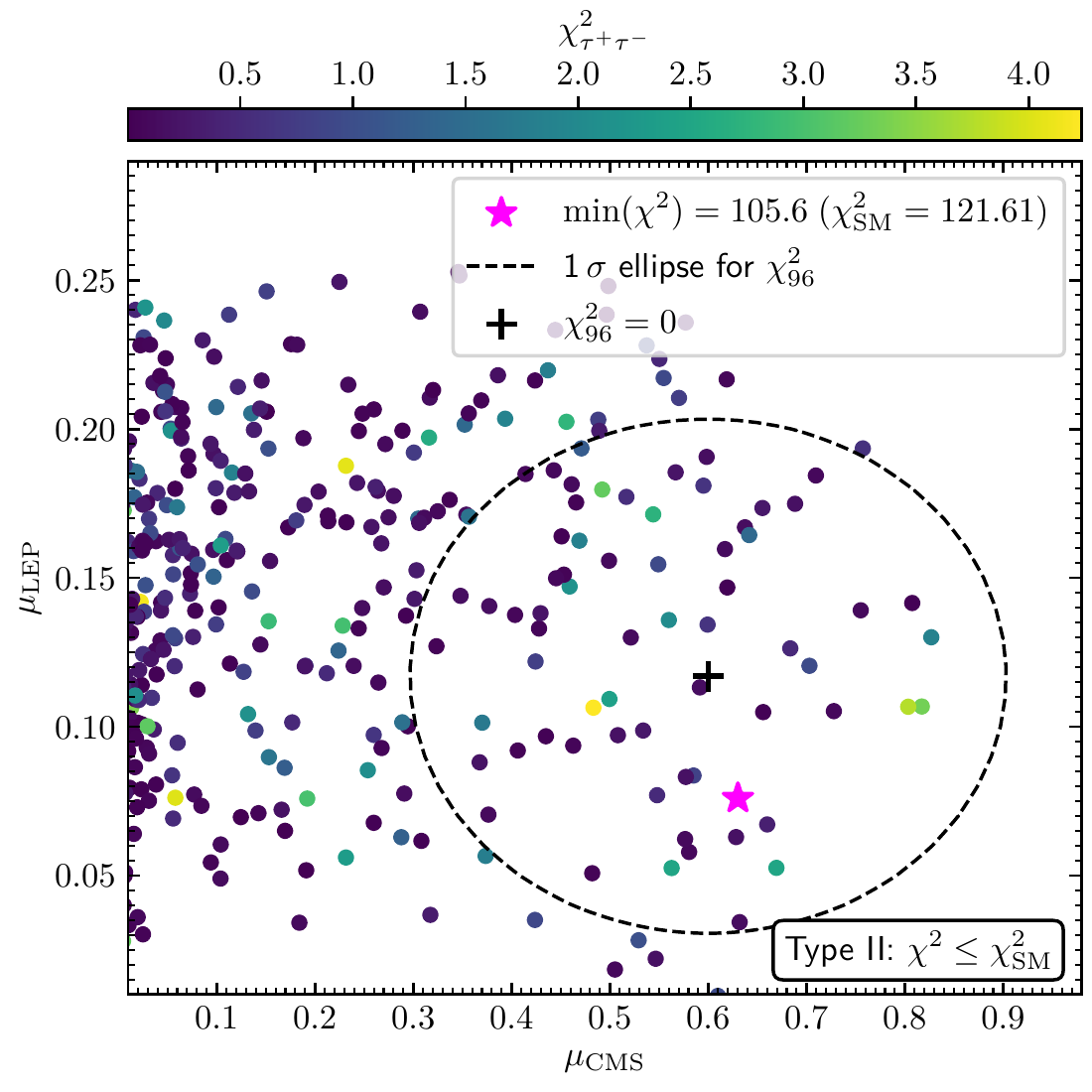}
\caption{\small
The $\mu_{\rm CMS}$--$\mu_{\rm LEP}$ plane
for the points of the high $\tan\beta$ scan in the
type~II of the N2HDM.
The black ellipse indicates the $1\sigma$ region of
$\chi^2_{96}$ with its center marked with
a black cross.
The best-fit point is highlighted with
a magenta star.
The colors of the points indicate
$\Delta \chi^2_{125}$
in the left plot,
and $\chi^2_{\tautau}$ in the right plot.}
\label{figll2}
\end{figure}

In the right plot \reffi{figll1} we show the same
parameter points, however with the color of the
points indicating the value of $\chi^2_{96}$.
One can see that a large fraction of points that lie
within the $1\sigma$ ellipse of $\chi^2_{\tautau}$
also have very small values of $\chi^2_{96}$.
Consequently, the parameter space covered in
our scan contains parameter points in which
the excesses at $96\gev$ are accommodated
simultaneously with the $\tautau$ excess
at $400\gev$.
One can see a slight correlation between
the cross section
$\sigma(g g \rightarrow A \rightarrow \tautau)$
and $\chi^2_{96}$. Larger values of the cross section
correspond to, on average, lower
values of $\chi^2_{96}$.

In \reffi{figll2} we show the 
results of our scan
in the plane $\mu_{\rm CMS}$--$\mu_{\rm LEP}$.
In the left plot one can see that
$\Delta \chi^2_{125}$, indicated
by the colors of the points, ranges from values
of $\approx 3.5$ to 23. In the scan in the low
$\tan\beta$ regime of type~II also smaller values
closer to zero could be found (see
right plot of \reffi{figttII}).
Thus, the scenario in the high $\tan\beta$ regime
might be associated with larger deviations of
the properties of the SM Higgs boson at $125\gev$
compared to the SM prediction.
However, for the points with the
smallest values of $\Delta \chi^2_{125}$
the deviations of the signal rates of $h_{125}$
compared to the SM prediction
are much below the current
experimental uncertainties, and
also more precise future measurements
at the HL-LHC might not be sufficient
to entirely probe this scenario.
One can also see in the right plot of
\reffi{figll2} that many points with low values
of $\chi^2_{\tautau}$ lie within the
$1\sigma$ ellipse regarding $\chi^2_{96}$.
The best-fit point, indicated by the magenta
star, is 
close to the central point of
the ellipse. We conclude that the type~II
N2HDM is perfectly capable of accommodating
the excesses at $96\gev$ in combination
with the $\tautau$ excess for values
of $\tan\beta \approx 10$.

Since we saw in \refse{fullII} that for the high
$\tan\beta$ region the CP-odd Higgs boson can also
provide an explanation for the excess in
the $Z h$ final state, we show
the corresponding signal cross
sections for the points of this scan in
\reffi{figll3}. As before, the subset
of points that are in
the wrong-sign Yukawa coupling regime
provide relatively large contributions
to the excess, as can be seen in the right
plot of \reffi{figll3},
while the other parameter points do not
give rise to a signal in the
$A \to Zh$ channel.
Overall, the values for the cross sections
are slightly smaller here than what was
found in \refse{fullII}. The reason is
that here $A$ has an additional decay mode
to a $Z$ boson and the lightest Higgs boson
$h_1$ at around $96\gev$.

\begin{figure}
\centering
\includegraphics[width=0.4\textwidth]{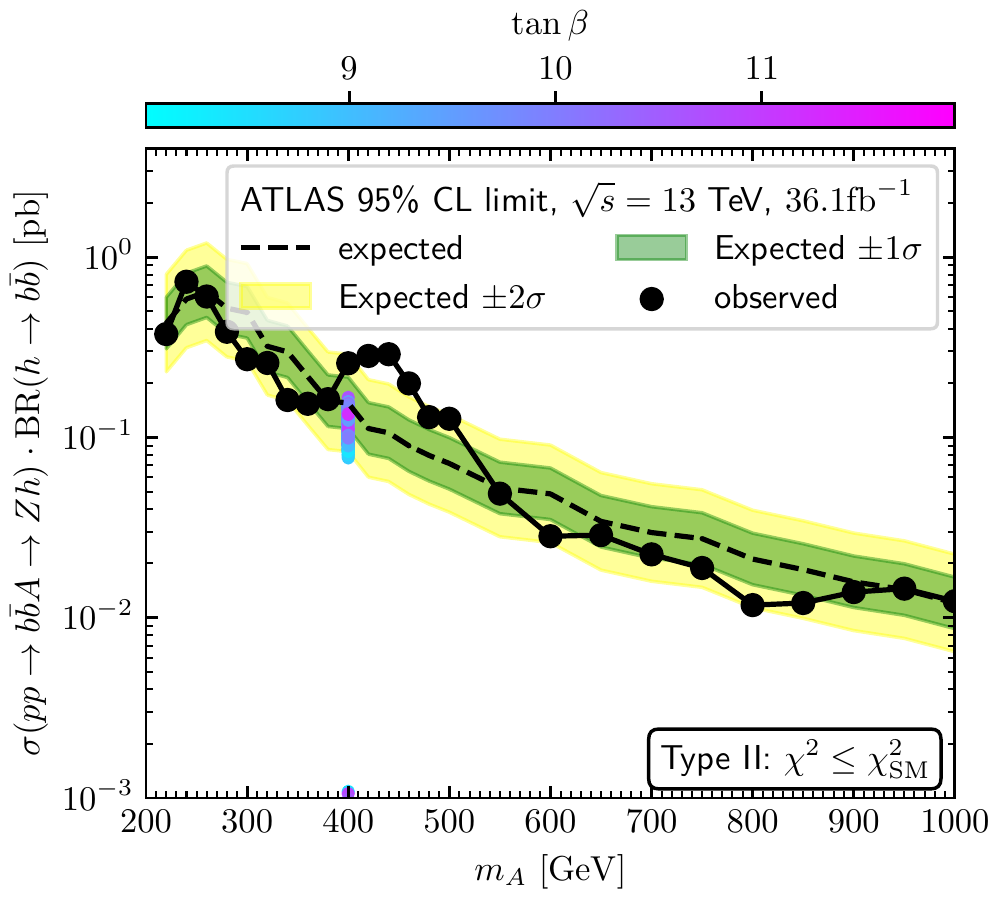}~
\includegraphics[width=0.4\textwidth]{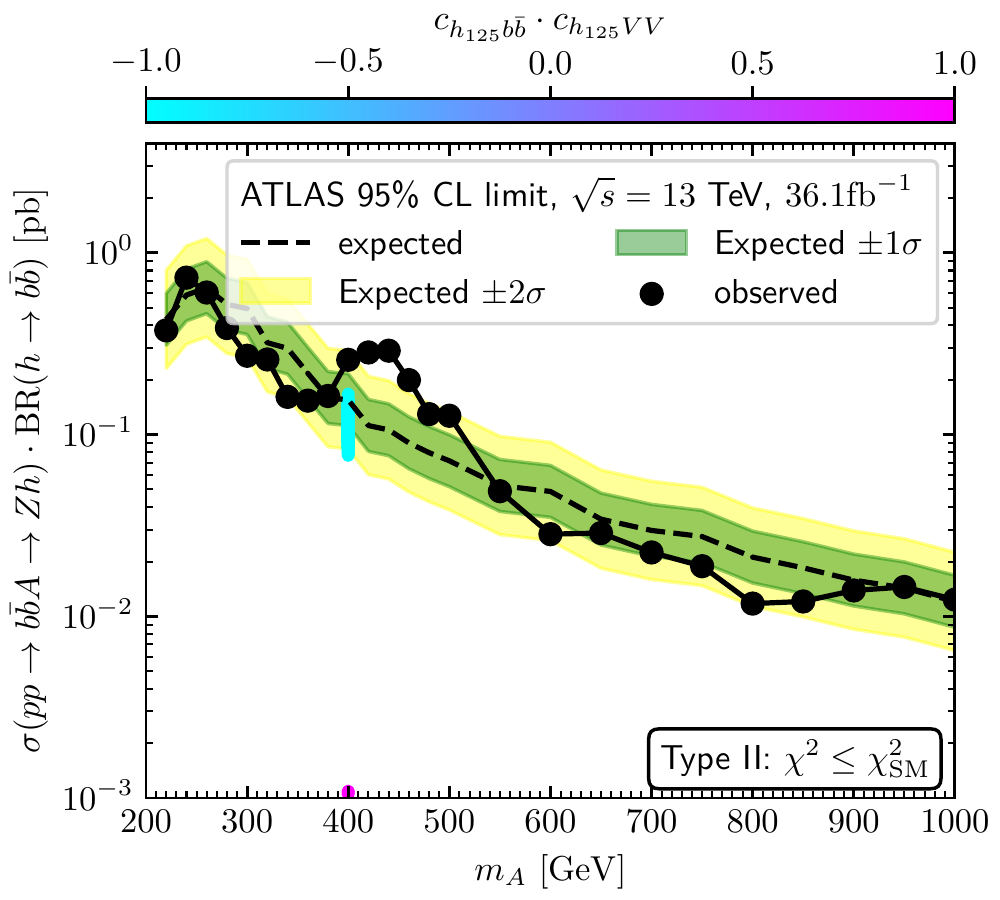}
\caption{\small Predicted rate for
$\sigma(b \bar b \to A \to Z  h_{125}) \times \br(h_{125} \to b \bar b)$
in comparison with the expected and observed $95\%$ confidence level
upper limits obtained 
by ATLAS~\cite{Aaboud:2017cxo}. The colors of the points indicate
the value of $\tan\beta$ (left) and 
the value of the product
$c_{h_{125} b \bar b} \cdot c_{h_{125} VV}$ (right).}
\label{figll3}
\end{figure}

\section{NMSSM interpretation}
\label{sec:nmssm}

Our analysis in the previous section has shown that the N2HDM of type~II
provides an attractive framework for accommodating either the 
$t \bar t$ or the $\tautau$ excess at $400\gev$ simultaneously with the
observed excesses at $96\gev$.
In SUSY models, due
to the holomorphy of the superpotential,
the presence of two Higgs doublet fields
with opposite hypercharge, usually denoted as
$H_u$ and $H_d$, is required for
the mass generation of both up- and down-type
fermions. Identifying $H_d = \Phi_1$ and
$H_u = \Phi_2^*$, there is a 
correspondence
between the Yukawa structure of the (N)2HDM type~II
and SUSY extensions of the SM.
Thus, it is an interesting question whether
the excesses can also be realized in SUSY
models. It should be noted in this context that SUSY yields further restrictions
on the 
Higgs sector compared to non-supersymmetric models with
a similar structure of the extended Higgs sector.

The simplest SUSY extension of the SM is the 
Minimal Supersymmetric Standard Model
(MSSM)~\cite{Nilles:1983ge,Haber:1984rc}.
The Higgs sector of the MSSM comprises only
the two Higgs doublet fields $H_{u,d}$ introduced
above.
In the MSSM the quartic scalar couplings
of the Higgs sector are at lowest order
functions of the gauge couplings due
to SUSY relations.
The rigid structure of the
scalar potential leads to the fact that
deviations w.r.t.\ the SM
of the couplings of the lightest
Higgs boson $h_1 = h_{125}$
scale with the inverse of the
mass parameter
$M_A$~\cite{Carena:2001bg}
of the CP-odd Higgs boson.
Consequently, it is possible to
suppress such deviations in the decoupling
limit $M_A \gg M_Z$, in which $h_{125}$ is
aligned with the SM vev and effectively
indistinguishable from a SM Higgs boson, whereas
for $M_A \lesssim 1\tev$ deviations at the
level of $1$--$10\%$ for the coupling
coefficients $c_{h_1 f \bar f}$ are
present~\cite{Gunion:2002zf}.\footnote{In principle,
there is also a possibility of achieving
a SM-like Higgs boson in the so-called
alignment-without-decoupling limit of
the MSSM~\cite{Carena:2013ooa}.
However, this scenario relies on
an accidental cancellation between tree-level
and loop contributions. Moreover,
taking into account LHC constraints, it cannot
be realized in the $\MA$--$\tb$ regime relevant
for the excesses~\cite{Carena:2014nza,Bahl:2018zmf},
and we therefore
do not consider
it in the following discussion.}
Hence, the signal-rate measurements
of 
$h_{125}$ at the LHC,
which at the present level of accuracy show no significant deviations from
the SM prediction,
can be cast into a lower limit on $M_A$.
Taking into account the most recent LHC
results~\cite{Aad:2019mbh,CMS:2020gsy},
the current limit was found to be
$M_A \gtrsim 600\gev$, (largely) independently of the value of
$\tan\beta$~\cite{Bahl:2018zmf,Bahl:2020kwe}.
Consequently, the MSSM cannot account for
the presence of a CP-odd Higgs boson state around $400\gev$
without 
being in strong tension with the signal rates
of $h_{125}$, and a SUSY interpretation of the observed excesses 
therefore needs to be based on non-minimal SUSY models.

The Next-to Minimal Supersymmetric Standard Model
(NMSSM)~\cite{Maniatis:2009re,Ellwanger:2009dp}
extends the Higgs sector
of the MSSM by a
complex singlet scalar field.
The presence of this singlet field is motivated
in particular
by the so-called $\mu$-problem of the
MSSM~\cite{Ellis:1988er}, in which a dimensionful parameter
$\mu$ is present in the superpotential.
While this parameter would naturally
be expected to be of the order of
the ultraviolet cutoff of the model, it must be
of the order of the EW scale to allow for
a phenomenologically viable Higgs sector.
In the $Z_3$ symmetric NMSSM this problem is
absent, because the $\mu$ term is forbidden
by the global symmetry. Instead, it is generated
dynamically
when the scalar component of the singlet
superfield acquires a vev,
and the $Z_3$ symmetry is
broken spontaneously
(see \citere{Ellwanger:2009dp} for details).
Besides the fact that the singlet field
is complex in the NMSSM,
leading to the presence
of a second CP-odd Higgs boson in addition to the
CP-even scalar singlet state,
and that the singlet field is charged
under a $Z_3$ and not a $Z_2$ symmetry,
the Higgs sector
and the Yukawa sector
resemble the one of the N2HDM type~II introduced
in \refse{sec:defn2hdm}. Thus,
from the point of view of the Higgs phenomenology,
the NMSSM is a promising model for investigating
the possible realization of the
excesses around $400\gev$ in the context
of SUSY.

We briefly
summarize some of the key features of
the NMSSM that are relevant for
our analysis. We will focus in particular on the
Higgs sector of the model. The superpotential
of the $Z_3$ symmetric NMSSM is given by
\begin{equation}
W_{\mathrm{NMSSM}} =
W_{\mathrm{MSSM}, \cancel{\mu}} +  \lambda \ \hat{s} \ \hat{H}_u
    \cdot \hat{H}_d +
\frac{1}{3} \ \kappa \ \hat{s}^3 \ ,
\label{superpot}
\end{equation}
where $W_{\mathrm{MSSM}, \cancel{\mu}}$ denotes the
superpotential of the MSSM except the aforementioned
$\mu$-term. 
The second term is the portal coupling between the gauge singlet
superfield $\hat{s}$ and the Higgs doublet
superfields $\hat{H}_{u,d}$ (superfields are
denoted here by a hat).
The (effective) $\mu$-term arises from this term
when the scalar component
of $\hat{s}$ acquires the vev $\langle s \rangle = v_S$,
$\mu := \la v_S$.
The third term proportional to $\kappa$ is the
singlet self-coupling and, again once
$v_S$ is non-zero, gives rise to bilinear mass
terms both for the scalar and the fermionic
component of $\hat{s}$.
The complete Higgs potential is then derived from
the usual $F$- and $D$-terms
in combination with the terms arising from
the soft SUSY-breaking Lagrangian.

The fermionic superpartners of the Higgs fields
(called Higgsinos and singlino),
together with the superpartners of the gauge bosons
(called gauginos) give rise to five massive neutral
fermion states (called neutralinos) and two massive
charged fermions (called charginos).
The gauginos obtain their masses via the soft
SUSY-breaking gaugino mass parameters
$M_1$, $M_2$ and $M_3$, while the higgsinos obtain their masses
  from the $\mu$ parameter.
Compared to the MSSM, the only additional
fermion is the fifth neutralino
(singlino).
Moreover, extending the matter sector
of the SM, the NMSSM contains the scalar
partners of the leptons (called sleptons
and sneutrinos)
and quarks (called squarks).

The superpotential as defined in
\refeq{superpot} conserves
$R$~parity~\cite{Farrar:1978xj}.
As a result, the lightest SUSY
particle (LSP) is stable.
In case the LSP forms a sizable part
of the dark matter abundance,
bounds from direct detection
experiments exclude a substantial
part of the NMSSM parameter space.
Since we are primarily interested here
in the phenomenology of the Higgs sector,
we will not include these limits in our analysis.
Accordingly, the NMSSM can be considered to be a low-energy
effective model of a more complete
theory in which $R$~parity violating effects
prohibit the existence of a stable SUSY
particle, but in which the Higgs phenomenology
remains effectively unchanged. This can be realized if,
for instance, the $R$~parity violating operators
are non-renormalizable and suppressed by the
inverse of a high energy scale such as the
Planck scale, or if another symmetry suppresses
the amount of $R$~parity breaking.
An example for the latter is the so-called
$\mu$-from-$\nu$ supersymmetric standard model
($\mu\nu$SSM)~\cite{Bratchikov:2005vp,Munoz:2009an}
(see \citere{Lopez-Fogliani:2020gzo} for a
recent review).
It was shown that besides
the fact that there are
more than one gauge singlet
scalar fields present in the $\mu\nu$SSM,
the phenomenology of the SM-like Higgs boson
can be practically unchanged w.r.t.\ the
NMSSM~\cite{Biekotter:2017xmf,Biekotter:2019gtq,
Kpatcha:2019qsz}.

\subsection{The Higgs sector: alignment without decoupling}
\label{sec:align}
In the NMSSM an alignment limit exists
in which no decoupling of the BSM Higgs bosons and no 
large cancellations between lowest-order contributions and
radiative corrections to scalar masses
and their couplings are required~\cite{Carena:2015moc}.
We briefly describe here the necessary
conditions that have to be fulfilled in order
to account for the presence of a Higgs boson
resembling the discovered particle state
at $125\gev$ even for $M_A \approx 400\gev$.
We will also demonstrate that only for low values
of $\tan\beta$ the alignment conditions
can be satisfied. Thus, we expect 
that a description of the $t \bar t$ excess can be realized
in the exact alignment limit of the NMSSM, while
a description of the $\tautau$ excess will require departures from
the alignment limit.

In the following, we will denote the CP-even
Higgs bosons either with $h_{1,2,3}$, which
is the mass-ordered notation, such that
$m_{h_1} < m_{h_2} < m_{h_3}$, or we
will use $h_{125}$ for the discovered
Higgs boson, $h_S$ for the Higgs boson with
dominant singlet component and $H$ for the
``heavy'' CP-even Higgs boson.
The latter notation is useful because in the
alignment limit the singlet field
does not mix with $h_{125}$.
For the two CP-odd Higgs bosons
we either use the mass-ordered notation
$A_{1,2}$ or, in order to make connection to the
notation of the MSSM, we write $A_S$ for
the singlet-like state and $A$ for
the MSSM-like CP-odd Higgs boson. As for the
CP-even scalars, the latter notation does
not imply a mass ordering of the particles.

In the alignment limit one of the doublet
fields is aligned in field space with the
Higgs doublet vev $v$ (with $v^2 = v_1^2 + v_2^2$).
Then the tree-level couplings
of the corresponding Higgs state, $h_{125}$,
acquire
their SM values.
This can be achieved
if the mass matrix of the Higgs bosons
is diagonal in the Higgs basis, i.e.,
after a rotation of the $2\times 2$ submatrix
of the doublet components by the
angle $\beta$. In order to ensure that the non-diagonal entry of the
mass matrix between the two doublet
fields vanishes, as a first condition the following relation has to be
fulfilled~\cite{Carena:2015moc},
\begin{equation}
\lambda^2 =
\frac{m_{h_{125}}^2 - M_Z^2
    \cos 2 \beta}{v^2 \sin^2\beta} \ .
\label{aligncond1}
\end{equation}
This condition was derived
taking into account the dominant one-loop
corrections to the diagonal mass term
of $h_{125}$, stemming from the (s)top-sector.
As long as $\tan\beta$ is not much larger than $10$,
the other
one-loop corrections
(entering also the non-diagonal mass terms)
are suppressed by a
factor $\mu/M_S$.
Since we will choose $M_S \gg \mu$
in our numerical discussion, they
can safely be neglected.
The same is true also for corrections beyond the one-loop level
in the considered parameter region, where the prediction for the mass of the
SM-like Higgs boson is largely dominated by the tree-level contribution (see
the discussion below).
Given the values for the SM vev
$v \approx 246\gev$, the mass of the
$Z$~boson $M_Z \approx 91\gev$
and ${m_{h_{125}} \approx
125\gev}$, the first alignment condition
determines $\lambda$ as a function of $\tan\beta$.
A second alignment condition
arises from the fact that
also the non-diagonal
mass matrix entry relating the
state $h_{125}$ and the
gauge singlet field $s$ has to vanish
in the alignment limit.
This translates into the requirement~\cite{Carena:2015moc}
\begin{equation}
\frac{M_A^2 \sin^2 2 \beta}{4 \mu^2} +
\frac{\kappa \sin 2 \beta}{2 \lambda} = 1 \ .
\label{aligncond2}
\end{equation}
Here, $M_A^2$ is the squared 
CP-odd
mass parameter
defined by
\begin{equation}
\MA^2 =
\frac{\mu \left(
A_\lambda + \kappa v_S
\right)}{\sin\beta \cos\beta} \ ,
\label{defma}
\end{equation}
where $A_\lambda$ is the soft SUSY-breaking trilinear
coupling corresponding to the $\lambda$ term in the superpotential
of \refeq{superpot}.
At tree level $\MA$ is equal to the
mass of the MSSM-like CP-odd Higgs boson $A$,
so that it is a useful input parameter
for our analysis.
In order to make a distinction between $M_A$ and
the loop-corrected physical masses of the
CP-odd Higgs bosons, we will denote the latter
with the lower-case letter $m$ in the following, i.e.\ as
$m_{A_{1}}$, $m_{A_{2}}$ or $m_{A_{S}}$, $m_A$.
Since $M_A$ will be fixed to
$M_A \approx 400\gev$ in order to
account for the $t \bar t$ or the $\tau^+\tau^-$ excess, and taking
into account that the value of $\lambda$
is given by \refeq{aligncond1} for a fixed
value of $\tan\beta$, the second
alignment condition can be regarded
as a 
relation for $\kappa$ in terms of $\mu$ and $\tan\beta$.

\begin{figure}
\centering
\includegraphics[height=7cm]{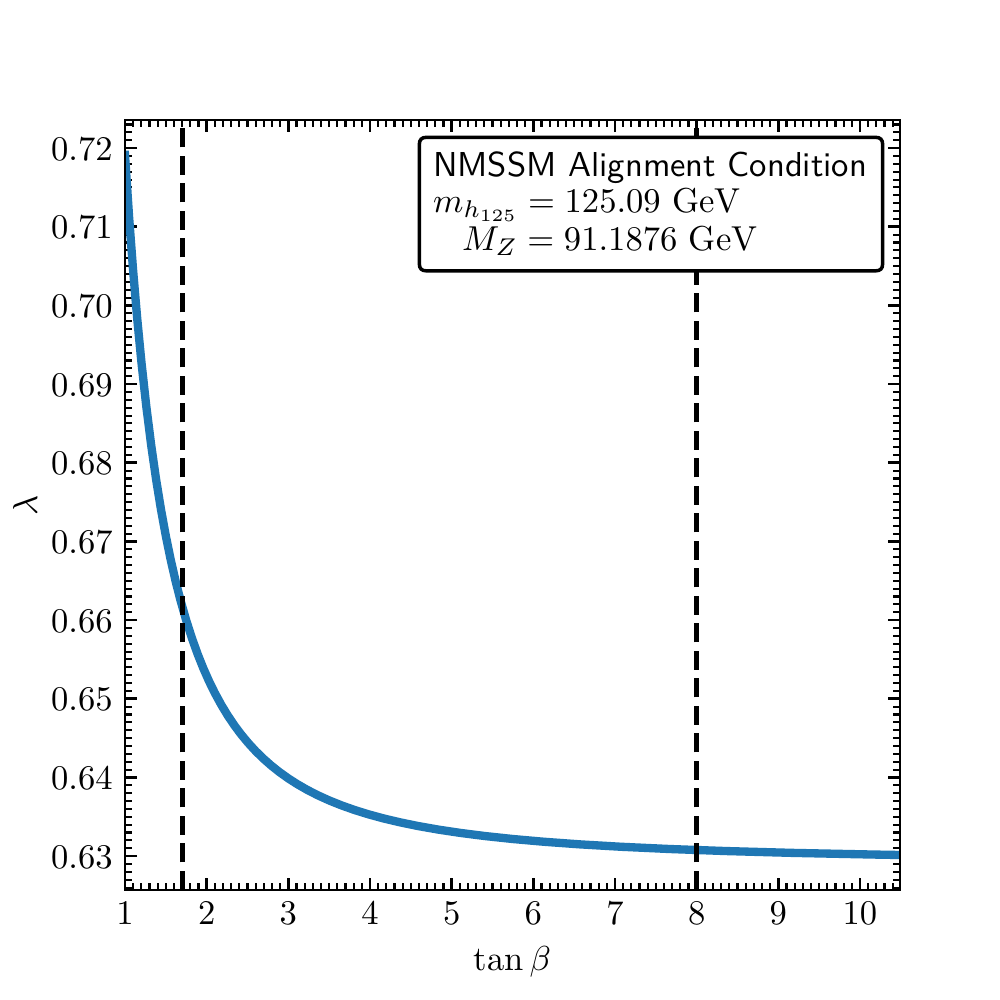}~
\includegraphics[height=7cm]{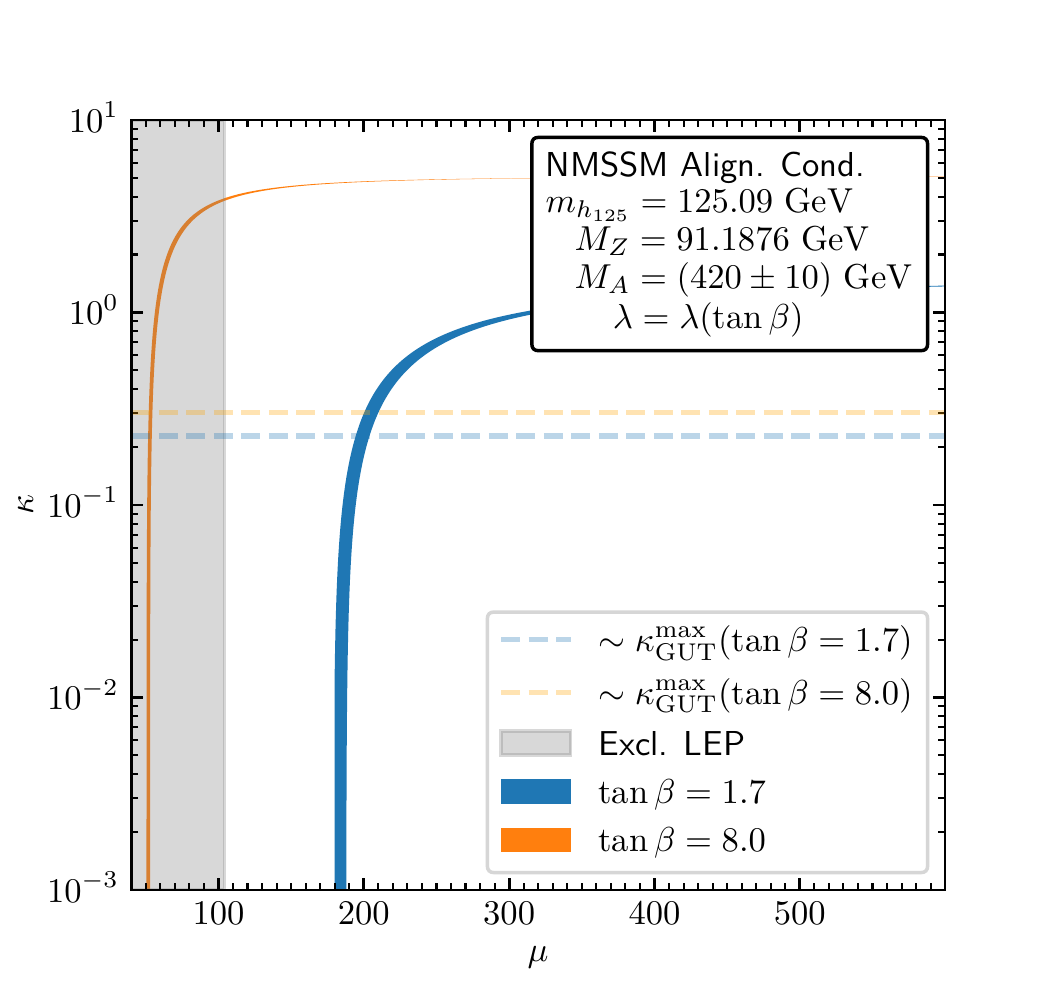}
\caption{Left: $\lambda$
as a function of $\tan\beta$ 
obtained from
the first alignment condition given
in \refeq{aligncond1}.
The black dashed lines indicate the values
$\tan\beta = 1.7$ and $\tan\beta = 8.0$ used
in our benchmark scenarios (see text).
Right: $\kappa$ as a function of $\mu$
obtained from
the second alignment condition given
in \refeq{aligncond2}
for $\tan\beta = 1.7$ (blue) and
$\tan\beta = 8.0$ (orange), and assuming
that $\lambda$ is chosen such that it
satisfies the first alignment condition. The gray shaded region
is excluded by LEP searches for charginos~\cite{Abdallah:2003xe}. 
The dashed lines indicate the maximum
values of $\kappa$ satisfying the condition
that there are no Landau poles below the
GUT scale given in \refeq{kappagut}.}
\label{figsaligncond}
\end{figure}

In the left plot of \reffi{figsaligncond}
we show the values of $\lambda$
as a function of $\tan\beta$ 
obtained from
\refeq{aligncond1}. One can see that relatively
large values of $\lambda > 0.6$ are required.
The two black dashed vertical
lines indicate the $\tan\beta$ values that we will choose as
benchmark scenarios to realize the $t \bar t$
or the $\tautau$ excesses in our numerical discussion
in \refse{sectanbetaeinssieben} and
\refse{sectanbetaacht},
respectively.
Using the corresponding values for $\lambda$,
we show in the right plot of
\reffi{figsaligncond} $\kappa$
as a function of $\mu$
for the two values of $\tan\beta$.
We also indicate with a gray shaded region
the values of $\mu \lesssim 104\gev$
for which the mass of the lightest chargino is below the limits 
that were obtained from chargino searches
at LEP~\cite{Abdallah:2003xe}.
Here we have assumed that $M_2$ is sufficiently larger than $\mu$, such
that the light chargino mass is determined by $\mu$.
One can see that for $\tan\beta = 8$
(orange line) the second alignment condition
can only be satisfied for 
rather large
values of the singlet self-coupling
$\kappa$. 
The required values
are roughly an order of magnitude larger
than the maximum value $\kappa^{\rm max}_{\rm GUT}$
(indicated by the orange dashed line)
based on the condition that there should be no
Landau poles below the GUT scale,
approximately given by~\cite{Miller:2003ay}
\begin{equation}
\kappa^2 +
\lambda^2 \leq 0.7^{\;2} \ .
\label{kappagut}
\end{equation}
We thus conclude that for values of
$\tan\beta$ required to realize the
$\tautau$ excess one cannot
satisfy the alignment conditions
for $\kappa$ values that are in agreement with \refeq{kappagut}.
On the other hand, for $\tan\beta = 1.7$,
which is the value
used for the benchmark scenario 
addressing the $t \bar t$ excess
(see \refse{nmssmconstraints}),
one can see that for $\mu \lesssim 200\gev$
and $\kappa \lesssim \kappa_{\rm GUT}^{\rm max}
= 0.23$ (blue dashed line in \reffi{figsaligncond}) a region of
parameter space in the alignment limit can be explored
that is in agreement with the LEP limits and the perturbativity constraint
of \refeq{kappagut}.
For the description of the $\tautau$ excess at $\tan\beta = 8$ in 
\refse{sectanbetaacht} we will allow for a departure from the alignment limit
in order to satisfy \refeq{kappagut}.

\smallskip
Focusing on the low $\tan\beta$ regime and thus the $t\bar t$
excess,
one can 
easily estimate the structure of
the Higgs boson spectrum. Obviously, one has
to require $M_A \approx 400\gev$ such
that the CP-odd Higgs boson $A$ can play the role
of the particle explaining the $t \bar t$
excess. Moreover, the tree-level mass of the charged
Higgs bosons $H^\pm$ is related to $M_A$ via the relation
\begin{equation}
M_{H^\pm}^2 = M_A^2 +
M_W^2 - \frac{1}{2} v^2 \lambda^2 \ .
\label{masshpmnmssm}
\end{equation}
For the considered
values of $\lambda$,
the charged Higgs bosons are a few~GeV lighter than the CP-odd state~$A$.
The CP-even doublet state~$H$ is also
close in mass to the $A$~boson as long
as its singlet component is small.
For the squared mass of
the SM-like Higgs boson 
at tree level an upper bound is obtained that is
approximately given by
\begin{equation}
(m_{h_{125}}^{(0)})^2 \lesssim
M_Z^2 \cos^2 2 \beta +
\frac{\lambda}{2}
v^2 \sin^2 2 \beta
\ .
\label{treemass}
\end{equation}
In the alignment limit this inequality
is nearly exhausted, and values
of $m_{h_{125}}^2 \approx 125\gev$
or even slighly larger are obtained for small $\tb$ and large $\la$.
Thus, no large
radiative corrections to $m_{h_{125}}$
are required in order to obtain
a physical mass of $125\gev$.
Finally, the masses of the singlet-like
particle states~$h_S$ and~$A_S$
are controlled by $\kappa$ and the
corresponding soft trilinear parameter
$A_{\kappa}$. Due to the small values
of $\kappa$ in order to satisfy
\refeq{aligncond2} (see also
\reffi{figsaligncond}), these two
Higgs bosons have masses substantially
below $M_A$. The dependence of their
masses on $A_\kappa$ is opposite
and approximately given by
\begin{equation}
m_{h_S}^2 \approx
\frac{\kappa \mu}{\lambda} A_\kappa
+ \frac{\kappa^2 \mu^2}{\lambda^2} \ ,
\quad
m_{A_S}^2 \approx - \frac{3 \kappa \mu}{\lambda}
A_\kappa  \ ,
\label{signletsmassesAkap}
\end{equation}
where for simplicity possible mixing contributions have not been spelled
out in \refeq{signletsmassesAkap} (those mixing contributions are included in
our numerical analysis below).
Thus, depending on the value of $A_\kappa$
both $h_S$ and $A_S$ can be lighter
than $125 \gev / 2$,
giving rise to both a lower and an upper
limit on
$A_\kappa$ in order to avoid
large values of $\br ( h_{125} \to h_S h_S, A_S A_S)$.
Moreover, the singlino mass term is given by
\begin{equation}
m_{\tilde{s} \tilde{s}} =
\frac{2 \kappa \mu}{\lambda} \ .
\label{chargmass}
\end{equation}
A mixed singlino-higgsino neutralino has
a mass of $m_{\widetilde{\chi}^0_1}
\approx m_{\tilde{s} \tilde{s}}$
at tree level.
For $m_{\widetilde{\chi}^0_1} < m_{h_{125}} / 2$ large
values of $\br(h_{125} \to \widetilde{\chi}^0_1 \widetilde{\chi}^0_1)$
can spoil the SM-like properties
of $h_{125}$. Thus, \refeq{chargmass} translates
into a lower limit on the possible values of $\mu$.
Further experimental constraints and 
their impact on the
parameter space
will be discussed in \refse{nmssmconstraints}.

\subsection{Experimental constraints}
\label{nmssmconstraints}

The experimental constraints that we take
into account in our NMSSM analysis are
focused on the collider phenomenology.
In particular, as
discussed above, 
we do not intend to reproduce the observed
dark matter relic abundance, nor
do we 
apply
constraints from dark matter direct
detection experiments.
Furthermore, we do not exclude parameter
points based on
constrains from flavor observables, as
the theoretical predictions of these
observables in SUSY models depend on
various different sectors of the model,
while the focus of our analysis is the Higgs
sector phenomenology.
The impact of constraints of this kind on the parameter space
investigated in our analysis would depend on the parameter settings of other
BSM contributions that are not relevant for Higgs physics.
While we do not impose those constraints 
for excluding parameter points, in our numerical
discussion we will point out possible tensions with constraints from
observables beyond the Higgs sector.
Tensions with observables from the flavor sector are expected in
particular for
relatively small values of $m_{H^\pm} \lesssim M_A$ 
and low
values of $\tan\beta$~\cite{Domingo:2015wyn}, see the discussion above.
It should be noted that we treat the mentioned upper limits on possible BSM
effects from other sectors in the same way as the $4.2\,\sigma$ discrepancy
between the experimental results for the anomalous magnetic
moment of the muon $(g-2)_\mu$~\cite{Bennett:2006fi,Abi:2021gix} 
and the SM prediction~\cite{Aoyama:2020ynm}.
While in principle SUSY models are capable of explaining this 
discrepancy~\cite{Martin:2001st,Stockinger:2006zn}
(see also \citere{Chakraborti:2021dli} for an analysis of the EW MSSM
sector, \citere{Abdughani:2021pdc} for a specific NMSSM analysis,
\citere{Heinemeyer:2021opc} for an analysis in the above discussed
$\mu\nu$SSM, and \citere{Athron:2021iuf} for a recent review), 
since the Higgs-boson sector is only marginally involved
we also do not apply this constraint favoring a non-zero BSM contribution.

As 
done for the N2HDM analysis,
we confronted all parameter points with
the current 
limits from
searches for additional Higgs bosons
using \texttt{HiggsBounds}.
There are several collider searches
which are especially relevant for the
low-$M_A$ region of the NMSSM. We briefly
summarize the most important ones here:

\textit{(i)} For low values of $\tan\beta \lesssim 2$,
the LHC searches for $H^\pm$ decaying into
a $tb$ pair exclude parameter regions with
$m_A \approx m_{H^\pm} \approx 400\gev$ in the (N)2HDM
type~II~\cite{ATLAS:2021upq}.
In the alignment limit of the NMSSM,
in which such values of $\tan\beta$ allow for
the realization of the $t \bar t$ excess,
one finds $m_{H^\pm} < M_A$ because
of the relatively large values of $\lambda$
(see \refeq{masshpmnmssm}).
Thus, compatibility with the experimental
limits on the charged Higgs rates requires a
reduction of ${\br(H^\pm \rightarrow t b)}$ compared
to the (N)2HDM prediction which can occur if additional decay
channels into SUSY particles are kinematically open. Nevertheless,
even for a considerably suppressed
$\br(H^\pm \rightarrow t b)$
the LHC charged Higgs boson
searches~\cite{Aaboud:2018cwk,Sirunyan:2020hwv,
ATLAS:2021upq} belong to the most constraining
BSM Higgs boson searches
for the analysis dedicated to the $t \bar t$
excess.

\textit{(ii)} 
The results for the searches in the $t \bar t$ and $\tautau$
final states are not only relevant in view of a possible description of the
observed excesses, but the limits obtained from those searches are also
important regarding their compatibility with the other Higgs bosons of the
model, namely $h_S$, $H$ and $A_S$.
For small values of $\tan\beta$,
searches for the additional 
Higgs bosons in the $t \bar t$ final state
are relevant~\cite{Sirunyan:2019wph,Sirunyan:2019wxt}.
For larger values
of $\tan\beta$, 
the limits for the $\tautau$
final state 
have the largest impact~\cite{Sirunyan:2018zut,Aad:2020zxo}.

\textit{(iii)} For singlet like states $h_S$ and
$A_S$ in the vicinity of $ 125\gev$ or
below, the searches for the SM Higgs boson
at LEP~\cite{Abbiendi:2002qp,Barate:2003sz,Schael:2006cr}
and the Tevatron~\cite{Group:2012zca} are
important. At the LHC, the 
most sensitive search for
those light additional Higgs bosons is the diphoton search
in which CMS found the excess at around 
$96\gev$~\cite{Sirunyan:2018aui}, see \refse{sec:excesses96}.
Outside of the mass interval of this excess,
the CMS searches 
place important limits on the
presence of the singlet-like scalars.
The corresponding ATLAS limits are substantially
weaker in almost all mass regions~\cite{ATLAS:2018xad}.

\textit{(iv)} The presence of
rather light singlet
states $h_S$ and $A_S$, in particular
as predicted in the alignment limit,
opens up the possibility of Higgs
cascade decays ${H(A) \rightarrow A_S(h_S)\,Z}$
and ${H^\pm \rightarrow h_S/A_S\,W^\pm}$.
CMS and ATLAS searched for the former processes
in multilepton and $l \bar l b \bar b$
final states~\cite{Khachatryan:2016are,
Aaboud:2018eoy,Sirunyan:2019wrn}.
In our analysis, only the constraints
from the searches regarding
the signature $H(A) \to A_S (h_S) \, Z$ were
capable of excluding parameter points.
On the hand,
for our analysis the presence of the 
decay mode
${H^\pm \rightarrow h_S/A_S\,W^\pm}$
turned out to be 
relevant,
because additional $H^\pm$ decay channels
into BSM scalars have the potential to further reduce
$\br(H^\pm \rightarrow t b)$ and help
to avoid the constraints from charged
Higgs boson searches in the $tb$ final
state discussed above.
The experimental signatures mentioned
above were also investigated
under the assumption
that the Higgs boson in the
final state is $h_{125}$,
giving rise to exclusions of
parts of the here analyzed
parameter space~\cite{Sirunyan:2019xls}.
This happens when
a second BSM Higgs boson $h_S$ or $A_S$
is present in the
mass window $(125 \pm 10)\gev$, yielding
another contribution that has to be added
to the one of $h_{125}$, such
that there can be a relevant enhancement
of the experimental signature. This case will be discussed
in more detail below.
Finally, also purely scalar cascade decays
of the form $H(A) \to h_{125}\,h_S(A_S)$
are relevant, where CMS searched for such a
signature assuming that $h_{125}$ decays
into a $\tautau$ pair and the other (BSM)
final state particle decays into a
$b \bar b$ pair~\cite{CMS:2021yci}.

In addition to the presence of additional Higgs
bosons, also SUSY particles can be
in the reach of the LHC or previous colliders. The corresponding
search limits
yield further constraints
on the parameter space. We list below the
most relevant searches and their impact
on our analysis:

\textit{(i)}
In our analysis in the alignment limit, the
tree-level mass of $m_{h_{125}}^{(0)} \approx 125\gev$
is determined by the alignment conditions
(see \refse{sec:align}) for low
$\tan\beta$ values,
which places a limit on the size of the radiative corrections to the mass
of the SM-like Higgs boson. As a consequence,
the SUSY breaking scale $M_S$,
and therefore the stop masses $m_{\widetilde{t}}$, cannot
be too large (see \citere{Carena:2015moc} for
more details).
This implies that
the LHC stop searches
can be relevant, as they provide (under several assumptions)
a lower limit on $m_{\widetilde{t}}$.
Taking into account the
uncertainties on the predictions
for the Higgs-boson mass(es) in the
NMSSM~\cite{Staub:2015aea,Drechsel:2016htw,Slavich:2020zjv},
we use an interval
of $m_{h_{125}} = (125 \pm 4)\gev$ in our
analysis. In order to not exceed this
mass interval given the value of $\tan\beta=1.7$ 
used in the analysis regarding the $t \bar t$
excess in \refse{sectanbetaeinssieben},
we find that values of
$M_S \approx m_{\widetilde{t}} \lesssim 1.2\tev$
are required, which is of the order of the
current experimental lower limit on the stop
masses in simplified
scenarios~\cite{CMS:2021eha}. 
Here it should be noted that
in our analysis, due to the
compressed electroweakino spectrum and the
presence of Higgs bosons with masses much
below $M_S$,
the actual experimental lower limit on the stop
masses could be substantially weaker
than $1 \tev$.
For larger values of $\tan\beta$, as required for
the $\tautau$ excess discussed in
\refse{sectanbetaacht}, the tree-level enhancement
of $m_{h_{125}}$ given by the term
$\sim \lambda^2$ (see \refeq{treemass})
is suppressed, such that in this case
larger radiative corrections
are needed to achieve a value of
$m_{h_{125}} \approx 125\gev$.
Consequently, for the $\tautau$ excess
larger values of $M_S$ are
required and LHC searches for stops
have no significant impact.

\textit{(ii)} As already mentioned in
\refse{sec:align}, LEP searches for charginos yield
lower limits on the chargino masses
of up to
$m_{\widetilde{\chi}^\pm_1} >
104\gev$~\cite{Abdallah:2003xe}.
In order to ensure compatibility with the
LEP bounds,
we demand $m_{\widetilde{\chi}^\pm_1} >
104\gev$ in our analysis.

\textit{(iii)} LHC searches for light neutralinos
in the context of the NMSSM
are very challenging when
the chargino and
neutralino spectra are compressed.
In addition, the searches for the
neutralinos suffer from the fact that also
the background estimation depends on the
precise form of the spectrum of the SUSY
particles. Therefore, a model interpretation
of a particular experimental search has to
be done not only for each SUSY model
separately, but also for each parameter point
within a certain model. Besides,
neutralino searches critically
depend on whether $R$~parity is
assumed to be conserved, or not. Taking the
above mentioned considerations into account,
we cannot apply general exclusions on the
parameter space of the NMSSM from neutralino
searches in our analysis. The only exception
are exclusion bounds based on
the presence of a neutralino with a mass
below $m_{h_{125}} / 2$, since this
can give rise to large values of
$\br(h_{125} \to \widetilde{\chi}^0_1 \widetilde{\chi}^0_1)$
(see \refse{sectanbetaeinssieben}),
which are excluded by global constraints
on the signal rates of the $h_{125}$.

Finally, as we did in the N2HDM analysis,
we perform a $\chi^2$ test regarding the
signal rates of $h_{125}$ using
\texttt{HiggsSignals} (which as discussed above excludes large values of
  $\br(h_{125} \to \widetilde{\chi}^0_1 \widetilde{\chi}^0_1)$).
An important difference w.r.t.\ the N2HDM arises from
the fact that the Higgs-boson mass could
be chosen as an input parameter in the
N2HDM, whereas $m_{h_{125}}$ is predicted
in the NMSSM as a function of other model
parameters. The model predictions 
are affected by a theoretical uncertainty in particular
for larger values of $\lambda$ due to 
radiative corrections beyond the one-loop
level (see 
\citere{Dao:2021khm} for a recent account of
this subject).
This is why, as already mentioned
above, we allow for an uncertainty of
$\Delta m_{h_{125}}^{\rm theo} = 4\gev$ in our
analysis.
In order to make sure that the mass uncertainty
does not give rise to unacceptably large modifications of
the predicted branching ratios for the
decays of $h_{125}$
into SM particles (which could occur if the branching rations are
evaluated with mass values that are several GeV away from $125
\gev$, but still within the $\pm 4 \gev$ interval),
we recalculated the corresponding decay widths
by requiring $m_{h_{125}} = 125\gev$ for all points in which
the prediction of \texttt{NMSSMTools} for
$m_{h_{125}}$ lies within the mentioned
uncertainty band. This recalculation is especially
relevant for $\br(h_{125} \rightarrow WW^*,ZZ^*)$,
which have a very sensitive dependence on $m_{h_{125}}$.
For the recalculation of the branching ratios
we made use of the effective coupling coefficients
provided by \texttt{NMSSMTools}, by which the
partial decay widths for $h_{125}$
as predicted by the SM were rescaled.
The values for the SM decay widths were taken from
\citere{deFlorian:2016spz}.

\subsection{Benchmark scenario with
\texorpdfstring{\boldmath{$\tan\beta = 1.7$}}{tb17}}
\label{sectanbetaeinssieben}

As discussed in \refse{sec:align}, the
alignment-without-decoupling limit
of the NMSSM is a theoretically
well motivated scenario for
realizing the $t \bar t$ excess in the context
of SUSY.
Thus, we will investigate in
this section the possibility of accommodating
the $t \bar t$ excess in this limit taking into
account the various different collider constraints
mentioned in the previous section.
As in the N2HDM analysis, we consider only
the CP-odd Higgs boson $A$ as the origin of the excess,
accounting for the fact that the contribution
of CP-even states is much smaller compared
to the one of a CP-odd state.
Moreover, since we restrict our discussion to the CP-conserving case
the contributions of CP-even and CP-odd states do not interfere with each other.
In addition, we pointed out that in this limit
the presence of a second light CP-even Higgs boson
arises naturally.
Thus, although not being the main focus of
this analysis, for the
subset of points that feature a particle
candidate in the relevant mass range for the
LEP and the CMS excesses
(see \refse{sec:excesses96})
we will check in a second step
whether also a simultaneous realization of
these excesses in combination with
the $t \bar t$ excess
is possible.

\smallskip
As can be seen from the first alignment
condition shown in \refeq{aligncond1}, the 
choice of the
parameter $\tan\beta$ plays a key role. Starting from the
value of $\tan\beta$, and given $M_A \approx 400\gev$,
the remaining parameter values can either be
derived or scanned over. We
begin by choosing an initial value of
$\tan\beta = 1.7$ for our scan. This value
arises from the following considerations.
As described in \refse{sec:align}, in the
alignment limit the singlet CP-odd
state $A_S$ is substantially lighter than
the doublet state $A$. Accordingly, also
the mixing of these states is small. Then
one finds that ${c_{A t \bar t} \approx 1 / \tan\beta}$
(as in the N2HDM type~II),
such that regarding the
$t \bar t$ excess even smaller values of
$\tan\beta \approx 1$ would be 
favored.
However, the presence of the charged Higgs
bosons $H^\pm$, having a mass slightly below
$M_A$ in the alignment limit, yields
a lower bound on the possible values
of $\tan\beta$ based on LHC searches
for charged scalars (see \refse{nmssmconstraints}).
Our numerical analysis has revealed that,
by taking into account the possibility
that $H^\pm$ can partially decay
into SUSY particles and lighter
neutral Higgs bosons plus a $W$ boson,
values of $\tan\beta = 1.7$
provide points that are compatible with the constraints
arising from the LHC searches, while also
allowing for sizable values of $c_{A t \bar t}$.

Given the chosen value for $\tan\beta$, one derives
a value of $\lambda = 0.6617$ from
\refeq{aligncond1}. Taking into account that
loop corrections yield a (physical)
mass $m_A$ of the CP-odd Higgs boson that is
slightly below the (tree-level) input parameter $M_A$,
we chose to generate parameter points with
$410\gev \leq M_A \leq 430 \gev$, in steps of
$5\gev$. Given a value for $M_A$ and the ones
for $\tan\beta$ and $\lambda$, we used the
second alignment condition shown in \refeq{aligncond2}
to obtain values of $\kappa$ as a function of $\mu$.
The parameter points were then generated by
scanning over $\mu$ and $A_\kappa$ in steps of $1\gev$.
We covered the parameter ranges for which
points could be found that are in agreement with
the experimental and theoretical constraints.
The range
of $\mu$ has a lower limit (for each value
of $M_A$) arising from the restriction that
none of the neutralinos should become substantially lighter
than $125 / 2\gev$, since otherwise the decay
$h_{125} \to \widetilde{\chi}^0_1
\widetilde{\chi}^0_1$ would become
kinematically allowed, spoiling agreement of
the properties of $h_{125}$ with the LHC measurements.
An upper limit on $\mu$ is given by
the aforementioned constraints from the
charged Higgs boson searches. These constraints can only
be evaded if there are sizable branching
ratios for the decays $H^\pm \rightarrow
\widetilde{\chi}^0_{1} \widetilde{\chi}^\pm_{1},
h_{S} W^\pm$ and/or $A_{S} W^\pm$.
However, the branching ratios of these decays
get reduced for increasing values of $\mu$, as
the masses of the final state particles,
in particular the mass of the
Higgsino-like chargino $\widetilde{\chi}_1^\pm$,
increase with $\mu$.
Upper and lower limits on $A_\kappa$,
in dependence of the other parameters,
are given by the fact that either $h_S$
or $A_S$ becomes lighter than $125 /2 \gev$
(see \refeq{signletsmassesAkap}).
In this case the large value of $\lambda$ gives rise to
unacceptably large values of
$\br(h_{125} \to h_S h_S / A_S A_S)$.
In addition, we find that all points with
$A_S \lesssim 100\gev$ are excluded due to
constraints from the search for the signature
$A \to A_S\,h_{125}$~\cite{CMS:2021yci}.

\begin{table}[t]
\centering
\def\arraystretch{1.5}
\setlength\tabcolsep{4.5pt}
\footnotesize
\begin{tabular}{ccccccccccc}
$M_S$ & $M_1$ & $M_2$ & $M_3$ & $A_{\widetilde{f}}$ &
    $M_A$ & $\tan\beta$ & $\lambda(\tan\beta)$ & $\mu$ &
        $\kappa(\mu,\tan\beta,M_A)$ & $A_\kappa$ \\ 
\hline
1200 & 140 & 180 & 2000 & 0 & $[410,430]$ &
    1.7 & 0.6617 & $[182,201]$ & $[0.047,0.191]$ &
        $[-498,-116]$
\end{tabular}
\caption{\small
Parameter values for the scan in the
alignment limit of the NMSSM with
$\tan\beta = 1.7$ 
for investigating a possible realization of
the $t \bar t$ excess (see text).
}
\label{catttableparas}
\end{table}

We summarize the parameter ranges of the
scan in \refta{catttableparas}. We emphasize that
the values of $\lambda$ and $\kappa$
are derived from the alignment conditions
for $\tan\beta = 1.7$, $410\gev \leq
M_A \leq 430\gev$ and $182 \gev \leq \mu
\leq 201\gev$.
The range of $\mu$ is given by the
experimental constraints related to
$\br(h_{125} \to \widetilde{\chi}^0_1
\widetilde{\chi}^0_1)$ (lower end) and
$H^\pm$ searches (upper end), and
the range of $A_\kappa$ is given by
the experimental constraints related
to $\br(h_{125} \to h_S h_S, A_S, A_S)$
(as explained already before).
In \refta{catttableparas} we also
give the values used for the parameters of
the soft SUSY-breaking sector. All soft scalar
masses are set equal to the SUSY breaking scale
$M_S = 1200\gev$. This value was chosen in order
to allow for a SM-like Higgs boson mass of
$125\gev$ without potentially being in conflict
with experimental constraints from stop
searches (see \refse{nmssmconstraints}).\footnote{We
set the masses of all squarks equal to $M_S$ in our
analysis for simplicity.
For the squarks of the first and the
second generation these values are potentially
excluded by LHC searches. However, the masses of
these quarks have no impact on the mass of $h_{125}$,
such that they could be increased without any impact on
the discussion here.}
The soft trilinear couplings of the
sfermions $A_{\widetilde{f}}$ are set to zero.
This is relevant for the stops in order to
avoid large radiative corrections to $m_{h_{125}}$,
which would yield too large values for $m_{h_{125}}$.
The gaugino mass parameters are set to $M_1 = 140\gev$,
$M_2 = 180\gev$ and $M_3 = 2000\gev$. The values of
$M_1$ and $M_2$ lead to the presence of gaugino-like
neutralinos and charginos with masses below $\approx 200\gev$. 
For this choice several decays of the kind $H^\pm \rightarrow
\widetilde{\chi}^0_{i} \widetilde{\chi}^\pm_j$
can become relevant, which play a role in avoiding the
constraints from charged Higgs boson searches.
As discussed above,
we checked that the lighter chargino mass is above the
lower limit from LEP constraints. The value for
the gluino mass $M_3$ is large enough to 
be compatible with the limits
from direct searches given the compressed
neutralino-chargino spectrum,
independently of the fact whether
$R$~parity is assumed to be conserved or
not~\cite{ATLAS:2021twp}.

Taking into account the experimental constraints
listed in \refse{nmssmconstraints},
we found an allowed parameter
region that is shown in
\reffi{paraspace17}. The left plot shows the $\mu$--$\kappa$ plane,
where the color coding indicates $\MA$,
whereas the right plot shows the $\kappa$--$A_\kappa$ plane, where the
color coding indicates $m_{h_S}$.
Here, we applied the same condition as in the N2HDM analysis, demanding
that $\chi^2 \leq \chi^2_{\rm SM}$, where the
total $\chi^2$ is defined in \refeq{eqchisqtt}.
It should be noted that the contribution
of $\chi^2_{96}$ is included in $\chi^2$,
even though it is not the main focus of this analysis.
For parameter points that feature a
CP-even scalar in the mass interval
$94\gev \leq m_{h_1} \leq 98\gev$,
we calculate $\chi^2_{96}$ given
the predicted signal strengths of $h_1$,
whereas for the other points we set
$\chi^2_{96} = \chi^2_{\mathrm{SM},96}$.
Thus, only a 
subset of points
has a contribution of $\chi^2_{96}$ below the SM one.
However, even though the formal definition
of $\chi^2$ is identical, one should keep
in mind that in the NMSSM analysis the
physical mass $m_A$ of the CP-odd Higgs boson is
not an input parameter, such that it can
differ slightly from the value $m_A = 400\gev$
used in the N2HDM. This leads to the fact
that in the NMSSM $\chi^2_{t \bar t}$ is
a function of the
three predicted quantities $c_{A t \bar t}$, $\Gamma_A$
and $m_A$ 
(see also \refse{sec:excesses}).
As before,
the result for the SM is given by
$\chi^2_{\mathrm{SM}, t \bar t} = 13.98$.
As a consequence
of the alignment conditions, we found
that the properties
of $h_{125}$ are very well in agreement with the
experimental measurements. Even parameter
points in which the decay of $h_{125}$ into
two neutralinos is kinematically allowed were
found to predict $\chi^2 \leq \chi^2_{\rm SM}$.
A more detailed discussion of the properties of
$h_{125}$ can be
found in \refap{secnmssmh125}.

\begin{figure}
\centering
\includegraphics[height=6.2cm]{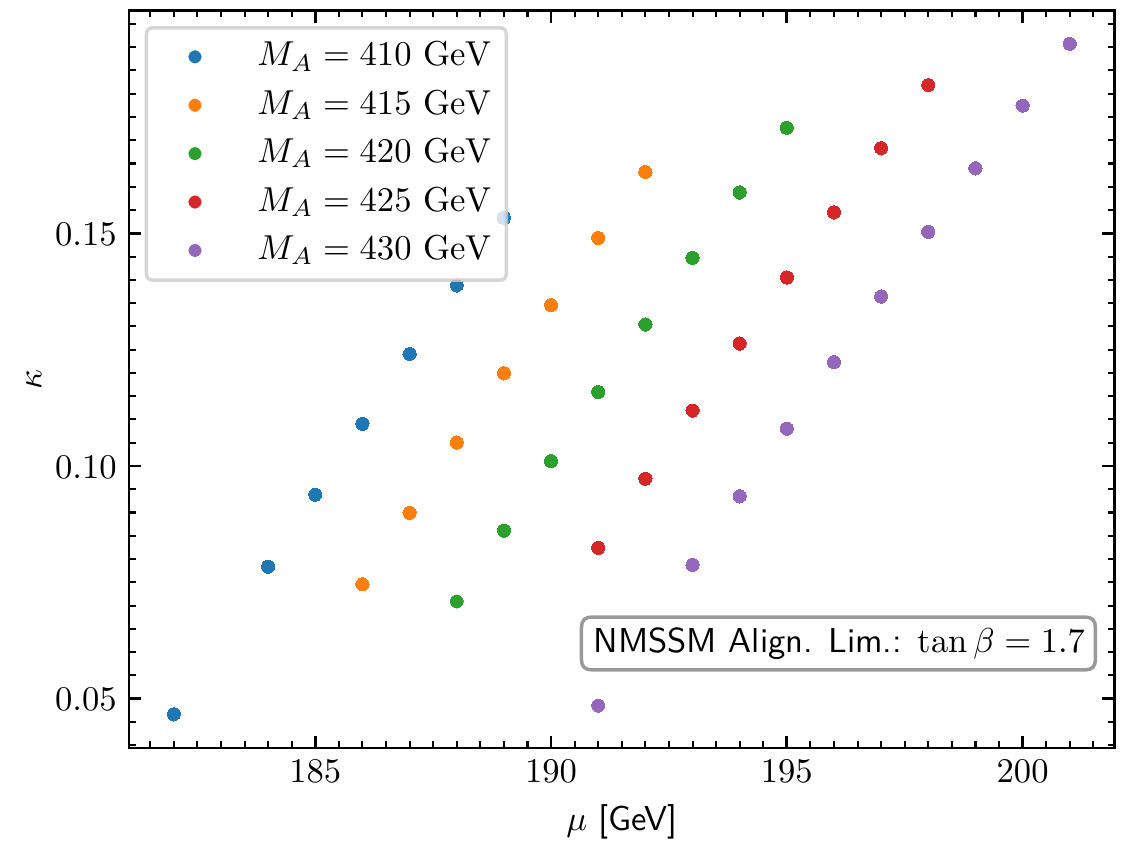}~
\includegraphics[height=6.2cm]{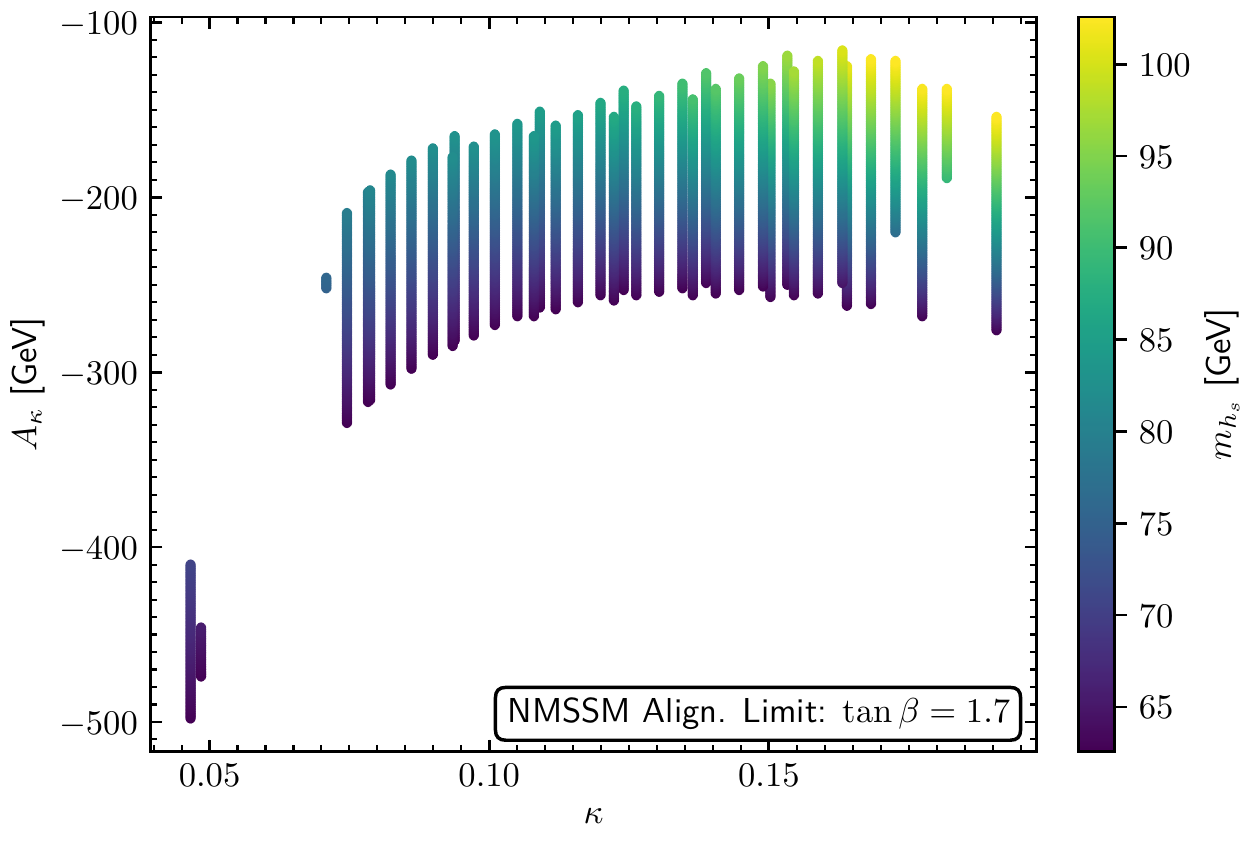}
\caption{NMSSM parameter points with
$\chi^2 \leq \chi^2_{\mathrm{SM}}$
in the $\mu$--$\kappa$ plane (left)
and in the $\kappa$--$A_\kappa$
plane (right). The colors of the points
indicate the value of the tree-level
mass parameter $M_A$ (left) and the
mass of the CP-even singlet-like Higgs
boson $m_{h_S}$ (right).
}
\label{paraspace17}
\end{figure}

In the left plot of \reffi{paraspace17}
one can see that the allowed range of $\kappa$ is below
the limit from the GUT condition shown in
\refeq{kappagut} in most cases. This indicates
that the considered scenarios correspond
to model realizations that may be well defined
up to the GUT scale. The parameter range of $\mu$ is
close to the EW scale, 
which corresponds to the parameter region of the NMSSM
that is favored by naturalness arguments~\cite{Baer:2012up, King:2012tr},
in which no large
fine tuning (corresponding to a ``little hierarchy problem'')
is required in the EW sector of
the model. Concerning the colors of the points, indicating $M_A$,
one should note that each
point in the plot corresponds to a subset
of parameter points with different values of $A_\kappa$.
One can see that for increasing values of $M_A$
also larger values of $\mu$ and $\kappa$ are required
in order to fulfill the second alignment condition.

In the right plot of \reffi{paraspace17} one
can see that only points with negative
values for $A_\kappa$ are allowed.
For values of $A_\kappa > -100\gev$ we find
that the condition $\chi^2 < \chi^2_{\rm SM}$
is only fulfilled for a small number of points,
which then are excluded by LHC searches involving
the light CP-odd Higgs boson $A_S$ with $m_{A_S} \lesssim 148\gev$.
In particular, the CMS search for the signature
$A \to A_S\,h_{125}$ excludes all such points
with $m_{A_S} \lesssim 100\gev$, and points with
slightly larger values of $m_{A_S}$ are excluded
by searches for $H \to A_S\,Z$ (see the discussion
in \refse{nmssmconstraints}).
On the other hand, $m^2_{h_S}$ receives additional contributions
proportional to $\kappa v_S$ that can compensate the
negative contribution $\sim A_\kappa$, so that
$m_{h_S}^2 > 0$ even for values of $A_\kappa \approx -500\gev$
(see \refeq{signletsmassesAkap}).
Nevertheless, one can see that, given a certain
value for $\kappa$, a lower bound on $A_\kappa$ is
found 
where $m_{h_S}$ (indicated by the color coding)
decreases below $\approx 125 / 2\gev$, giving rise to unacceptably
large values of $\br(h_{125} \rightarrow h_S h_S)$.
In addition, even though $A_S$ is overall heavier
if $A_\kappa$ has large negative values, for the lower values of
$\kappa$ it  can still be relatively light,
with masses just above or even below $125\gev$.
Hence, for small values of $\kappa$ the searches
for $A_S$ already mentioned above become
relevant again, and
give rise to the fact that none of the points
with $0.05 \leq \kappa \leq 0.07$ pass the
\texttt{HiggsBounds} test. Also the two
remaining isolated lines of points at the lower end
of the $\kappa$ range, which correspond to the two points at the lowest
$\kappa$ values in the left plot of
\reffi{paraspace17},
barely escape the constraints
from the searches for $H \to A_S Z$.
As discussed above, points with
$m_{h_S} \approx 96\gev$ can contribute to the
LEP and CMS excesses. These points are found for larger $\kappa$
and smaller $|A_\kappa|$. Given the
condition $\chi^2 \leq \chi^2_{\rm SM}$ in
our scans, for these points somewhat larger
values of $\Delta \chi^2_{125}$
are allowed as compared to
the points for which no candidate at $96\gev$
is present.\footnote{It should be remembered
that due to the uncertainty in
the prediction of 
the Higgs-boson masses, we calculate
$\chi^2_{96}$ as described in \refse{sec:excesses96} for
points with $94\gev \leq m_{h_S} \leq 98 \gev$,
while for the other points we set
$\chi^2_{96} = \chi^2_{\mathrm{SM}, 96}$.}

\begin{figure}
\centering
\includegraphics[width=0.48\textwidth]{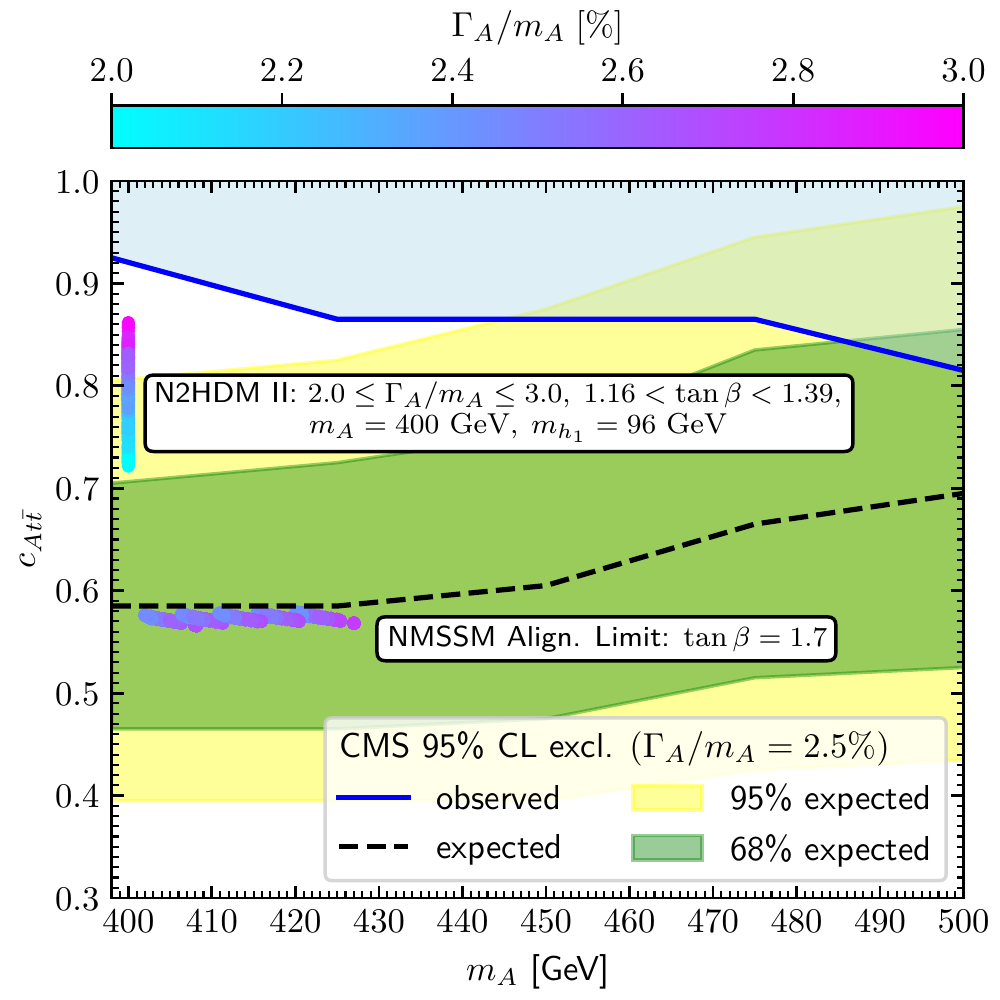}~
\includegraphics[width=0.48\textwidth]{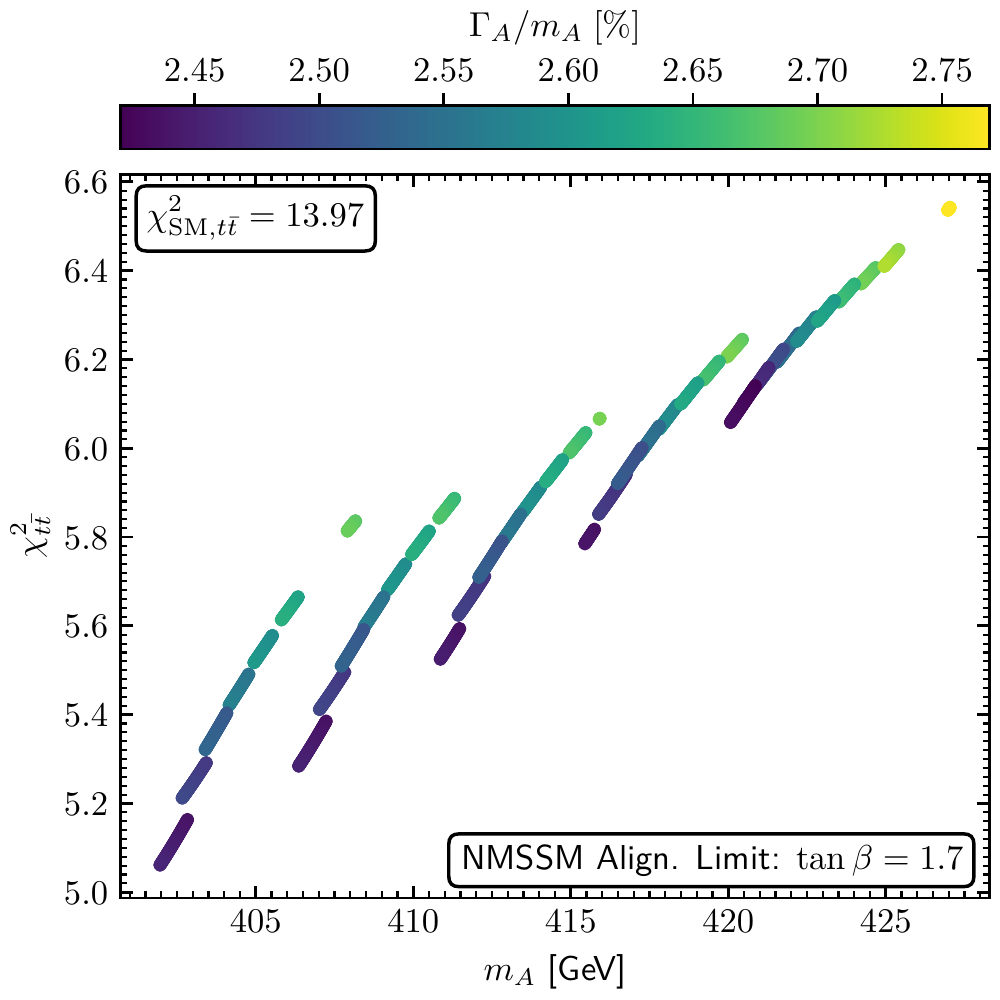}
\caption{NMSSM parameter points
with
$\chi^2 \leq \chi^2_{\mathrm{SM}}$
in the $m_{A}$--$c_{A t \bar t}$ plane (left) and the
$m_{A}$--$\chi^2_{t \bar t}$
plane (right). The colors of the points
indicate the values of $\Gamma_{A}/m_{A}$ in \%.
Left:
Also shown are the observed (blue)
and expected (black dashed) upper limits
at the $95\%$ C.L.\ as well as the 
$1\sigma$ (green) and $2\sigma$ (yellow)
regions around the expected exclusion limit assuming
$\Gamma_{A}/m_{A} = 2.5\%$, as
published in \citere{Sirunyan:2019wph},
and the N2HDM parameter points from the
scan discussed in \refse{numlowII}
with $\chi^2 \leq \chi^2_{\mathrm{SM}}$
and $2.0\% \leq \Gamma_A / m_A \leq 3.0\%$.}
\label{cattnmssm}
\end{figure}

We now turn to the discussion of
the description of
the $t \bar t$ excess, which is the
main motivation for the scan
presented here. On the left-hand
side of \reffi{cattnmssm} we show the
value of $c_{A t \bar t}$ of the parameter
points in dependence of the (physical) mass $m_A$.
The color coding shows $\Ga_A/m_A$ in~\%.
Since the singlet admixture of $A$ is
very small, $c_{A t \bar t}$ is given
to a very good approximation by the
inverse of $\tan\beta$, such that
$c_{A t \bar t} \approx 1 / 1.7 \approx 0.59$.
Consequently, the points are all located
within a tight branch at roughly
this value of $c_{A t \bar t}$, slightly below
the experimental best-fit value of
$c_{A t \bar t} \approx 0.8$ assuming
$m_A = 400\gev$ and
$\Gamma_A / m_A = 2.5\%$.
This results in
a description of
the excess at a confidence level of
about $1.5\,\sigma$ for the smallest values of $m_A$,
as further discussed below.\footnote{We
wish to stress that the fact that the
NMSSM parameter points happen to lie along
the curve for the expected limit is a coincidence
from which no direct information about the
fit result can be derived.}
The mass of the CP-odd Higgs boson $A$ ranges
between $400\gev$ and $430\gev$, where
the range is defined by the scan
range used for the tree-level
parameter $M_A$.
In order to facilitate the
comparison between the N2HDM and
the NMSSM we also show the subset of
N2HDM parameter points featuring
$2.0\% \leq \Gamma_A / m_A \leq 3.0\%$
from the scan performed in \refse{numlowII}.
The N2HDM points
lie at larger values of $c_{A t \bar t} \approx 0.8$,
reflecting the fact that the N2HDM allows for
an even better description of the excess.
The reason for this is that
one cannot further reduce the value of $\tan\beta$
in the NMSSM without violating constraints from
the charged Higgs boson searches. In the N2HDM
these can be avoided by increasing the mass
$m_{H^\pm} \gg 400\gev$. In the NMSSM this is not
possible, because of the 
relations among the Higgs-boson
masses, as explained in
\refse{nmssmconstraints}.
Thus, also the maximum value of $|c_{A t \bar t}|$
that is possibly realized remains below
the measured best-fit value ($c_{A t \bar t} \approx 0.8$).
The collider phenomenology
of the $H^\pm$ state in the NMSSM
will be discussed in more detail below.

It is also important to note that the
$t \bar t$ excess is most pronounced
for the smallest values of $m_A$
analyzed in the experimental analysis.
In \reffi{cattnmssm} this is reflected by the fact
that the smaller
$m_A$, the larger is the difference
between the observed (blue) and expected
(black dashed) exclusion limit.
It should be remembered here
that $\chi^2_{t \bar t}$ is defined
as the $\chi^2$
difference compared to the best fit value, where the latter
was found for $m_A = 400\gev$,
$\Gamma_A / m_A = 4.5 \% $ and $c_{A t \bar t}
= 0.94$ (see also \reffi{Chisqtt}).
Thus, even though the theoretical
predictions for the values of
$c_{A t \bar t}$ are independent of $m_A$,
the improvement of the $\chi^2_{t \bar t}$
values in comparison to $\chi^2_{\mathrm{SM}, t \bar t}$
is substantially larger for
values of $m_A \approx 400\gev$, where the smallest
$\chi^2_{t \bar t}$ values found are
$\chi^2_{t\bar t} \approx 5$, i.e.\ somewhat worse than in the
N2HDM (see above).
This is clearly visible
in the right plot of \reffi{cattnmssm},
in which we show the values of $\chi^2_{t \bar t}$
in dependence of $m_A$: 
the smallest values of
$\chi^2_{t \bar t} \approx 5$
(corresponding to a fit at about
$1.4 \, \sigma$ confidence level)
are obtained for the points
with the lowest values of $m_A$.\footnote{We
perform a linear interpolation to
obtain a value of $\chi_{t \bar t}$
as a function of $m_{A}$, $\Gamma_{A} / m_{A}$, and
$c_{A t \bar t}$ for each parameter point,
interpolating between the values
considered in the experimental analysis.}
Thus, even though
the NMSSM offers a substantially
better fit to the data than the SM, the NMSSM analysis
does not reach the same level of fit quality as
the N2HDM analysis, where values of
$\chi^2_{t \bar t} < 1$ could
be achieved (see, for instance,
\reffi{figGEN2}).

In addition, the total width
$\Gamma_A$ of the CP-odd Higgs boson, indicated
by the color coding in units of
$m_A$ in \reffi{cattnmssm},
correlates with $m_A$,
where for each branch displayed in the right plot of \reffi{cattnmssm}
(each branch corresponds to a different value of $M_A$)
smaller values of $m_A$ give rise to smaller values of $\Ga/m_A$.
This correlation has 
its origin in the phase space
factor for the $t \bar t$ decay width,
which grows with increasing values of $m_A$
and overcompensates the factor $m_A$ in the denominator.
Since the experimental analysis was carried out
for different hypotheses on $\Gamma_A / m_A$,
$\chi^2_{t \bar t}$ depends on $m_A$ both directly and via its impact on the
ratio $\Gamma_A / m_A$.
On the other hand, one can
see that the overall variation of
$\Gamma_A / m_A$ of the parameter points is small compared to the
step sizes of the different width hypothesis
used in the experimental
analysis, as shown in \reffi{Chisqtt}.
Thus, given the experimental uncertainties,
it would have been sufficient to only
take into account the expected and observed
exclusion limits (and the resulting
$\chi^2_{t \bar t}$ values) obtained
under the assumption $\Gamma_A / m_A = 2.5\%$
(this is the reason why the
left plot of \reffi{cattnmssm}
is displayed for $\Gamma_A / m_A = 2.5\%$).

\begin{figure}
\centering
\includegraphics[height=6.cm]{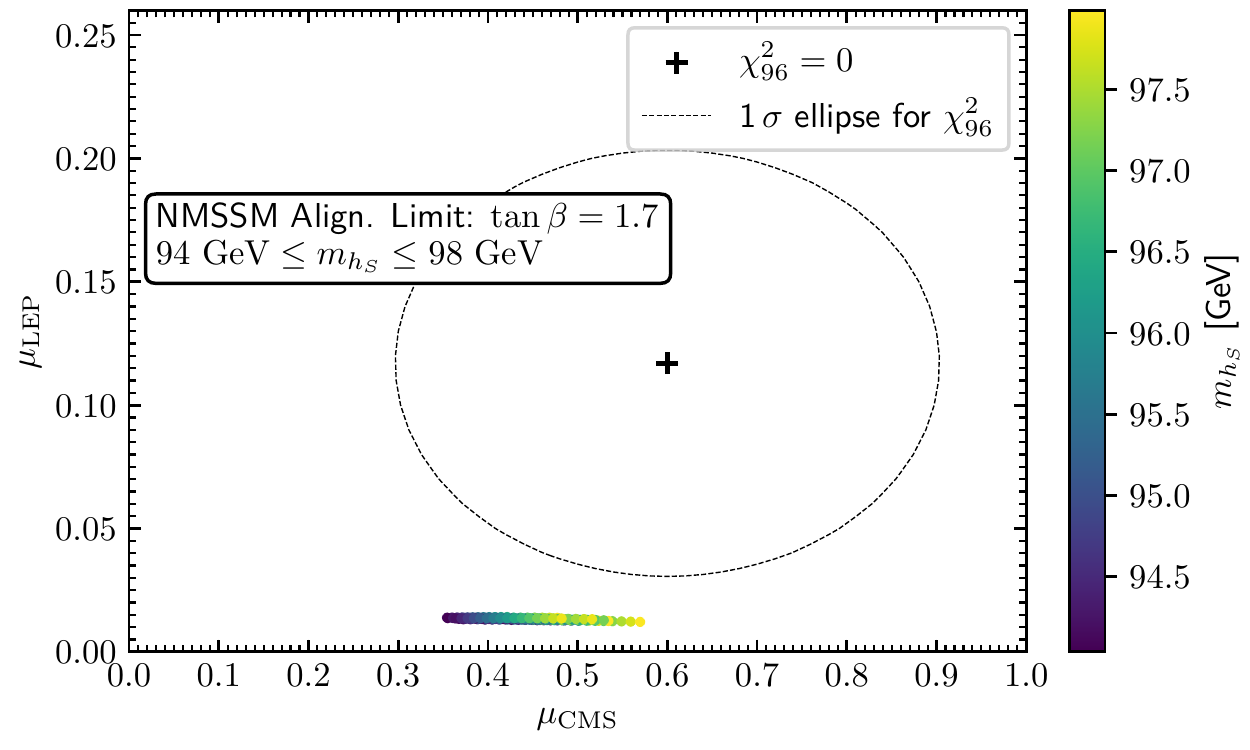}
\caption{NMSSM parameter points with
$\chi^2 \leq \chi^2_{\mathrm{SM}}$
and for which
$94\gev \leq m_{h_S} \leq 98$
in the
$\mu_{\rm CMS}$--$\mu_{\rm LEP}$
plane. The colors of the points
indicate the values of
$m_{h_S}$.
}
\label{mucmsnmssm}
\end{figure}

\smallskip
As discussed at the beginning of this section,
the alignment conditions lead to the presence of a light
CP-even Higgs boson $h_S \ (=h_1)$,
which is almost entirely singlet-like.
As a result, $h_S$
can be a candidate for 
describing
the excesses at $\approx 96\gev$.
We show in
\reffi{mucmsnmssm} the parameter points
that fulfill the additional constraint
$94\gev \leq m_{h_S} \leq 98\gev$ in
the plane of $\mu_{\rm CMS}$ and
$\mu_{\rm LEP}$, where the color coding indicates $m_{h_S}$.
We calculate the signal strengths via
\begin{equation}
\mu_{\rm LEP} \approx c_{h_S VV}^2 \
\frac{\br (h_S \to b \bar b)}
{\br_{\rm SM} (H \to b \bar b)} \ , \quad
\mu_{\rm CMS} \approx c_{h_S t \bar t}^2 \
\frac{\br (h_S \to \gamma \gamma)}
{\br_{\rm SM} (H \to \gamma \gamma)} \ ,
\end{equation}
hence making use of the fact
that the cross section ratios
can be approximated via the respective
effective coupling coefficients squared.
One can see that the CMS excess can easily be accommodated.
On the other hand, in our analysis the LEP excess cannot
be described. This was expected
as a consequence of the dominantly singlet-like character
of $h_S$ and the resulting suppression
of the Higgsstrahlung production cross
section.\footnote{Scenarios
in the NMSSM and the $\mu\nu$SSM were found
in which both excesses at $\approx 96\gev$
can be realized at the level of $1\sigma$,
relying on the fact that a sizable coupling
of $h_S$ to vector bosons can be present
via a mixing with $h_{125}$~\cite{Cao:2016uwt,
Domingo:2018uim,Choi:2019yrv,Biekotter:2017xmf,Biekotter:2019gtq}.
Such scenarios rely on
decoupling effects in order to agree with
the experimental constraints on $h_{125}$. As a consequence, there
is a strong tension with the low values of $m_A
\approx 400\gev$ that we focus on in the present analysis.}
Regarding the CMS excess, the obtained relatively large
values of $0.35 < \mu_{\rm CMS} < 0.6$
have their origin in an enhancement
of the diphoton branching ratio of the state $h_S$,
which reaches values that are an order of magnitude
higher than the SM prediction.
This enhancement compensates the
suppression of the gluon fusion production
cross section that is approximately given
by $c_{h_S gg}^2 \approx 0.07$.

The large values of $\br(h_S \to \gamma \gamma)$ are caused by two
different contributions.
Firstly, the light chargino, which is
close in mass to $h_S$, provides a (positive) BSM
contribution to $c_{h_S \gamma \gamma}$.
Even more important, however, are the
patterns of the effective couplings of $h_S$ to top and bottom quarks. In
our scan we find for the coupling coefficients
$0.220 < |c_{h_S t \bar t}| < 0.237$ and
$0.074 < |c_{h_S b \bar b}| < 0.105$.
The dominant component for the
decay width of the diphoton decay is the
diagram with top quarks in the loop.
Thus, this partial decay width scales with $c_{h_S t \bar t}^2$.
The dominant component of the total width of $h_S$
is given by $\Ga(h_S \to b \bar b)$, such that the total width
is approximately proportional to $c_{h_S b \bar b}^2$.
As a result, we observe an enhancement of
$\br(h_S \to \gamma \gamma)$ by
factors of $4.5 < c_{h_S t \bar t}^2 /
c_{h_S b \bar b}^2 < 7.3$,
which can be further increased by a possible
chargino contribution.
Accordingly, the CMS excess can be described with an almost
singlet-like Higgs boson.

\begin{figure}
\centering
\includegraphics[height=7cm]{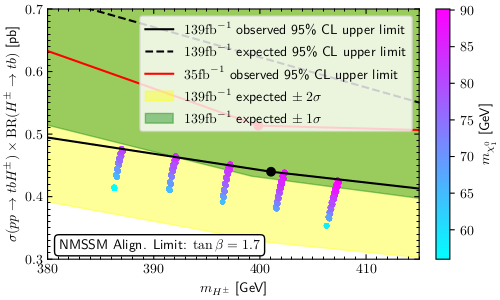}
\caption{NMSSM parameter points with
$\chi^2 \leq \chi^2_{\mathrm{SM}}$
in the
$m_{H^\pm}$--${\sigma(pp \to t b H^\pm) \times
\br(H^\pm \to tb)}$
plane. The colors of the points
indicate the values of
$m_{\widetilde{\chi}^0_1}$.
Also shown are the current observed (black)
and expected (black dashed) upper limits
at the $95\%$ C.L.\ and the expected
$1\sigma$ 
(green) and $2\sigma$ (yellow)
exclusion regions obtained by ATLAS~\cite{ATLAS:2021upq}.
The red line indicates
the previous ATLAS result for the observed
upper limit using
$35\,\mathrm{fb}^{-1}$~\cite{Aaboud:2018cwk}.
}
\label{hpmnmssm}
\end{figure}

\smallskip
Besides the candidates for describing the discussed
excesses, there is another
very prominent feature of the scenario
presented here that offers promising
possibilities for future
collider searches. As was already mentioned
in \refse{nmssmconstraints},
in the alignment limit the CP-odd Higgs boson $A$
at $\approx 400\gev$ is accompanied by slightly
lighter charged Higgs bosons $H^\pm$.
Due to the small value of $\tan\beta$, 
the coupling of $H^\pm$ to top quarks is relatively large.
Experimental upper limits
were obtaind on the product
$\sigma(pp \to t b H^\pm) \times \br(H^\pm \to tb)$
from LHC searches performed both by ATLAS~\cite{ATLAS:2021upq}
and CMS~\cite{Sirunyan:2019arl}.
The currently most stringent limits
for the mass region relevant for our scenario
was reported by ATLAS in \citere{ATLAS:2021upq},
using the full $13\tev$ Run~2 data set with
$139\ifb$. We show in \reffi{hpmnmssm} the
parameter points of our scan with the
corresponding cross section times
branching ratio on the
vertical axis and the charged Higgs boson
mass on the horizontal axis. We also include
the ATLAS observed and expected upper limits
at the $95\%$ C.L. Since these limits are
included in the most recent version of
\texttt{HiggsBounds}, which was used to
check for the constraints from direct
searches of BSM Higgs bosons, our set
of parameter points lies below the
observed limit.

As discussed in
\refse{nmssmconstraints},
those experimental limits could be evaded
if decay modes of $H^\pm$ to SUSY
particles are kinematically open.
In \reffi{hpmnmssm} the color
of the points indicates the values
of the lightest neutralino
mass, $m_{\widetilde{\chi}^0_1}$.
One can observe that the signal rate
of $H^\pm$ decreases with decreasing
$m_{\widetilde{\chi}^0_1}$. This indicates
that in particular the decay modes
$H^\pm \to \widetilde{\chi}^0_1 \widetilde{\chi}^\pm_{1,2}$ 
play an important role in this context, weakening
the collider bounds from
the searches ultilizing the $tb$
final state.\footnote{For the
allowed parameter points we
found values of $3.8\% < \mathrm{BR}(H^\pm
\to \widetilde{\chi}^\pm_1 \widetilde{\chi}^0_1)
< 14\%$.}
Nevertheless, even with such a reduction of the $tb$ mode 
our analysis shows that the signal rate is still rather close to
the current observed upper limit. 
As a consequence, there are good prospects 
that future charged Higgs boson
searches at the LHC and the HL-LHC, both in the $tb$ final state and via
dedicated searches exploring decays into final states involving BSM particles,
will be able to probe this scenario.

Another possiblity for testing the
scenario presented here is given
by the precise measurements of
low energy observables. As was mentioned
in \refse{nmssmconstraints},
these are particularly
sensitive to the presence of the
relatively light charged Higgs bosons
and electroweakinos.
We used \texttt{NMSSMTools} to obtain
predictions for flavour physics observables,
including their theoretical uncertainties,
and the anomalous magnetic moment of the muon.
We find that most of the
flavour observables are
predicted to be in agreement with
the experimentally measured values,
despite the presence of $H^\pm$
below $400\gev$.
However, there are a few observables that show
deviations at the $\approx 2\sigma$
level. As an illustrative example,
we briefly state the largest discrepancies
for the best-fit point of this scan.
Similar values are found
for the other points of the scan.
For the $\br(b \to s \gamma)$ decay,
the predicted range for the best-fit point
within the estimated theoretical uncertainties
is $3.67 \cdot 10^{-4} \leq
\br_{\rm theo}( b \to s \gamma) \leq 4.82 \cdot
10^{-4}$~\cite{Buras:2002vd,Chetyrkin:1996vx,
Ciuchini:1997xe,Ciuchini:1998xy,Bobeth:1999ww,
Czakon:2006ss,Gambino:2001au,Czakon:2015exa},
which lies just above the experimental
$2\sigma$ interval $3.02 \cdot 10^{-4} \leq
\br_{\rm exp}( b \to s \gamma) \leq 3.62 \cdot
10^{-4}$~\cite{Amhis:2014hma}.
A similar discrepancy, however this time
in the opposite direction, is found
for the decay $B_d \to \mu^+\mu^-$,
where the predicted range $1.34 \cdot 10^{-11} \leq
\br_{\rm theo}(B_d \to \mu\bar\mu) \leq 9.15 \cdot
10^{-11}$~\cite{Buras:2002vd,Misiak:2015xwa,
Bobeth:2013uxa,Bobeth:2001jm} is $\approx 2\sigma$
smaller than the experimental range
$11 \cdot 10^{-11} \leq
\br_{\rm exp}(B_d \to \mu\bar\mu) \leq 71
\cdot 10^{-11}$~\cite{Amhis:2014hma}.
Here the substantially larger experimental
uncertainties should be kept in mind.

\smallskip
Summarizing
the results of
the scan presented here, we 
find that the alignment-without-decoupling
limit of the NMSSM
is 
a theoretically well
motivated framework
that is capable of describing, at least approximately, the observed
$t \bar t$ excess in the context
of SUSY. In addition, by choosing the
singlet scalar self-coupling $A_\kappa$
appropriately, a simultaenous 
description
of the CMS excess at about $96\gev$
is possible (but not of the LEP excess). 
We have found that the latter scenario is compatible with the experimental
results for the 
properties of the $h_{125}$ state
even for the case where the
decay of $h_{125}$
into BSM particles,
for instance into two neutralinos,
is kinematically open.

\subsection{Benchmark scenario with
\texorpdfstring{\boldmath{$\tan\beta = 8$}}{tb8}}
\label{sectanbetaacht}

In this section we
analyze a parameter region
in the NMSSM aiming towards a possible description
of the $\tautau$ excess.
As already discussed in \refse{sec:align}, this requires
larger values of $\tan\beta$ as compared to the
scan in \refse{sectanbetaeinssieben},
which leads to the fact that
the alignment-without-decoupling
conditions cannot be 
met anymore for perturbative values
of $\lambda$ and $\kappa$.
This is why we 
consider here the usual
decoupling limit, i.e.\ larger values of $M_A$, in order to ensure
that the $h_{125}$
properties are in agreement with the
signal rate measurements at the LHC.
Hence, we use 
a wider range of $M_A$ in our scan, extending up to $M_A = 560\gev$.
Physical masses of the neutral BSM
Higgs bosons of about $400\gev$ are
then achieved via mixing of the doublet states
with the singlet states.
Moreover, in the decoupling limit the
deviations from the alignment limit scale with the factor
$\Delta c_{h_{125} VV} = M_Z^2 \sin{4\beta}/(2 M_A^2)$~\cite{Carena:2001bg}
at tree level,
such that also the larger value of $\tan\beta = 8$
(compared to $\tan\beta = 1.7$ in
\refse{sectanbetaeinssieben}) already
facilitates SM-like behavior
of $h_{125}$
even for $M_A \approx 400\gev$.\footnote{One
finds $\sin{4\beta} = -0.85 (-0.48)$ for $\tan\beta = 1.7 (8.0)$.}
However, for the coupling of $h_{125}$
to down-type fermions the decoupling behavior is delayed
for larger values of $\tan\beta$, so that
sizable deviations of $c_{h_{125} b \bar b}$
and $c_{h_{125} \tautau}$ from the respective SM value
can occur (see e.g.\ \citere{Carena:2001bg}).

\begin{table}
\centering
\def\arraystretch{1.5}
\footnotesize
\begin{tabular}{cccccccccccc}
$\tan\beta$ & $\lambda$ & $\kappa$ &
    $\mu$ & $M_A$ & $A_\kappa$ &
$A_t$ & $A_{b,\tau}$ & $m_{\widetilde{f}}^2$ &
    $M_1$ & $M_2$ & $M_3$ \\
\hline
$8$ & $0.36$ & $0.58$ &
    $[110, 170]$ & $[360, 560]$ & $-200$ &
$6200$ & $3000$ & $2500^2$ &
    $1000$ & $2000$ & $2700$
\end{tabular}
\caption{\small
Parameter values for the NMSSM scan
with $\tan\beta = 8$
for investigating a possible realization of
the $\tautau$ excess.
}
\label{tablenmssm8}
\end{table}

We show the set of input parameters for
our scan in \refta{tablenmssm8}.
By comparing to the left plot of \reffi{figsaligncond}
one can see that the value for $\lambda$
is far below the value 
required to fulfill
the alignment condition given in
\refeq{aligncond1}. In order to avoid a sizable mixing
of $h_{125}$ and $h_S$ we require that
$h_S$ is much heavier than $125\gev$
by using a large value of $\kappa = 0.58 > \lambda$.
In combination with the value $A_\kappa = -200\gev$
and the parameter range of $\mu$,
we find masses of $280\gev < m_{h_S} < 442\gev$,
$286\gev < m_{A_S} < 367\gev$,
$352\gev < m_H < 526\gev$
and $364\gev < m_A < 522\gev$.
The masses of the states with dominant doublet component
($H$ and $A$) are more closely related to the value
of $M_A$, and therefore can be even larger
than $500\gev$ in this scan.
In order to calculate $\chi^2_{\tautau}$ we
sum up the cross sections of all neutral scalars
in the mass interval $(400 \pm 40)\gev$,
where interference effects can be neglected
due to CP conservation and a sizable
mass splitting of $A_S/A$ and $h_S/H$
in each parameter point (see also the discussion below).
Thus, for the upper end of these ranges both singlet-like
states can 
contribute to the $\tautau$
excess, in addition to the doublet states
$H$ and $A$.
Here it should be noted that
the pairs $A$/$A_S$ and $H$/$h_S$ are mixed,
such that the classification into doublet 
and singlet states is only approximate.
Therefore, in the following
we will adopt the mass-ordered notation
$h_{2,3}$ and $A_{1,2}$, 
with $h_1 = h_{125}$.

Taking into account that
we want to exploit the mixing of
the singlet states with the heavy
doublet states in order to obtain
physical masses of $m_{A,H}\approx 400\gev$ even
for considerably larger
values of $M_A$, and also that the
singlet-like states can
contribute to the $\tautau$
excess for $m_{h_S,A_S} \lesssim 400\gev$,
it is apparent that in the scenario considered
here the excesses at $96\gev$
cannot be addressed. We emphasize
that attempting to fit the
excesses at $96\gev$ in combination with
the $\tautau$ excess is in any case
not very promising, since (as discussed above)
the description of the
$\tautau$ excess at about $400\gev$ in the NMSSM relies
on decoupling effects in order to obtain
a phenomenologically viable state $h_{125}$.
Demanding a CP-even singlet-like state at
$96\gev$ that is mixed with $h_{125}$
in order to provide a description of the observed
excesses would give rise to
unacceptably large
modifications of the couplings
of $h_{125}$ compared to the SM.\footnote{See
\citere{Cao:2016uwt,Domingo:2018uim} for discussions of 
fits that exclusively address the excesses at $96\gev$
in the NMSSM.}

The parameter
ranges as defined in \refta{tablenmssm8}
lead to the presence of
at least two neutral Higgs bosons with masses of
$\approx 400\gev$. As mentioned before, for $M_A > 500\gev$
the masses $m_{h_3}$ and $m_{A_2}$ can still
be 
close to $400\gev$
as long as large mixings with the lighter 
states are present.
The parameters related to the squark sector are
chosen to obtain a SM-like Higgs boson mass
of $\approx 125\gev$. For a value of $\tan\beta = 8$
the additional contribution to the tree-level
mass of $h_1$ proportional to $\lambda^2$ is
small. Thus, sizable radiative corrections to $m_{h_1}$
are required,
in analogy to the case of the MSSM. Compared to the
scan presented in \refse{sectanbetaeinssieben}, we therefore
increase the stop soft SUSY-breaking mass parameters
to $m_{\widetilde{t}_{L,R}}^2 = (2.5\tev)^2$.
Furthermore, the large value of $A_t = 6.2\tev$ yields
a large stop mixing, which further increases
the radiative corrections to the mass of $h_1$.\footnote{The
large value of
$|A_t / m_{\widetilde{t}}|$
can potentially lead to the presence
of color-breaking minima below the EW minimum,
and therefore to problems with vacuum
stability ~\cite{Casas:1995pd,Hollik:2018wrr}.
While the chosen value is near the border between the (phenomenologically
viable) region featuring a meta-stable vacuum and the region where the vacuum
would become unacceptably short-lived, we stress that a prediction for
$m_{h_{125}}$ in accordance with the experimental result
can also be achieved with values
of $A_t$ below $6\tev$ and/or larger stop masses.}
For completeness, we list in \refta{tablenmssm8} also
the values of the
remaining soft trilinear parameters $A_{b, \tau}$ and the
soft gaugino mass parameters $M_{1,2,3}$. However,
besides the value of $M_3 = 2.7\tev$ on which
$m_{h_1}$ depends mildly, these parameters are
of no relevance in the following discussion.

We performed a simple grid scan over the
parameters $\mu$ and $M_A$. While the former is
proportional to the singlet vev $v_S$ and thus
closely related to the mass scale of the singlet
states, the latter defines the approximate mass scale of the
heavy doublet states.
Since in this scan no Higgs boson at $\approx 96\gev$
is present, we used the definition of the total
$\chi^2$ as defined in \refeq{eqchisqttno96},
i.e.\ without $\chi^2_{96}$, in order
to quantify the quality of the fit for each
parameter point. We emphasize that for the chosen value of
$\tan\beta = 8$ no sizable contribution to the
$t \bar t$ excess is present. However, we
keep the quantity $\chi^2_{t \bar t}$
in $\chi^2$ for completeness,
as was done in \refse{sectanbetaeinssieben}
for the $\chi^2_{\tautau}$
contribution ($\chi^2_{\mathrm{SM},\tautau}
= 9.99$).\footnote{Since we find
$m_{A_1} < 365\gev$, we only take
into account the contribution of $A_2$ for
the calculation of $\chi^2_{t \bar t}$.}

In comparison to the N2HDM analysis of
the $\tautau$ excess, in which only
one particle gave rise to the excess,
in our NMSSM analysis 
we have always at least two Higgs bosons
with masses close to $\approx 400 \gev$.
For the calculation of
$\chi^2_{\tautau}$, as mentioned above, we
take into account the signal contributions
of all neutral BSM Higgs bosons within a
mass interval of $(400\pm 40)\gev$, corresponding
to a mass resultion of $10\%$.
We checked for all parameter points that
the mass differences of $A_1/A_2$ and
$h_2/h_3$ are much larger than the sums
of their total widths.
We thus can neglect interference effects 
(see the discussion in \citeres{Fuchs:2014ola,Fuchs:2016swt,Fuchs:2017wkq})
and simply sum the individual contributions. They are each
obtained as the product of the cross sections
for $b \bar b$ 
and $gg$ production
multiplied by the branching ratio for the decay
into a $\tau$~lepton pair.

\begin{figure}
\centering
\includegraphics[width=0.48\textwidth]{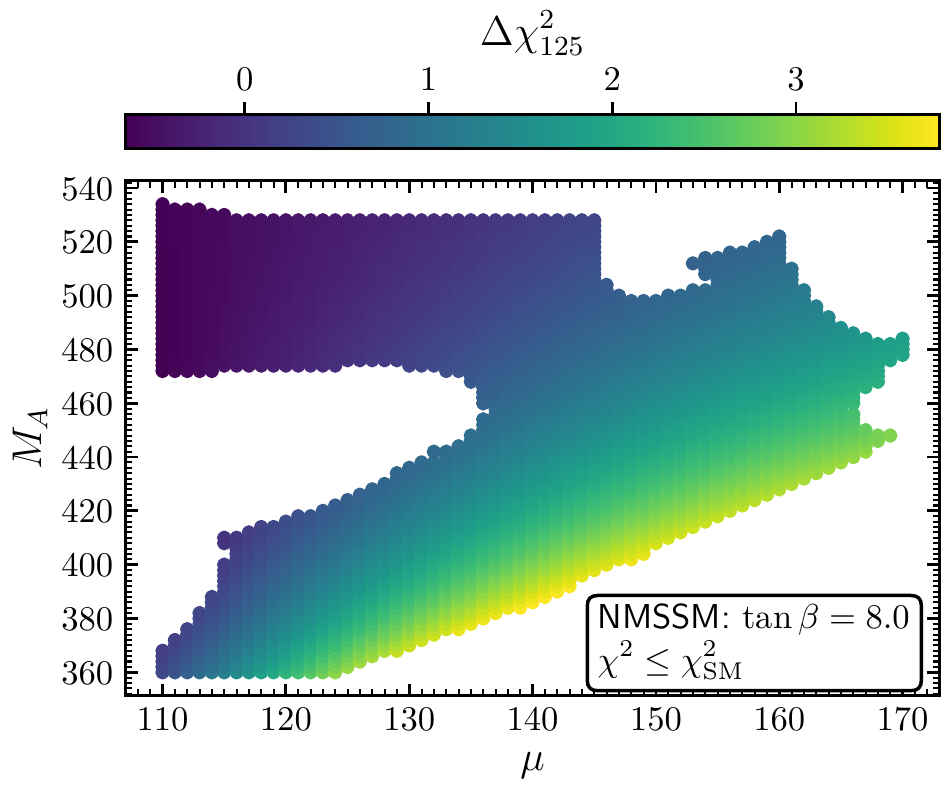}~
\includegraphics[width=0.48\textwidth]{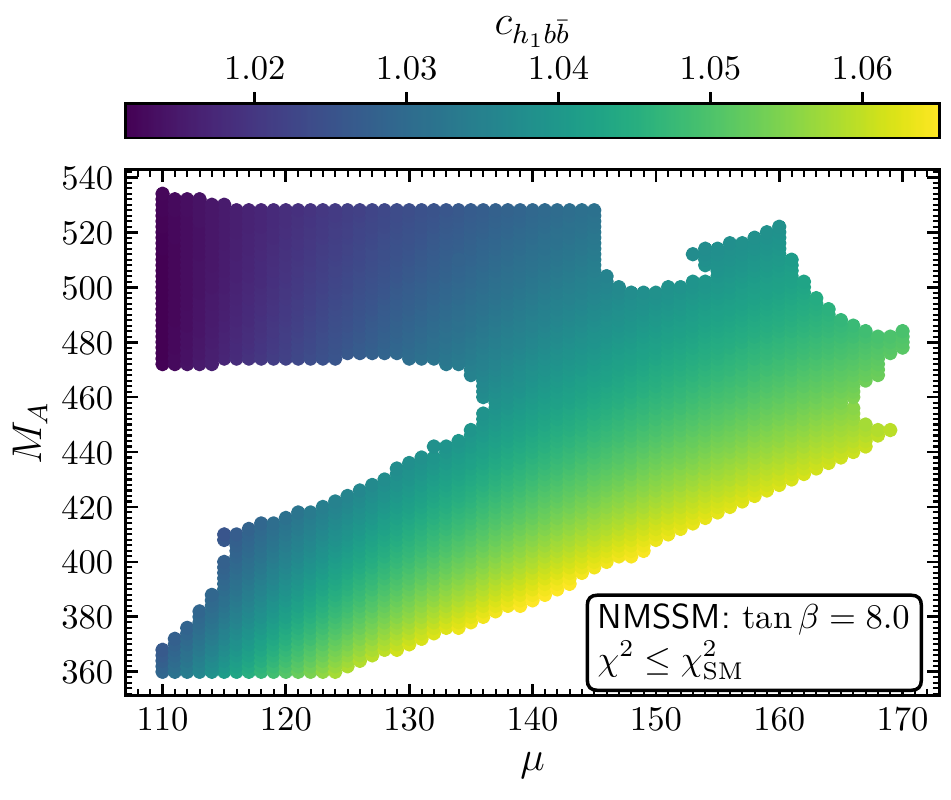}
\caption{NMSSM parameter points of the
scan (see \refta{tablenmssm8}) with $\tan\beta = 8.0$
in the plane $\mu$--$M_A$. All shown points have
$\chi^2 \leq \chi^2_{\rm SM}$ and 
pass the \texttt{HiggsBounds} test.
The colors of the points indicate
the values of
$\Delta\chi^2_{125}$ (left) and
$c_{h_1 b \bar b}$ (right).}
\label{nmssm8hs}
\end{figure}

We 
start the discussion of the scan
by presenting the parameter
ranges of $\mu$ and $M_A$ that pass
the experimental constraints.
In \reffi{nmssm8hs} we show the parameter points,
where $\mu$ is displayed on the horizontal and $M_A$ on the
vertical axis. In the left plot the colors of
the points indicate the value of $\Delta \chi^2_{125}$. 
One can see that all points
have very small values of
$\Delta \chi^2_{125} \lesssim 3.5$, 
and some points even yield a better description of the properties 
of $h_{125}$ than the SM (corresponding to negative values
of $\Delta \chi^2_{125}$).
Accordingly,
the signal rates of the $h_{125} (= h_1)$ state agree very well
with the experimental data. As discussed before, in
this scan we rely on the decoupling effects
in order to obtain a SM-like Higgs boson
at $\approx 125\gev$. This is reflected in the
fact that $\Delta \chi^2_{125}$ decreases
with increasing values of $M_A$.
One can also see that, regarding the values
of $\Delta \chi^2_{125}$, 
there is a slight preference for lower values of $\mu$.
Comparing to the
right plot of \reffi{nmssm8hs}, in which the colors
of the points indicate the value of
$c_{h_1 b \bar b}$, one can see 
that the overall pattern of deviations in $c_{h_1 b\bar b}$ is 
very similar to the one of $\Delta \chi^2_{125}$.
For the lowest values of $M_A$ for a given value of $\mu$
the coupling coefficient
$c_{h_1 b \bar b}$ deviates from
the SM prediction by up to $\approx 6\%$.
This leads to a small increase in the total
width of $h_1$, which in turn reduces the
branching ratios $\br(h_1 \to \gamma\gamma, WW^*)$, and thus
increases
$\Delta \chi^2_{125}$.
The deviations of $c_{h_1 t \bar t}$
and $c_{h_1 VV}$ (not shown here) remain below $1\%$
and are therefore much smaller than the ones
of $c_{h_1 b \bar b}$.
This is due to the fact
that those couplings show a faster decoupling with increasing
values of $\tan\beta$ and/or $M_A$,
while the decoupling behavior
of $c_{h_1 b \bar b}$ is delayed (see the discussion above).

The parameter points shown
in \reffi{nmssm8hs}
furthermore
passed the \texttt{HiggsBounds}
test.
As expected, for the present scan that is targeted to a description of 
the $\tautau$ excess we find that
for all the parameter points the most sensitive
constraint is the one from the
searches for neutral Higgs bosons in
the $\tautau$ final state performed
by ATLAS~\cite{Aad:2020zxo}.
The precise shape of the excluded regions
in \reffi{nmssm8hs} arises from a complicated interplay
between the masses of the Higgs bosons, the
mixing between $h_2$/$h_3$ and $A_1$/$A_2$,
and finally the number of Higgs bosons
that \texttt{HiggsBounds}
combines into a single signal 
that is confronted with the search limits.
For instance, the lower right triangular shaped
region is excluded because in this parameter region \texttt{HiggsBounds}
combines the signal contributions of all
four neutral BSM Higgs bosons into a single
signal, owing to the fact that, roughly speaking,
their mass differences
decrease
for decreasing values of the ratio
$M_A / \mu$. On the contrary, for larger
values of this ratio, the contributions
of just the two CP-even Higgs bosons $h_{2,3}$
with masses
$m_{h_{2,3}}$ above the region of $\approx 400\gev$ 
where the excess was observed
yield a signal rate that for many points 
is excluded by the limits from the search for a
$\tautau$ resonance.
This is the reason for the fact that the highest values of $M_A$ that are
displayed in \reffi{nmssm8hs} are excluded by \texttt{HiggsBounds}.

\begin{figure}
\centering
\includegraphics[height=0.85\textheight]{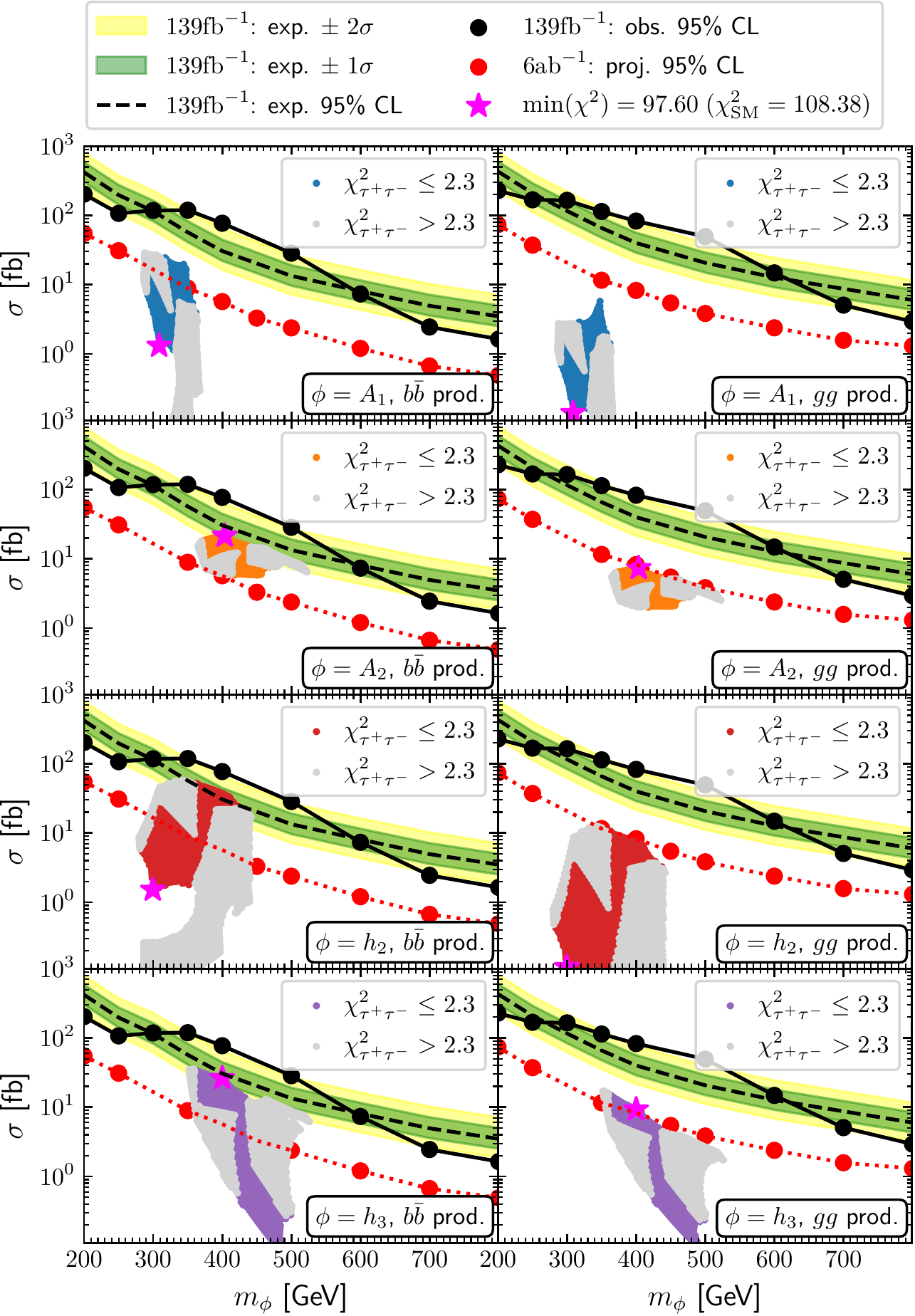}~
\caption{NMSSM parameter points of the
scan with $\tan\beta = 8.0$ (see \refta{tablenmssm8})
with $m_\phi$ on the horizontal axis
and $\sigma(b \bar b \to \phi \to \tautau)$
(left) and $\sigma(gg \to \phi \to \tautau)$
(right) on the vertical axis,
where $\phi = A_1, A_2,
h_2$ and $h_3$ in the first, second, third
and fourth row, respectively.
The colors of the points indicate whether
the parameter point lies inside (colored)
or outside (gray) of the $1\sigma$
ellipse regarding $\chi^2_{\tautau}$.
The best fit values are highlighted with
a magenta star.
The current expected and observed limit~\cite{Aad:2020zxo}
as well as a projection for the expected HL-LHC exclusion
limits~\cite{Cepeda:2019klc} are also shown.
}
\label{nmssm8xsll}
\end{figure}

In order
to get an idea of the composition of
the Higgs bosons that contribute to
the description of the $\tautau$
excess in this scenario, we show
in \reffi{nmssm8xsll} the individual values of
the signal cross sections for each neutral BSM Higgs boson
(i.e., the cross sections for $h_{125}$ are not displayed).
The colored points fit the $\tautau$
excess within the $1\sigma$ level, i.e.\
$\chi^2_{\tautau} < 2.3$, whereas
the gray points lie outside of the $1\sigma$
ellipse regarding $\chi^2_{\tautau}$.
In each row of \reffi{nmssm8xsll}, we also
show the expected (black dashed line with green and yellow bands)
and observed (black dots) exclusion limits
from the ATLAS search
using the full Run~2 dataset~\cite{Aad:2020zxo}.
In addition, the red points indicate
the expected HL-LHC exclusion limits taking
into account $6\iab$~\cite{Cepeda:2019klc}.
For each parameter point
there are always at least two ($h_2$, $A_2$),
sometimes three particles
($h_{2,3}$, $A_2$) 
that contribute to the excess within the considered 
mass interval of $(400 \pm 40) \gev$.
For some points, even $A_1$ has a mass
of $m_{A_1} > 360\gev$. However, its signal
contribution is small compared to the other
Higgs bosons as can be seen in the top
row of \reffi{nmssm8xsll}. For $b \bar b$ production,
as shown on the left-hand side of
\reffi{nmssm8xsll}, the most significant
contribution is given by $h_2$ (third row).
For the $gg$ production mode there
is no parameter point for which a single
Higgs boson individually produces a sufficiently
large signal, but signal rates of
up to about $20\ifb$ are achieved when
summing over the contributions of
two or more Higgs bosons within
the required mass window
(see also the discussion below).
Due to the value of $\tan\beta = 8$
chosen for this scan, we obtain cross sections
for the $b \bar b$ production mode that
are roughly twice as large as for the
$gg$ production mode.
This is well in line with
the signal interpretation of the
$\tautau$ excess, 
so that one can expect
a good fit to the data. 
The best fit point is highlighted in
\reffi{nmssm8xsll} with a magenta star.
For this point, 
the signal contribution describing the observed excess arises from
the two Higgs bosons $A_2$ (second row)
and $h_3$ (fourth row)
with masses of $m_{A_2} \approx m_{h_3}
\approx 400\gev$.
The remaining two BSM Higgs bosons
$A_1$ (first row) and $h_2$ (third row)
are roughly $100\gev$
lighter and have much smaller cross
sections for this parameter point and therefore do not contribute
to the signal. The best fit point
has $\chi^2 = 97.60$, which is substantially
below the SM value of $\chi^2_{\rm SM} = 108.38$.
The difference of these values arises almost
entirely from the values
$\chi^2_{\tautau} = 0.29$ and
$\chi^2_{\mathrm{SM}, \tautau} = 9.99$,
respectively.\footnote{The remaining
difference comes from
$\chi^2_{t \bar t} - \chi^2_{\mathrm{SM}, t\bar t}
\approx  12.90 - 13.97
\approx -1$.} It is also important to note
that 
with the projected sensitivity the HL-LHC will be very well suited 
for probing the considered scenario. It will either very significantly 
confirm or rule out the $\tautau$ excess. In the latter case, it would exclude
the parameter region corresponding to the signal interpretation in our NMSSM
analysis (see in particular the left plot in the second row of
\reffi{nmssm8xsll}).
On the other hand,
in case the observed excess is indeed due to 
one ore more BSM Higgs boson(s) at $\approx 400\gev$ the HL-LHC has excellent
prospects for discovering those new states.

\begin{figure}
\centering
\includegraphics[width=0.48\textwidth]{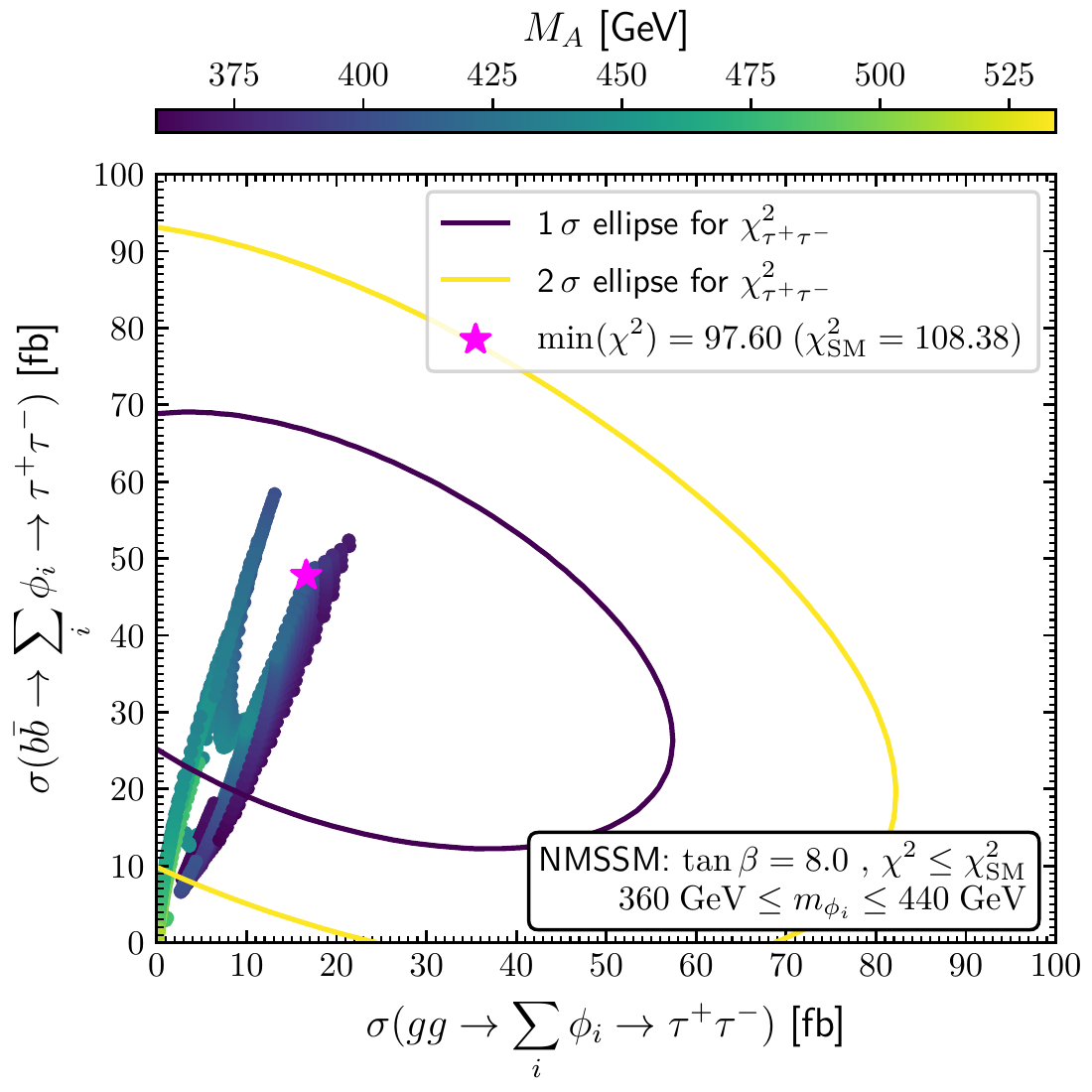}~
\includegraphics[width=0.48\textwidth]{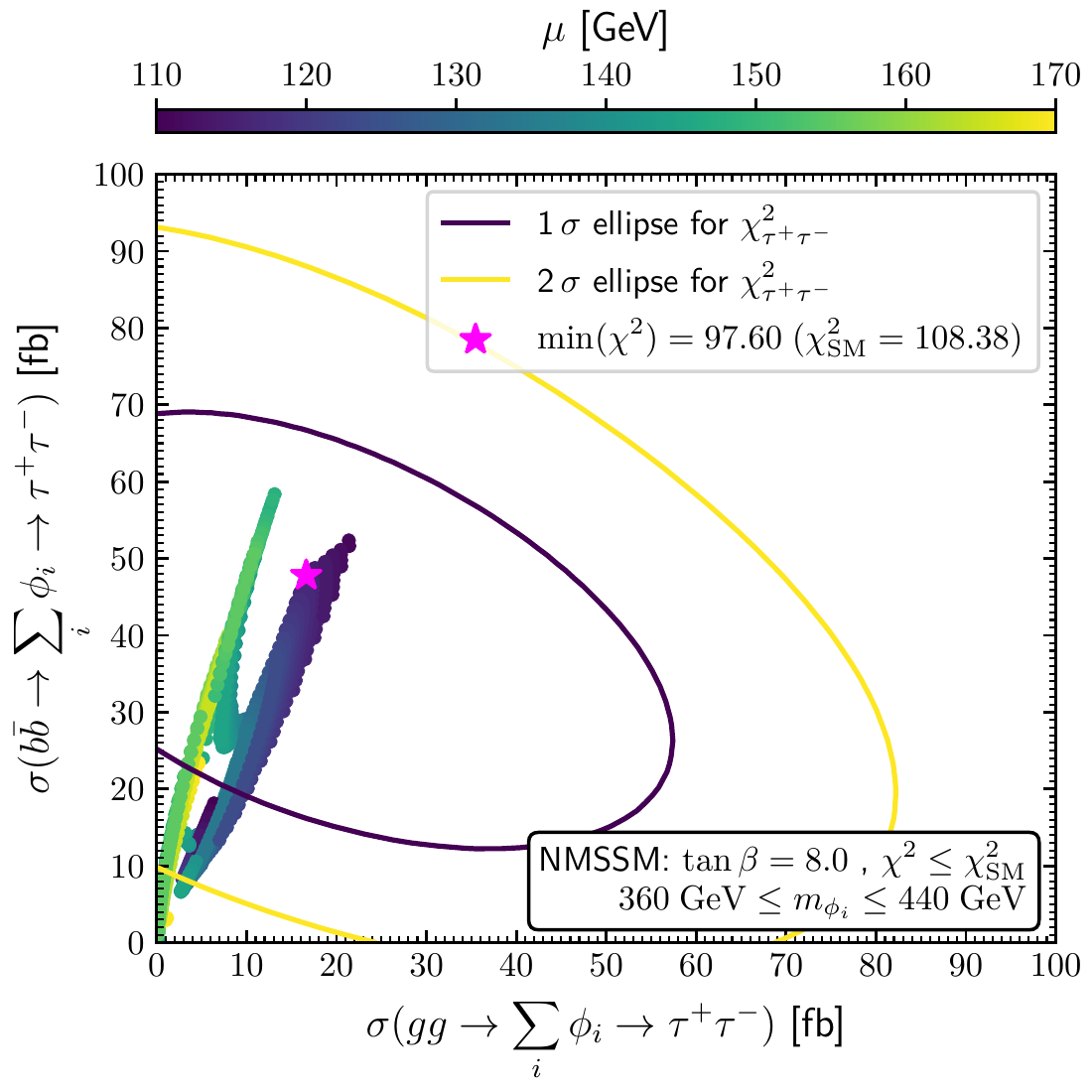}
\caption{NMSSM parameter points of the
scan with $\tan\beta = 8.0$ (see \refta{tablenmssm8}):
signal cross sections for the $\tautau$
excess for the $gg$ production mode on the
horizontal axis and the $b \bar b$ production mode
on the vertical axis. The colors of the points
indicate the value of $M_A$ (left) and $\mu$ (right).
The dark blue
and the yellow contours indicate the $1\sigma$
and $2\sigma$ ellipse of
$\chi^2_{\tautau}$, respectively~\cite{Aad:2020zxo}.
The best-fit point is indicated with a magenta star.}
\label{nmssm8llchsq}
\end{figure}

In \reffi{nmssm8llchsq} we show the parameter
points with $\chi^2 \leq \chi^2_{\rm SM}$ in
the plane of the cross sections for the
two production modes of the $\tautau$
excess. One should keep in mind that in order to
obtain $\chi^2_{\tautau}$ we took into
account the contributions of all BSM Higgs
bosons within the mass intervall
$(400 \pm 40)\gev$, which is of the order of
the step sizes of $50\gev$ of the different mass
hypotheses of the experimental analysis~\cite{Aad:2020zxo}.
The colors of the points indicate the values
of the NMSSM parameters that were varied
in this scan, where $M_A$ is displayed in the left and
$\mu$ in the right plot. The best fit point is indicated by a magenta star,
and the contours indicate the $1\sigma$
and $2\sigma$ ellipses of
$\chi^2_{\tautau}$~\cite{Aad:2020zxo}.
One can see that the majority of points lies
within the $1\sigma$ ellipse of
$\chi^2_{\tautau}$. Taking into
account the scan ranges, the points
in the $1\sigma$ ellipse cover the whole
range of $\mu$. However, for $M_A$ we find
that only values up to $M_A \lesssim 460\gev$
yield a description of the excess at the
$1\sigma$ level. For larger values the
particles $h_{2,3}$ and $A_2$ become
heavier than $440\gev$, such that only
a smaller overall signal is achieved.
In the left plot of
\reffi{nmssm8llchsq} one can furthermore
observe that there is a slight tendency towards
smaller values of $M_A$. Similarly,
a tendency towards smaller $\mu$ values
can be observed in the right plot.
These slight preferences are however
statistically
insignificant 
in view of
the current experimental uncertainties.

\section{Prospects and outlook}
\label{sec:prosp}
As a final step of our analyses
of the N2HDM and the NMSSM with respect to
possible interpretations of the observed excesses, we briefly discuss the most 
relevant future experimental searches and measurements which can probe those
scenarios. In fact, we would like to point out that there are good prospects
for either strengthening the evidence for a possible signal or ruling out the
discussed scenarios in the near future.

For the low $\tan\beta$ region, the searches for
$A \rightarrow t \bar t$ will obviously be important.
The most promising channels are
gluon fusion production, in which the excess was found,
so far only utilizing  the first year Run~2
data corresponding to $35.9\ifb$, as well
as the production of $A$
in association with two top quarks, leading to a
final state of four top quarks. The latter is 
already published taking into account the full
Run~2 data by both ATLAS and
CMS~\cite{ATLAS:2020hrf,Sirunyan:2019wxt}, and
it is interesting to note that ATLAS measured
a cross section for four-top production roughly
a factor of two larger than the SM
prediction~\cite{ATLAS:2020hrf}. In addition, in the N2HDM
the parameter region in the wrong
sign Yukawa coupling regime will further be probed
via the process ${gg \rightarrow A \rightarrow Z h_{125}}$,
where the ATLAS analysis using the full Run~2
data set puts important constraints on
the parameter space already.
For the scenarios including
the singlet-like Higgs boson at $\approx 96\gev$, also
the 
searches with $A$ decaying into a $Z$
boson and another BSM Higgs boson can be important.
Moreover, the searches for the charged
Higgs bosons in the $tb$ final state are very
promising. In the NMSSM, 
the current limits
exclude
parameter points with valuess for $c_{A t \bar t}$
of the size of the experimental best-fit values,
while in the N2HDM the charged Higgs bosons
can be somewhat heavier,
so that the parameter region yielding the best description of the 
$t \bar t$ excess is not affected by the current limits.
For the case of the NMSSM also dedicated searches exploring decays of the
charged Higgs boson into final states involving BSM particles can be promising
(see also the discussion in \refse{sectanbetaeinssieben}).

For the high $\tan\beta$ region, i.e.\ $\tan\beta \approx 8$,
the parameter region corresponding to the signal interpretation will fully
be probed by the HL-LHC Higgs boson
searches in the $\tautau$ final state.
If the CMS search
including the full Run~2 dataset does not confirm the excess observed by ATLAS,
the cross section values that
are preferred by
the $\tautau$ excess in the corresponding
ATLAS search could already be excluded on the basis of the Run~2 data from both
collaborations.
On the other hand, if the observed excess is indeed a first indication of a
signal of one or more more BSM Higgs boson(s), 
the prospects for discovering these new
states in future runs at the LHC and the HL-LHC will be excellent.
Furthermore, the cascade decays $A \rightarrow Z h_{125}$,
where the $A$ boson is produced in the $b \bar b$ production mode, 
will also probe this scenario. It is important to note in this context that
in contrast to the ATLAS search in the $gg$ production
mode, the ATLAS search assuming $b \bar b$ production
has not yet been updated to include
the full Run~2 data set.

As already mentioned before,
independently of the value of $\tan\beta$ the presence
of 
relatively light charged Higgs bosons is a
common prediction of
the scenarios discussed
in this paper, where in the NMSSM the
charged Higgs bosons are even lighter
than the CP-odd Higgs boson at $400\gev$,
while in the N2HDM an upper bound of
$m_{H^\pm} \lesssim 750\gev$ applies because of
the theoretical constraints.
In the NMSSM the constraints from flavor physics do not result 
in a firm lower limit on $m_{H^\pm}$, since
in general the theoretical predictions for the
flavor observables in SUSY models
depend on various different sectors of
the model, and may be weakened without
changing the Higgs-boson phenomenology
discussed here.
In the N2HDM, on the other hand, we included a lower limit of
$m_{H^\pm} = 550\gev$, based on
flavor physics constraints~\cite{Haller:2018nnx}.
More recently, a new theoretical
calculation suggested a lower limit of $m_{H^\pm} >
800\gev$~\cite{Misiak:2020vlo}, which 
however is still
under debate in particular in view of the results
of~\citere{Bernlochner:2020jlt} pointing to possible underestimates
of theoretical uncertainties 
that could have the effect of even weakening the 
bound on $m_{H^\pm}$ below the value that we have adopted in our scan.
In any case it is obvious that new results concerning flavor
physics observables can have an
important impact on the investigated parameter space.

Regarding the compatibility of the
predicted signal rates of $h_{125}$ with
the experimentally measured values, our
results indicate that projected HL-LHC
measurements of the properties of
$h_{125}$ alone might not be sufficient
to exclude (or confirm) the scenarios
describing the excesses at $400\gev$.
The situation is somewhat more promising
when additionally the presence of a
singlet-like Higgs boson at $96\gev$
is considered. In this case, the mixing
of the singlet-like Higgs boson with
$h_{125}$, as required to simultaneously
accommodate the LEP and CMS excesses, gives rise
to modifications of the signal rates
of $h_{125}$ compared to the SM
predictions that could be observable
at the HL-LHC.

The NMSSM is additionally constrained
by experimental
searches targeted specifically to the
SUSY particles. For the scenario in
the alignment-without-decoupling limit,
it was found that the stops cannot be
much heavier than $1\tev$ in order to
obtain a theoretical prediction for
the mass of $h_{125}$ in agreement
with experiment. 
The HL-LHC has a high sensitivity for probing this
mass region.
Another possibility
in this context are searches for the
light electroweakinos, where in
particular searches focusing on
compressed electroweakino spectra
are promising. More exotic signals
that are present in our analysis arise
from the decays of the charged Higgs bosons
into pairs of a neutralino and a chargino,
which as mentioned above could be probed via dedicated searches.
Finally, a small fraction of the analyzed
parameter points feature a neutralino
with masses smaller than $62.5\gev$, such that
the decay of $h_{125}$ into a pair of
neutralinos is kinematically allowed.
Such points will be further probed by
direct searches for
invisible decays of $h_{125}$ and by the
global constraints from the signal-rate
measurements of $h_{125}$ at the
HL-LHC.

\section{Conclusions}
\label{sec:conclusion}

Various searches at the LHC for BSM Higgs bosons with masses above $125 \gev$ 
showed excesses over the background expectation
in the Run~2 data. As a remarkable feature, several of these excesses
were found around a mass scale $\mphi \approx 400 \gev$ 
of a hypothetical new Higgs boson $\phi$ (or more than one). 
In particular, we focused on possible interpretations of the two most
striking excesses:
CMS reported a local excess of more
than $3\,\sig$ in the channel
$A \to t\bar t$ in their first year Run~2 data~\cite{Sirunyan:2019wph},
while
ATLAS reported a local excess of about
$3\,\sig$ in the channel
$\phi \to \tautau$ in their full Run~2 data~\cite{Aad:2020zxo}.
In both cases the analysis of the other experiment for the same period
of data taking is not yet available.
In addition, a local excess of more than
$3\,\sigma$ was found in the ATLAS search for a heavy
resonance decaying into a $Z$ boson and $h_{125}$
assuming $b \bar b$-associated
production and utilizing $35\ifb$ of data~\cite{Aaboud:2017cxo}.
While the various excesses reach the level of
$3\,\sig$ local significance,
all stay below $3\,\sig$ global significance.
Also the searches at the LHC for BSM Higgs bosons with masses below
$125 \gev$ show an interesting excess at about
$96 \gev$ in the channel
$pp \to \phi \to \ga\ga$: CMS reported a local excess of
about $3\,\sig$ in their first
year Run~2 data~\cite{Sirunyan:2018aui} and a
similar deviation of~$2\,\sigma$ local significance
in their Run~1 data at a comparable
mass~\cite{CMS:2015ocq}. The ATLAS results based on the data of the
first two years of Run~2~\cite{ATLAS:2018xad} are not sensitive to
the excess observed at CMS.
Furthermore,
LEP reported a local excess of about $2\,\sig$
in the channel $e^+e^- \to Z\,\phi \to Z\,b\bar b$~\cite{Barate:2003sz}.

In this paper we have analyzed whether the observed excesses can be 
described by 
models comprising an extended Higgs sector, 
where we have concentrated on the Next-to-2HDM (N2HDM) and
the Next-to-MSSM (NMSSM) as minimal scenarios for accommodating the
experimentally observed patterns.
Concerning the excesses at about $96 \gev$ from CMS and LEP 
taken in isolation,
it is known that several models
can fit both excesses simultaneously at the level
of roughly $1\sigma$ or less, including the N2HDM
and the NMSSM.
In the present paper we have analyzed the question 
whether the excesses at about $400 \gev$ can be accommmodated in these models
and whether the parameter space that is preferred by those analyses in
addition permits a signal interpretation also of the excesses at about
$96 \gev$.
In our analysis we have taken into account 
the experimental constraints on the properties of the Higgs boson at
$125 \gev$, the limits from BSM Higgs-boson searches at the LHC and previous
colliders, constraints from electroweak precision data and flavor observables,
as well as theoretical constraints such as vacuum (meta)stability
and perturbativity.

We first investigated whether the N2HDM can fit the two excesses at
$400 \gev$ such that the relevant parameter region is in agreement with 
the theoretical and experimental constraints. While we find that 
both excesses cannot be fitted simultaneously, each of them can be
described with a very good $\chi^2$
by the type~II N2HDM, while complying with the
other constraints. The $t\bar t$ excess, independently of the N2HDM
type, can be described for
$1 \lesssim \tb \lesssim 2$,
whereas a description of the $\tautau$ excess in the N2HDM type~II
requires
$6 \lesssim \tb \lesssim 11$,
making them mutually incompatible.
In both cases the description of the excesses mainly occurs as a
consequence of the presence of a
CP-odd Higgs boson with $m_A \approx 400 \gev$, while the other Higgs
bosons are found to be not much heavier in those scenarios
owing to theoretical constraints.
Interestingly, even though not directly aimed for
in our analysis, a sizable fraction of the parameter
points that fit the $\tautau$ excess
additionally yield a contribution that would be compatible with a 
description of the $A \rightarrow Zh_{125}$ excess that
was found at a similar mass.

In a second step, we have demonstrated that 
for each of the parameter regions that are preferred by the excesses at about
$400\gev$
the N2HDM
can simultaneously also describe both excesses at
about $96 \gev$.
This happens through the presence of a 
singlet-like scalar at $96 \gev$ giving rise to signal rates that are
compatible with the excesses observed at CMS and at LEP.
More specifically,
in type~II either the $t \bar t$ or the $\tautau$ excess 
at about $400\gev$ can be
described in their respective $\tb$ range, where in the
latter case also a sizable contribution to the $Zh$
excess can be present.
In type~IV the $t \bar t$
excess is compatible with the $96 \gev$ excesses, where the CMS excess
can be described at the
level of roughly $1 \sigma$.

In the NMSSM, which has a type~II Yukawa sector,
as a consequence of the underlying symmetry relations of the model
one has much less freedom regarding the choice of parameters in the Higgs
sector as compared to the N2HDM.
Nevertheless, we have demonstrated that also the NMSSM
can fit each of the excesses at $400 \gev$ individually, while
complying with the
BSM Higgs-boson searches and Higgs-boson measurements 
as well as the other constraints.
Concretely, we discussed exemplary scenarios
in which we scanned over $\MA$, $\mu$, and in
case of low $\tb$ also over $\kappa$ and $A_\kappa$.
The $t\bar t$ excess is described
at the level of $1.5\, \sigma$
for low $\tb$ in the
alignment-without-decoupling limit,
which is a theoretically and, in view of the constraints from
the Higgs boson measurements, experimentally
well motivated scenario.
The $\tautau$ excess, on the other hand, can be described for larger
$\tb$ values somewhat away from the alignment limit, 
while the properties of the Higgs boson at about $125\gev$ are
nevertheless compatible with the experimental data within the current
uncertainties.

In the NMSSM parameter region which can give rise to the $t \bar t$
excess naturally a light singlet-like scalar is present. Requiring
its mass to be $\approx 96 \gev$, we demonstrated that 
the CMS excess at about $96 \gev$ 
can be described simultaneously with the $t \bar t$ excess at about $400 \gev$, 
whereas in this scenario hardly any signal contribution would have been
generated at LEP.

If the observed excesses indeed turn out to be first indications of one or
more BSM Higgs bosons, besides the channels featuring those excesses
particularly promising channels regarding a possible discovery of a BSM Higgs
boson would be
the search for charged Higgs bosons
with masses around $400 \gev$ 
and searches for the
heavier doublet-like Higgs bosons decaying to a $Z$ boson
and either the SM-like Higgs bosons or the CP even/odd
singlet-like scalar with masses below or slightly
above $125\gev$.
The scenarios describing the $\tautau$ excess
will entirely be probed by the $\phi \to \tautau$
searches at the HL-LHC.
Thus, there are very good prospects in the near future for clarifying the
tantalizing hints that have been observed in the Higgs searches via new results
from searches, more precise measurements of the properties of the Higgs boson
at $125\gev$ and further information, for instance from flavor physics.

\subsection*{Acknowledgements}

We thank
V.~Mart{\' i}n-Lozano for collaboration in the early stages of this work,
as well as F.~Campello
and
M.~Kado
for helpful discussions.
The work of S.~H.\ is supported in part by the
Spanish Agencia Estatal de Investigaci{\' o}n (AEI) and the EU Fondo Europeo de
Desarrollo Regional (FEDER) through the project FPA2016-78645-P and in part by
the “Spanish Red Consolider MultiDark” FPA2017-90566-REDC, in
part by the MEINCOP Spain under contract FPA2016-78022-P and in part by
the AEI through the grant IFT Centro de Excelencia Severo Ochoa SEV-2016-0597.
We acknowledge support by the Deutsche Forschungsgemeinschaft
(DFG, German Research Foundation) under Germany's Excellence
Strategy -- EXC 2121 ``Quantum Universe'' – 390833306.


\appendix

\section{Results for other Yukawa types}

\subsection{The excesses at $400\gev$ in type I--IV}
\label{n2hdmstrategy}
In the N2HDM with a
CP-odd Higgs boson $A$ at $400\gev$
the following requirements need to be fulfilled
in order to describe the observed excesses at about $400 \gev$:
\begin{description}
\item[\boldmath{$\hspace*{0.75cm}\quad t \bar t$}-excess:]
    $|c_{A t \bar t}| \gtrsim 0.5$
\item[\boldmath{$\quad \tautau$}-excess:]
    $|c_{A \tautau}| \gg |c_{A t \bar t}|$ to
    get sizable $\br(A \rightarrow \tautau)$
\item[{\color{white} \boldmath{$\quad \tautau$}-excess:}]
    $|c_{A b \bar b}| \gg |c_{A t \bar t}|$ to
    get $\sigma(b \bar b \rightarrow A) \gtrsim
        \sigma(gg \rightarrow A)$ .
\end{description}
Here $c_{A f \bar f}$ are the effective coupling
coefficients of $A$ to the SM fermions, 
which are defined
as the coupling of $A$ relative to
the one of a hypothetical SM Higgs boson of
the same mass,
see \refse{sec:excesses}.
In the N2HDM, the coefficients
are given depending on the type either by
$\tan\beta$ or $1 / \tan\beta$, as depicted
in \refta{catttable}.
Taking into account that for all four types
$|c_{A t \bar t}| = 1 / \tan\beta$ holds,
one can conclude that regarding the $t \bar t$ excess
all four types are equivalent.
A sufficiently large coupling
$|c_{A t \bar t}| \approx 1$ is obtained for values of $\tan\beta$ not
much larger than one.

\begin{table}[b]
\centering
\def\arraystretch{1.5}
\footnotesize
\begin{tabular}{l|ccc}
Yukawa type & $|c_{A t \bar t}|$ & $|c_{A \tautau}|$ & $|c_{A b \bar b}|$ \\
\hline
I & $1/\tan\beta$ & $1/\tan\beta$ & $1/\tan\beta$ \\
II & $1/\tan\beta$ & $\tan\beta$ & $\tan\beta$ \\
III & $1/\tan\beta$ & $\tan\beta$ & $1/\tan\beta$ \\
IV & $1/\tan\beta$ & $1/\tan\beta$ & $\tan\beta$
\end{tabular}
\caption{\small
Effective coupling coefficients of the CP-odd Higgs boson $A$
to top quarks, bottom quarks and $\tau$-leptons.}
\label{catttable}
\end{table}

The situation is different regarding the $\tautau$ excess.
Here, also the couplings of $A$ to bottom quarks and, obviously,
$\tau$ leptons are important.
In the following, we will separately discuss for each Yukawa
type whether it could realize the
$\tautau$ excess. We will also distinguish between
{the low $\tan\beta$ region,
$\tan\beta \approx 1$,
in which a simultaneous explanation
of the $t \bar t$ excess would be possible,
and the large $\tan\beta$ region,
$\tan\beta \approx 10$.

In the N2HDM type~I, the $\tautau$ excess
cannot be realized 
for any
value of $\tan\beta$. The reason is twofold: Firstly,
accommodating the observed pattern of
the excess requires that the $b \bar b$ associated
production cross section ${\sigma(b \bar b \rightarrow A)}$
is roughly of the same size or larger than
the $gg$ production cross section ${\sigma(g g \rightarrow A)}$.
The latter is mainly mediated via the diagram
with the top quark in the loop.
Due to the fact that the Yukawa coupling $Y_t$ is an order
of magnitude larger than $Y_b$, the condition
${\sigma(b \bar b \rightarrow A) \gtrsim \sigma(gg \rightarrow A)}$
requires that $|c_{A b \bar b}| \gg |c_{A t \bar t}|$,
see above. However, in type~I
$c_{A b \bar b} = c_{A t \bar t}$ holds.
In addition, $\br(A \rightarrow \tautau)$
is always tiny in type~I, because
$c_{A t \bar t} = c_{A \tautau}$,
and for $m_A = 400\gev$
the decay $A \rightarrow t \bar t$ is kinematically open.
Moreover, it was shown in \citere{Biekotter:2019kde}
that the excesses at $96\gev$ cannot be
realized in type~I. Thus, the only excess that can
be accommodated in type~I is the $t \bar t$ excess.

In type~II the situation is different. Here 
one has
${|c_{A b \bar b} / c_{A t \bar t}| = \tan^2\beta}$.
Thus, the suppression of $\sigma(b \bar b \rightarrow A)$ compared to 
$\sigma(gg \rightarrow A)$ due to $Y_b \ll Y_t$ 
can be compensated for $\tan\beta \approx 10$, so that a possible excess 
can occur also for the $b \bar b$ production mode.
Moreover,
${|c_{A \tautau} / c_{A t \bar t}| = \tan^2\beta}$
holds, and therefore
$\br(A \rightarrow \tautau)$
can be sizable due to the enhancement of $c_{A \tautau}$
while $\br(A \rightarrow t \bar t)$ is suppressed.
The question whether the $\tautau$ excess
can be correctly reproduced depends on whether there is
a range of $\tan\beta$ in which both the enhancement
of $\sigma(b \bar b \rightarrow A)$ and the one of
$\br(A \rightarrow \tautau)$ are of a size such
that the rates in both production modes are
compatible with the observed patterns.
This issue will be addressed in our scan together with the
question whether
both the $t \bar t$
and the $\tautau$ excess can be accommodated
simultaneously. 
Since 
it is possible in type~II to realize
the excesses at about $96\gev$~\cite{Biekotter:2019kde},
we will subsequently perform also a
dedicated scan with $m_A = 400\gev$ and $m_{h_1}
\approx 96\gev$ for this type.

In type~III the $\tautau$ excess cannot be accommodated
for any value of $\tan\beta$, because
${c_{A t \bar t} = c_{A b \bar b}}$,
so that the requirement
$\sigma(b \bar b \rightarrow A) \gtrsim \sigma(gg \rightarrow A)$
cannot be realized (as in type~I).
Also the excesses at $96\gev$ cannot be
accommodated in type~III~\cite{Biekotter:2019kde}.
Consequently, only the $t \bar t$ excess can be
explained in type~III.

Also in type~IV the $\tautau$ excess cannot
be realized. The coupling coefficient
$c_{A \tautau}$ scales with the inverse
of $\tan\beta$, such that $\tan\beta < 1$ would
be required to enhance $\br(A \rightarrow \tautau)$.
On the contrary, in analogy to the argument
given for type~II, the condition
${\sigma(b \bar b \rightarrow A) \gtrsim \sigma(gg \rightarrow A)}$
requires values of $\tan\beta \approx 10$. Thus,
depending on the value of $\tan\beta$,
either $\sigma(b \bar b \rightarrow A)$
or $\br(A \rightarrow \tautau)$ is too small.
Since type~IV is capable of explaining
the excesses at $96\gev$~\cite{Biekotter:2019kde},
there is the possibility to accommodate both
excesses at $96\gev$ together with the
$t \bar t$ excess at $400\gev$ in type~IV.

As a consequence of the above considerations we restrict our parameter
scans investigating exclusively
the excesses at $400 \gev$
in the N2HDM to type~II. For the other three types the
$\tautau$ excess cannot be fitted, and the discussion of the
$t \bar t$ excess would be qualitatively very similar to the case of type~II
(the only difference that could arise is that the
total width of $A$ can be different in type~I, III
and IV compared to type~II, leading to slightly
different preferred $\tan\beta$ values due to
different best-fit values of $c_{A t \bar t}$,
see \refse{sec:excesses}).
The results of our scan for 
type~II of the N2HDM 
are described in \refse{fullII}.
Concerning the question whether there is the
possibility of fitting both the $t \bar t$ excess and the 
$\tautau$ excess at $400\gev$ simultaneously,
based on our qualitative discussion above
we can already anticipate that there will be a tension
between the $\tan\beta$ values required to fit either
of them. Indeed, we find that
the excesses point towards different
regions of parameter space with either
$\tan\beta \lesssim 3$ for the $t \bar t$ excess
or $\tan\beta \gtrsim 6$ for the
$\tautau$ excess.

In a second step, we will analyze whether the excesses at about $96\gev$
can be realized in combination with either of the
excesses at about $400\gev$. 
Thus, we perform two different scans
with $\tan\beta_{\rm low} = [0.5,4]$ 
(regarding the $t \bar t$ excess at $400\gev$)
and $\tan\beta_{\rm high} = [6,12]$ 
(regarding the $\tautau$ excess at $400\gev$).
The results are detailed in \refse{numlowII}
and \refse{numhighII}. As discussed above, in type~IV one can
potentially fit the excesses at $96\gev$ in
combination with the $t \bar t$ excess for
$\tan\beta = \tan\beta_{\rm low}$.
A corresponding parameter scan is described
in \refap{numlowIV}.

\subsection{Higgs bosons at
\texorpdfstring{\boldmath{$96\gev$}}{96gev} and
\texorpdfstring{\boldmath{$400\gev$}}{400gev}
for low \texorpdfstring{\boldmath{$\tan\beta$}}{tb}
in type~IV}
\label{numlowIV}

We present here the results
for a
scan in the low $\tan\beta$ regime
in the type~IV (flipped) N2HDM.
The input parameters were chosen to be identical
to the scan in type~II as shown in \refta{n2hdmlowtbparas}.
We show the results of the scan in type~IV
in the $\tan\beta$--$c_{A t \bar t}$ plane in
\reffi{figttIVtt} and in the
$\mu_{\rm LEP}$--$\mu_{\rm CMS}$ plane in
\reffi{figttIV}.

\begin{figure}
\centering
\includegraphics[width=0.44\textwidth]{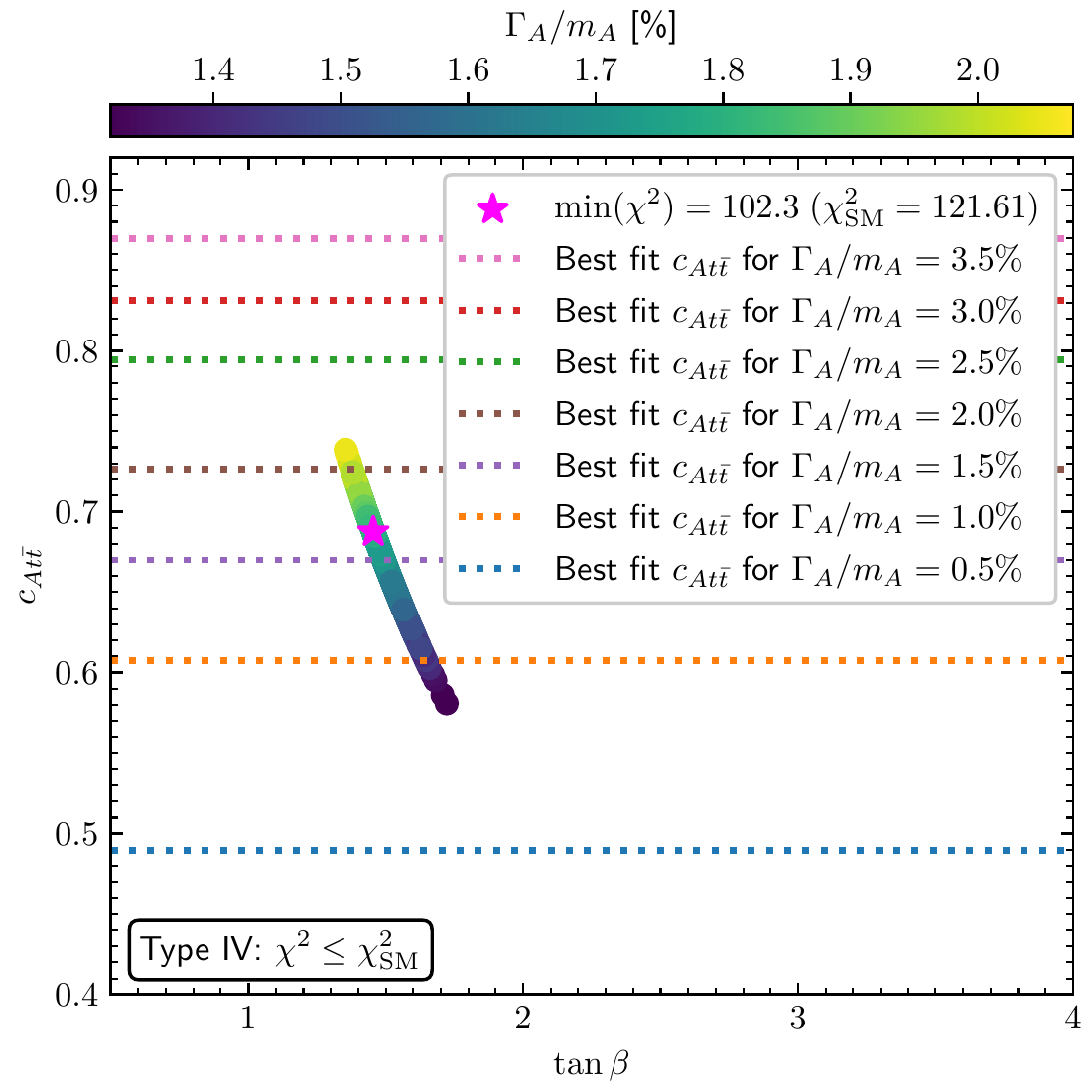}~
\includegraphics[width=0.44\textwidth]{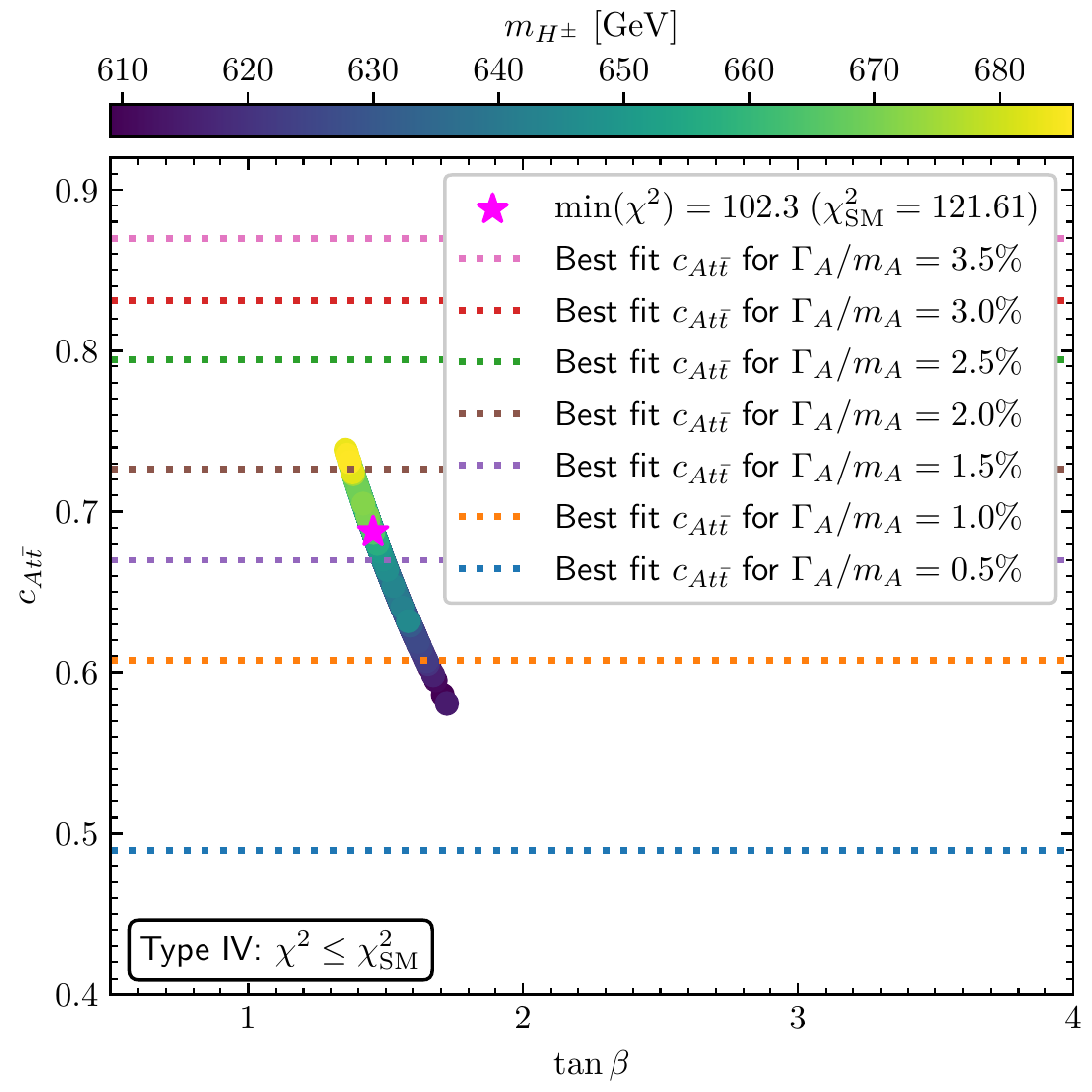}
\caption{\small
$c_{A t \bar t}$ in dependence of $\tan\beta$.
The colors of the points indicate the values
of $\Gamma_A / m_A$ in \% (left) and
the values of $m_{H^\pm}$ (right). The dashed horizontal
lines indicate the best-fit values of $c_{A t \bar t}$
for different width hypotheses in the
experimental analysis~\cite{Sirunyan:2019wph}.}
\label{figttIVtt}
\end{figure}

Comparing \reffi{figttIVtt} with
the corresponding type~II results
shown in \reffi{figttIItt}, one can see
that the values of $c_{A t \bar t}$ 
found in the scan
are considerably smaller
in type~IV. 
This is related to the smaller
predicted widths
of $A$ for the parameter points that
fulfill $\chi^2 \leq \chi^2_{\rm SM}$
in type~IV compared to type~II, as indicated
in the left plot of
\reffi{figttIVtt}. The experimental result
shows a less pronounced excess for
$\Gamma_A / m_A \approx 1.5\%\dots 2.0\%$~\cite{Sirunyan:2019wph},
and the minimum of
$\chi^2_{t \bar t}$
is shifted towards smaller couplings
(see also \reffi{Chisqtt}),
such that points with low values of
$\chi^2_{t \bar t}$
require smaller couplings. The smaller values of the
width $\Gamma_A$ in type~IV compared to type~II have their
origin in the modified leptonic couplings.
For values of $\tan\beta > 1$ the decay width for the decay
$A \rightarrow \tautau$ is suppressed
in type~IV, while it is
enhanced in type~II (see also \refta{catttable}).
In the right plot of \reffi{figttIVtt}, in which
the color coding indicates the value of $m_{H^\pm}$,
one can observe that $m_{H^\pm}$ is substantially
smaller than the upper limit of the scan range
$1\tev$ for all the parameter points.
The same result was found already in \refse{numlowII}
for type~II. The reason for the preference for
relatively small values of $m_{H^\pm}$ lies again in the theoretical
constraints on the absolute values of the quartic
couplings 
(where $m_A$ is kept fixed at $400\gev$), which are independent of
the Yukawa type.

Comparing \reffi{figttIV} to the results of type~II
(see \reffi{figttII}), one can see that the points
are substantially shifted towards lower values
of $\mu_{\rm CMS}$. As explained in
\citere{Biekotter:2019kde},
in type~IV $\br(h_1 \rightarrow \tautau)$
is enhanced in the
parameter space in which the CMS excess can
be accommodated, giving rise to smaller
values of $\br(h_1 \to \gamma\gamma)$.
While it is not possible to reach $\chi^2_{96}$ values close to zero,
corresponding to the center of the displayed ellipse,
several
points lie within the $1\sigma$ ellipse
of $\chi^2_{96}$ while in addition accommodating
the $t \bar t$ excess. This is indicated by the low
values of $\chi^2_{t \bar t}$ shown in the
left plot of \reffi{figttIV}.
The lowest values of $\chi^2_{t \bar t}$
are at the level of $\approx 1.1$. This is slightly larger than
the lowest values found in type~II ($\approx 0.25$), as shown in
\reffi{figttII}. This difference has its origin in
the smaller values of the width of $A$ in type~IV, which 
yield a worse fit to the data, as can be seen
in \reffi{Chisqtt}.
Taking the issues mentioned above into account,
we conclude that there is a slight tension
between fitting the CMS excess at around $96\gev$
and the $t \bar t$ excess at around $400\gev$
in type~IV,
but a realization of all three excess is possible
at the $1\sigma$ level of $\chi^2_{96}$.

\begin{figure}
\centering
\includegraphics[width=0.48\textwidth]{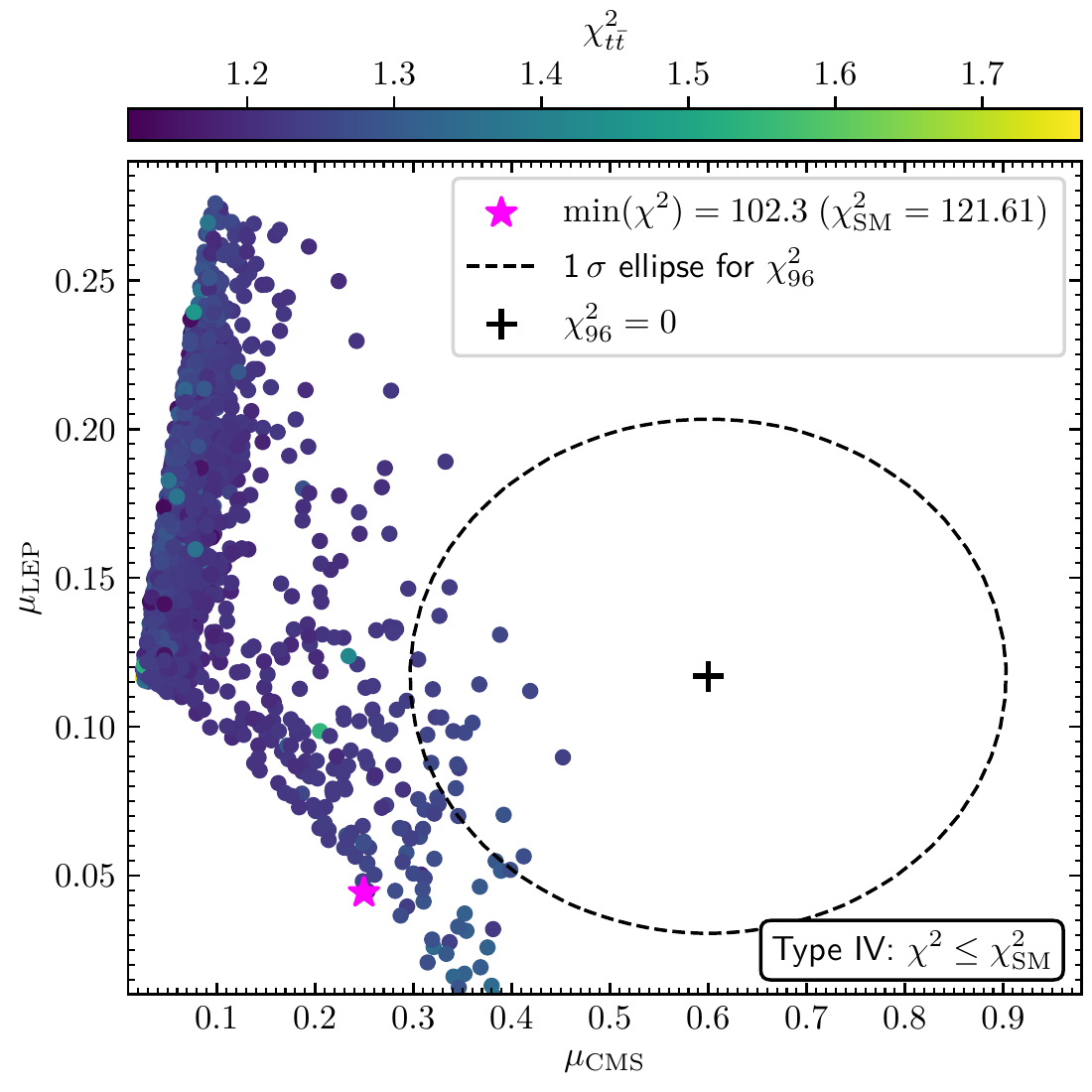}~
\includegraphics[width=0.48\textwidth]{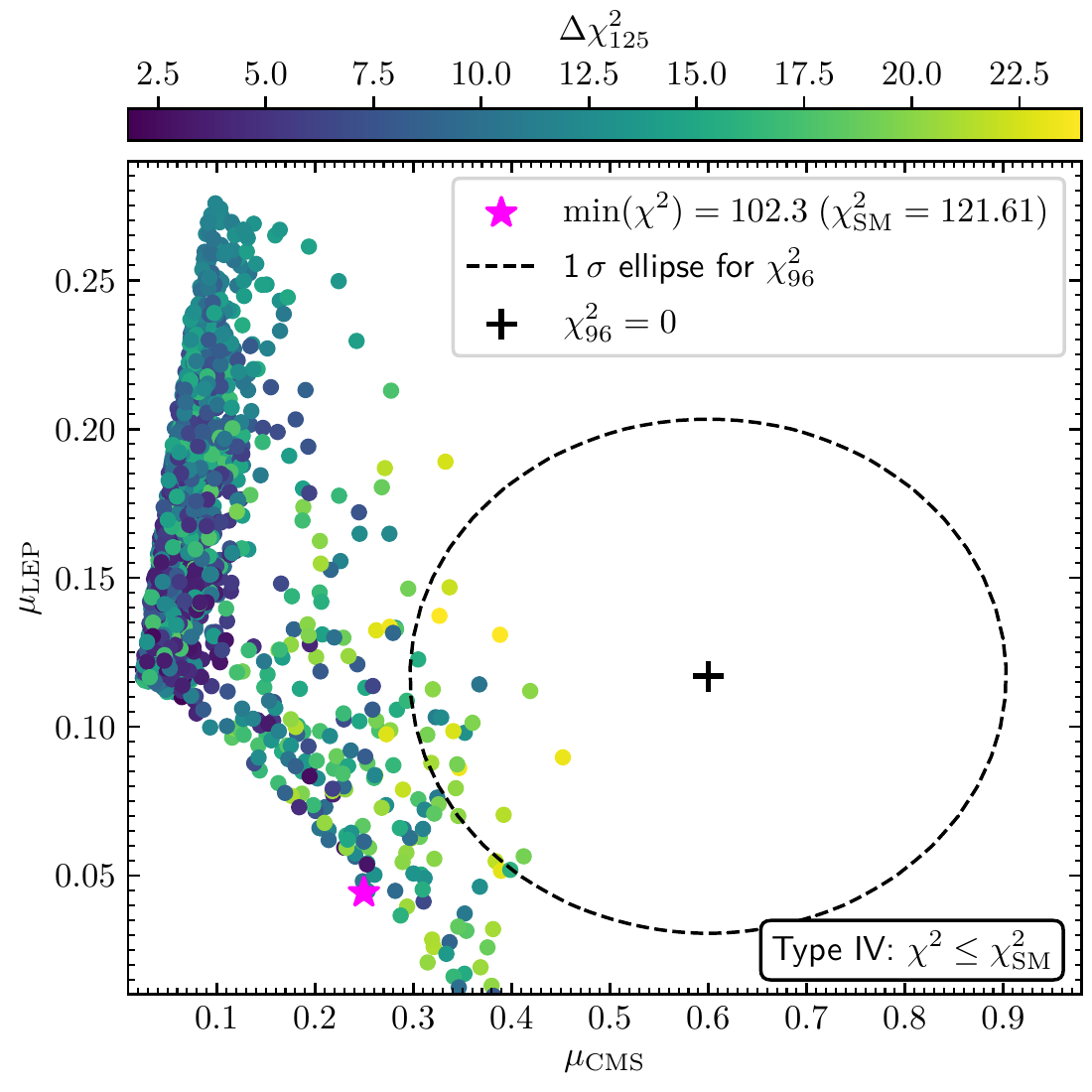}
\caption{\small The $\mu_{\rm CMS}$--$\mu_{\rm LEP}$ plane
for the points of the low $\tan\beta$ scan in the
type~IV of the N2HDM.
The black ellipse indicates the $1\sigma$ region of
$\chi^2_{96}$ with its center marked with
a black cross.
The best-fit point is highlighted with
a magenta star.
The colors of the points indicate
$\chi^2_{t \bar t}$ in the left
plot and $\Delta \chi^2_{125}$ in the right plot.}
\label{figttIV}
\end{figure}

Comparing the right plot
of \reffi{figttIV} to the one in \reffi{figttII},
one can see that the 
lowest values of $\Delta\chi^2_{125}$
found in the scan for type~IV are larger than the
ones for type~II.
This suggests that larger deviations of the
properties of the Higgs boson at $\approx 125\gev$
compared to the SM prediction can be expected
in type~IV.
The higher values for $\chi^2_{125}$
in combination with a worse fit to
the CMS excess also manifest themselves in the $\chi^2$ value
of the best fit point $\chi^2 = 102.3$, which
is substantially larger than the corresponding
value $\chi^2 = 97.93$ found in the type~II analysis.
Consequently, since
in both the type~II and the type~IV
analysis the condition $\chi^2 \leq \chi^2_{\rm SM}$
was required to be fulfilled for each point,
the total number of points within and near
the $1\sigma$ ellipse for $\chi^2_{96}$ in
\reffi{figttIV} compared to \reffi{figttII}
is substantially smaller.
In summary, our analysis in this section has revealed 
that for the low $\tan\beta$ region of the N2HDM 
type~II provides a better description of the observed data than type~IV.

\section{\texorpdfstring{\boldmath{$h_{125}$}}{h125}
in the NMSSM alignment-without-decoupling limit}
\label{secnmssmh125}
\begin{figure}[t]
\centering
\includegraphics[width=0.48\textwidth]{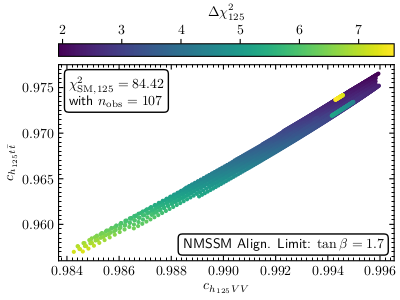}~
\includegraphics[width=0.48\textwidth]{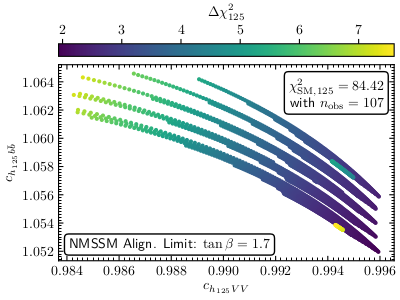}
\caption{NMSSM parameter points with
$\chi^2 \leq \chi^2_{\mathrm{SM}}$
in the
$c_{h_{125} VV}$--$c_{h_{125} t \bar t}$
plane (left) and the
$c_{h_{125} VV}$--$c_{h_{125} b \bar b}$
plane (right).
The colors of the points
indicate the values of
$\Delta\chi^2_{125}$.
}
\label{effcplshsm}
\end{figure}

In \reffi{effcplshsm} we show the effective
coupling coefficients of $h_{125}$,
with $c_{h_{125} VV}$ on the horizontal
axis and $c_{h_{125} t \bar t}$ and
$c_{h_{125} b \bar b}$ on the vertical
axis in the left and right plot, respectively.
One can see that, as 
a result of
solving \refeq{aligncond2}, 
$c_{h_{125} VV} \approx 1$ holds for
all the points, with deviations up to the 
level of $1.7\%$.
In addition, also
the couplings to the SM quarks are quite close to the
SM prediction, as a consequence of solving \refeq{aligncond1}
for $\lambda$. The deviations for both
$c_{h_{125} t \bar t}$ and
$c_{h_{125} b \bar b}$ are
at most of the size of $\approx 4$--$6\%$.
Consequently, most of the points have values
of $\Delta \chi^2_{125}$ rather close to zero, as
indicated by the color coding of the points,
and the properties of the particle state
at $125\gev$ are 
in good agreement
with the experimental constraints given the current
uncertainties.
We find $\Delta \chi^2_{125} \approx 2$
as lowest values, see the discussion below.
One can furthermore see two small clusters of
points 
at $c_{h_{125} VV} \approx 0.994$
for which the values of $\Delta \chi^2_{125}$
are considerably larger compared to
the surrounding points. These clusters arise from additional
decay channels of the state $h_{125}$, as will be discussed below,
and they appear distinctively in
\reffi{effcplshsm} because they correspond to the two
isolated points (lines of points) at the lower end
of $\kappa$ in the left (right) plot of \reffi{paraspace17}.
As mentioned before, a point is considered
to be allowed when the
condition $\chi^2 \leq \chi^2_{\rm SM}$
is fulfilled.
Thus, for points with $\Delta \chi^2_{125}$
considerably above zero, it is expected
that they 
provide predictions describing
at least one of the
excesses, such that the penalty of $\chi^2_{125}$
within the total $\chi^2$ is compensated
by either $\chi^2_{t \bar t} <
\chi^2_{\mathrm{SM},t \bar t}$,
$\chi^2_{96} < \chi^2_{\mathrm{SM},96}$, or both.

In order to 
further elucidate our results for $\Delta \chi^2_{125}$ and to
determine the origin of the larger
values of $\Delta \chi^2_{125}$ in the two
clusters mentioned
above, we show in \reffi{brs17} a selection of $h_{125}$ branching
ratios. On the left-hand
side, one can see that both $\br(h_{125} \to WW^*)$ and
$\br(h_{125} \to \gamma \gamma)$ agree with
the SM prediction (indicated by the
blue dashed lines) at the level of $\approx 5$--$10\%$.
For all the points the value of $\br(h_{125} \to WW^*)$
is slightly below the SM prediction. This has two reasons:
firstly, the coupling coefficient
$c_{h_{125} VV}$ ranges between
$0.983$ 
and $0.997$, as can be seen
in the left plot of \reffi{effcplshsm}, such
that a
corresponding suppression
is also expected for the
partial widths of the decays to $WW^*$ and $ZZ^*$.
Secondly, as shown in the right plot of
\reffi{effcplshsm}, the coefficients $c_{h_{125}b \bar b}$
are a few percent above the SM prediction
for all of the points. This leads
to a small enhancement of the partial decay width
to $b$~quarks, and therefore also
to an enhancement of the total width of
$h_{125}$, which in turn yields a further suppression of
$\br(h_{125} \to WW^*)$. Overall, the small deviations of
the branching ratios for $h_{125} \to VV^*,\gamma\gamma$
compared to the SM predictions give
rise to a penalty of $\Delta \chi^2_{125} \approx 2$--$6$.
The two isolated clusters of points mentioned above 
are also well visible in the left plot of \reffi{brs17}.
They have
an even slightly smaller BR for the decay to $WW^*$
than the rest of the points. This can be
understood from the right plot of \reffi{brs17}, where we show the
plane $\br(h_{125} \to \neu1\neu1)$--$\br(h_{125} \to \tautau)$ with
the color coding indicating again $\De\chi^2_{125}$. The blue
dotted lines indicate the SM prediction. Most of the points are found
at $\br(h_{125} \to \neu1\neu1) = 0$. However,
the two isolated clusters of points yield
a neutralino with a mass
$m_{\widetilde{\chi}^0_{1}} \lesssim 62\gev$,
such that the decay
$h_{125} \to \widetilde{\chi}^0_{1}\widetilde{\chi}^0_{1}$
becomes kinematically allowed. The corresponding partial decay width
contributes to the total decay width of
$h_{125}$ and reduces the branching ratios
for decays into SM particles. 
While for $h_{125} \to WW^*$ this results in a larger
deviation from the SM prediction, for the
decays $h_{125} \to \ga\ga, \tautau$
this brings the prediction closer to the SM value. Overall, depending on
the size of $\br(h_{125} \to \widetilde{\chi}^0_{1}\widetilde{\chi}^0_{1})$,
this yields 
values of $\Delta \chi^2_{125} \approx 5$--$8$.
In the right plot of
\reffi{brs17} we have also indicated the current upper
limit on \br($h_{125} \to {\rm inv.})$ as reported by
ATLAS~\cite{ATLAS:2020kdi}. 
For our analysis this bound does not yield a restriction since
the global experimental constraints
on the signal rates of $h_{125}$, as
tested by \texttt{HiggsSignals}, already
exclude all points potentially featuring values of
$\br(h_{125} \to \neu1\neu1) \gtrsim 0.07$.

\begin{figure}
\centering
\includegraphics[width=0.48\textwidth]{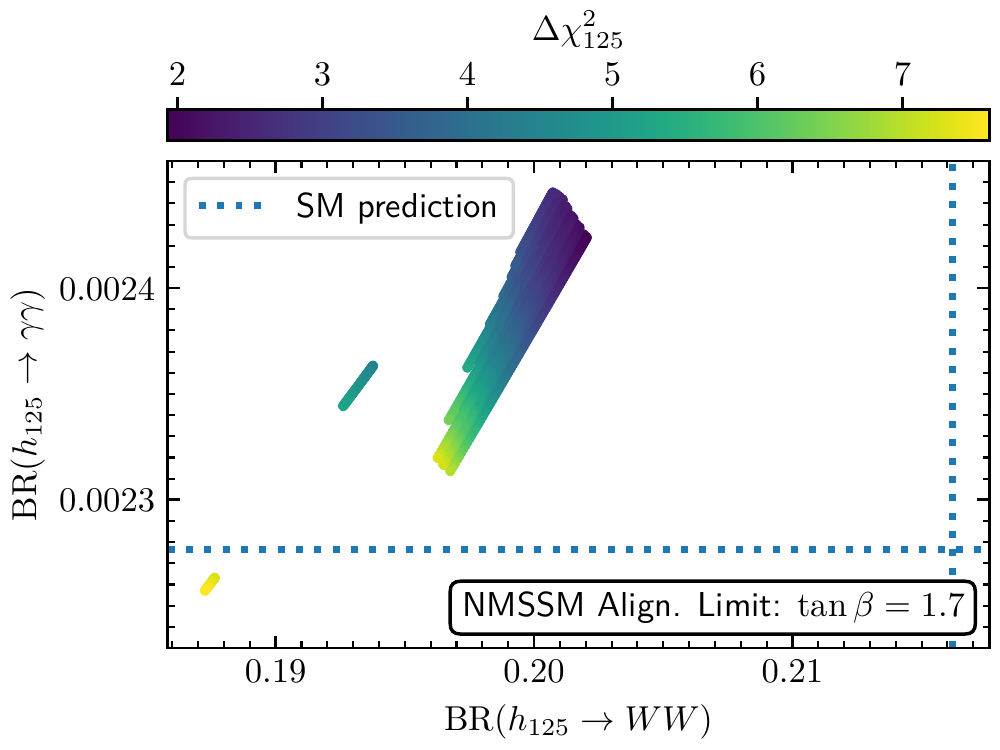}~
\includegraphics[width=0.48\textwidth]{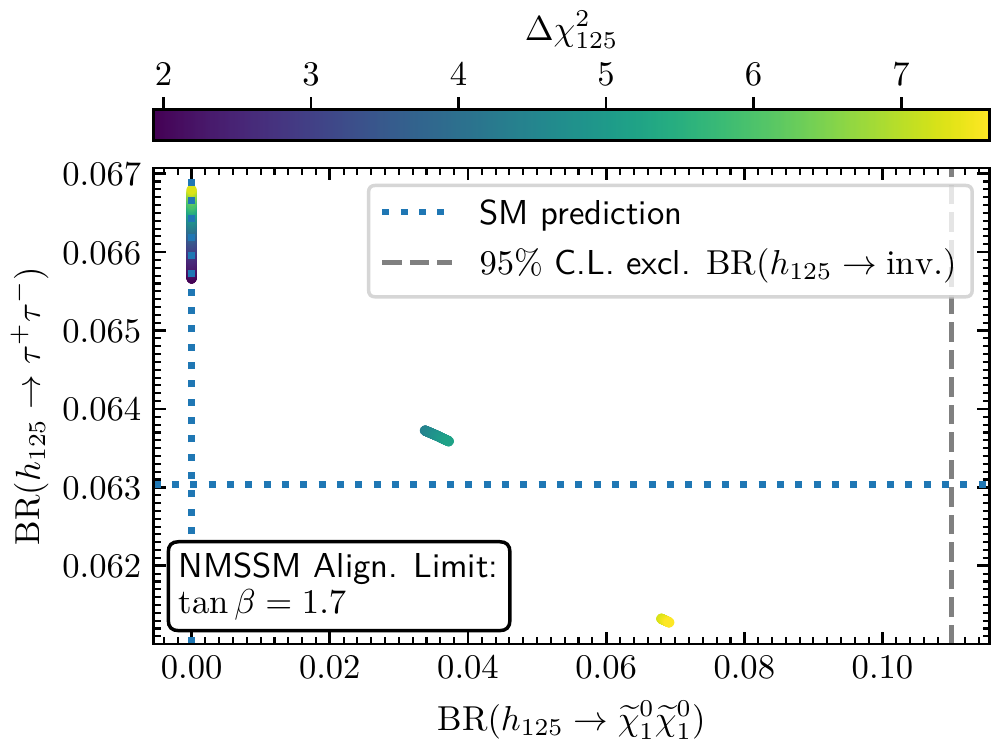}
\caption{NMSSM parameter points with
$\chi^2 \leq \chi^2_{\mathrm{SM}}$
in the plane
$\br(h_{125} \to WW^*)$--$\br(h_{125} \to \gamma \gamma)$
(left) and
$\br(h_{125} \to \widetilde{\chi}_1^0 \widetilde{\chi}_1^0)$--
$\br(h_{125} \to \tautau)$
(right).
The colors of the points
indicate the values of $\Delta\chi^2_{125}$.
The blue dashed lines indicate the SM predictions.
The gray dashed line in the right plot indicates the upper limit
for $\br(h_{125} \to \mathrm{inv.})$
at the 95\% C.L.\ reported in \citere{ATLAS:2020kdi}.
}
\label{brs17}
\end{figure}

\bibliographystyle{JHEP}
\bibliography{literature}

\end{document}